\documentclass[10pt,journal,compsoc]{IEEEtran}

\usepackage{amsmath,amsfonts}
\usepackage{graphicx}
\usepackage{array}
\usepackage{lipsum}
\usepackage{titlesec}
\usepackage{multirow}
\usepackage{tabularx}
\usepackage{ragged2e}
\usepackage{authblk}
\usepackage{soul}
\usepackage{url}
\usepackage{subcaption}

\usepackage{hyperref}
\hypersetup{hidelinks}

\usepackage{pifont}
\usepackage{footmisc}
\usepackage{xcolor}
\makeatletter
\renewcommand{\boxed}[1]{\text{\fboxsep=.2em\fbox{\m@th#1}}}

\makeatother
\DeclareUnicodeCharacter{2033}{"}
\titleformat{\paragraph}
{\normalfont\normalsize\bfseries}{\theparagraph}{1em}{}
\titlespacing*{\paragraph}
{0pt}{3.25ex plus 1ex minus .2ex}{1.5ex plus .2ex}
\begin{document}
\title{
The Metaverse: Survey, Trends, Novel Pipeline Ecosystem \& Future Directions\\
}

\author[1,*]{H. Sami}
\author[2,*]{A. Hammoud}
\author[3]{M. Arafeh}
\author[3]{M. Wazzeh}
\author[4]{S. Arisdakessian}
\author[5,8]{M. Chahoud}
\author[5,8]{O. Wehbi}
\author[5]{\newline M. Ajaj}
\author[5,6]{A. Mourad}
\author[7]{H. Otrok}
\author[4]{O. Abdel Wahab}
\author[7]{R. Mizouni}
\author[1,7]{J. Bentahar}
\author[3]{C. Talhi}
\author[2]{Z. Dziong}
\author[7]{\newline E. Damiani}
\author[8]{M. Guizani}

\affil[1]{Concordia Institute for Information Systems Engineering, Concordia University, Montréal, QC, Canada}
\affil[2]{Department of Electrical Engineering, École de Technologie Supérieure, Montréal, Canada}
\affil[3]{Department of Software and IT engineering, École de Technologie Supérieure, Montréal, Canada}
\affil[4]{Department of Computer and Software Engineering, Polytechnique Montr\'eal, Montreal, Canada}
\affil[5]{Cyber Security Systems and Applied AI Research Center, Department of CSM, Lebanese American University, Beirut, Lebanon}
\affil[6]{Division of Science, New York University, Abu Dhabi, UAE}
\affil[7]{Center of Cyber-Physical Systems (C2PS), Department of EECS, Khalifa University, Abu Dhabi, UAE}
\affil[8]{Department of ML, Mohamed Bin Zayed University of Artificial Intelligence, Abu Dhabi, UAE}

\IEEEtitleabstractindextext{%
\begin{abstract}
\justifying
The Metaverse offers a second world beyond reality, where boundaries are non-existent, and possibilities are endless through engagement and immersive experiences using the virtual reality (VR) technology. Many disciplines can benefit from the advancement of the Metaverse when accurately developed, including the fields of technology, gaming, education, art \& culture, socialization, commerce, and businesses. Nevertheless, developing the Metaverse environment to its full potential is an ambiguous task that needs proper guidance and directions. Existing surveys on the Metaverse focus only on a specific aspect and discipline of the Metaverse and lack a holistic view of the entire process. Moreover, most surveys refrain from providing detailed guidance about the development process of the metaverse, including its impact on technologies, businesses, existing challenges, and potential research directions due to their lack of a macro and micro perception of such a topic. 
To this end, a more holistic, multi-disciplinary, in-depth, and academic and industry-oriented review is required to provide a thorough study of the Metaverse development pipeline and fill the gap in existing Metaverse surveys. To address these issues, we present in this survey a novel multi-layered pipeline ecosystem composed of (1) the Metaverse computing, networking, communications and hardware infrastructure, (2) environment digitization, and (3) user interactions. For every layer, we discuss the components that detail the steps of its development. Also, for each of these components, we examine the impact of a set of enabling technologies and empowering domains (e.g., Artificial Intelligence, Security \& Privacy, Blockchain, Business, Ethics, and Social) on its advancement. In addition, we explain the importance of these technologies to support decentralization, interoperability, user experiences, interactions, and monetization. Our presented study highlights the existing challenges for each component, followed by research directions and potential solutions. To the best of our knowledge, this survey is the most comprehensive and allows users, scholars, and entrepreneurs to get an in-depth understanding of the Metaverse ecosystem to find their opportunities and potentials for contribution.
\end{abstract}

}

\maketitle

\section{Introduction}
\footnotetext{\hspace*{-0.8em}\textsuperscript{*}Equal contribution.}

\IEEEPARstart{T}he Second life is the expression used to describe what the Metaverse has to offer. The Metaverse is a shared virtual space benefiting from the new waves of technologies to gather all the aspects of the physical world into virtual entities and bring them to life \cite{wang2022survey}. It is the common space for combining all the attempts to empower virtualism, where a user represented as an avatar, can create homes or other spaces, and can interact with the virtual environment and other users \cite{oh2023social}. A verse, short of universe, refers to the virtual environment created technologies, such as virtual games or simulations. The Metaverse was first mentioned in a novel in 1992 \cite{a4} where it referred to a virtual world or verse for people to escape. Due to the advancements in Virtual Reality (VR) technology, computer vision, computing power, networking infrastructure,and communications, the term Metaverse got resurrected \cite{novak2022introducing}. With a focus on social connections and experiences, the Metaverse is empowered by the new generation of virtual and augmented reality (VR/AR) enabled through wearable devices \cite{kozinets2022immersive}. Although people are already familiar with the concept of the virtual world through some applications, the Metaverse is offering instead a shared virtual space with a combined set of applications and advanced technologies in one place. The Metaverse can benefit from the Digital Twin (DT) concept \cite{mihai2022digital}, where a virtual replica of the physical world is constructed and manipulated by reading data from sensors or Internet of Things (IoT) devices. Using DT, it is possible to understand and optimize the complexity of physical entity.

\begin{figure*}
    \centering
    \includegraphics[scale=0.22]{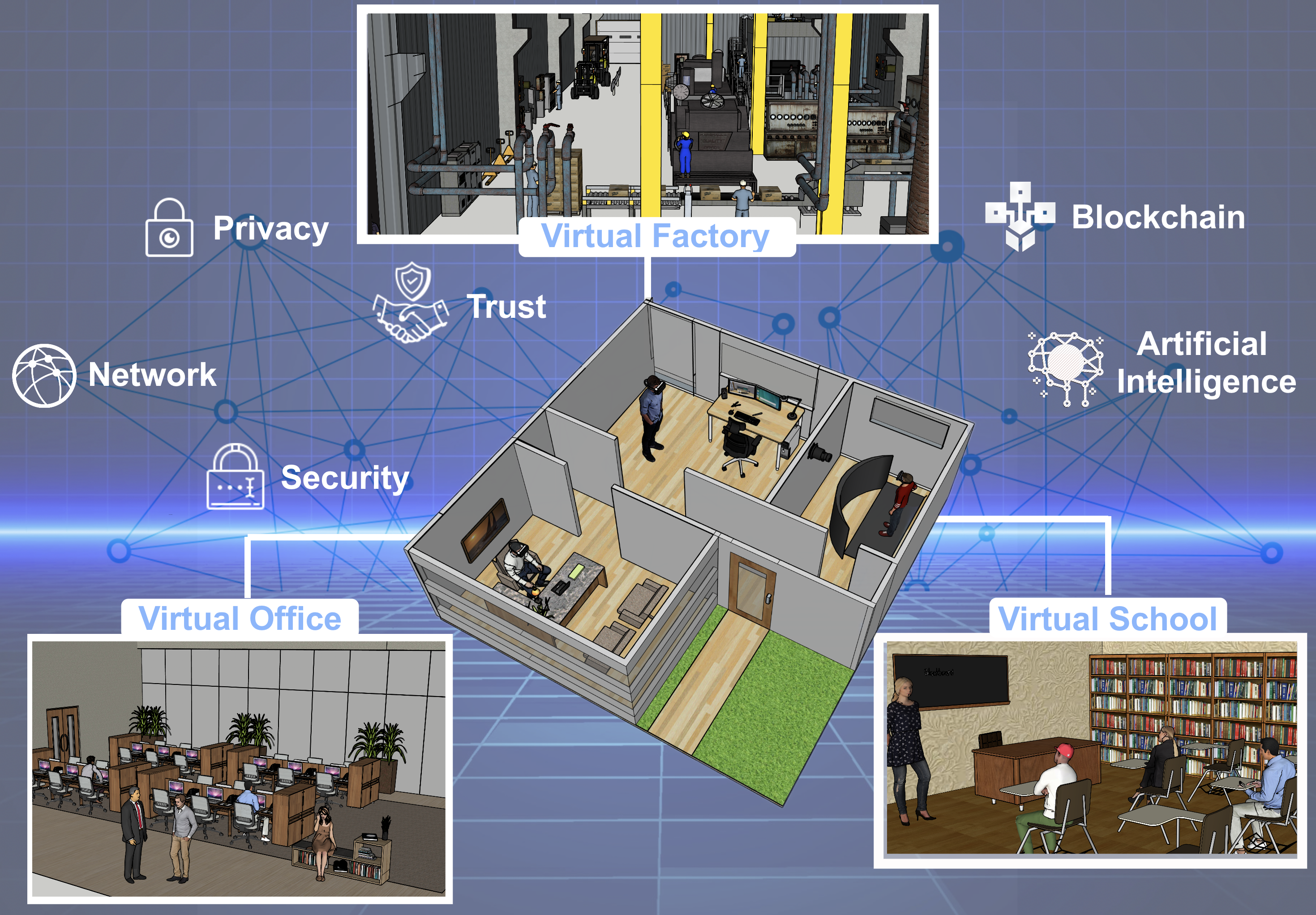}
    \caption{Multi-Disciplinary Metaverse Use-cases}
    \label{fig:intro_meta}
\end{figure*}

During the recent COVID-19 pandemic, many things could have been easier if the Metaverse had been accessible to everyone \cite{bibri2022metaverse}. Using an avatar, a user participates in different activities that mirror their preference to do things without leaving the room. Examples of practical Metaverse applications include virtual schools, teleworking in a company, or performing factory work. These environments empower higher quality education, ease of communications, and the ability to perform more complicated work in less effort and time. In Fig. \ref{fig:intro_meta}, we illustrate a view of the Metaverse experience between three people sharing the same house but engaging in different virtual environments with various objectives. 
Building virtual schools in the Metaverse is promising for the future of education by designing a digital copy or DT of schools for teachers and students to join \cite{tlili2022metaverse}. Students attend classes with instructors and colleagues virtually while being able to discover places and perform scientific experiments. This motivates students to learn and widen their capabilities through a live and engaging experience with the topic of study. Some demonstrations of use cases can be supported by visual aids with live interactions. For instance, demonstrations of virtual operations in medical schools help save time and effort for the professors and students while obtaining the needed experience \cite{garavand2022metaverse}.
In another context, a factory DT can be developed and deployed in the Metaverse using physical sensors and actuators that are bound to the factory \cite{guo2019modular}. Performing operations using the replicated factory's equipment in the virtual world is reproduced in the physical world in real-time. The verse can also be shared among employees who cannot attend the company physically and instead do their job virtually.

Despite the promising opportunities for businesses, most applications are still concepts and under development. There are many challenges to address before fully realizing the Metaverse. On the level of a single environment or entity within the Metaverse (i.e., one virtual verse), there still exist challenges in the computing and networking infrastructures that can affect the building and rendering of an environment in real-time \cite{tang2022roadmap}. Furthermore, security, privacy, and trust are the main concerns of the users to start accepting the idea of integration in the Metaverse \cite{wang2022survey}. In addition, different Metaverse environments can be built to resemble various user experiences depending on the interest and activity to perform. Subsequently, a user represented as an avatar should have access to the different virtual spaces and can interact with others and objects in real-time without delays. In order to address these requirements, the infrastructure should be ready with optimized architectures, sufficient resources, and advanced technologies. Synchronizing between various environments in a smart and secure manner is another major challenge. Thus, a clearly structured study is essential for the development pipeline of Metaverses while building and combining these environments into a globally distributed paradigm.

With many attempts to survey the current advancements in the Metaverse, a well-defined structure to develop, render, interact, and manage the content is still widely needed in the research community. Such a structure can highlight the main characteristics of the Metaverse, as well as the needs and challenges facing it in various sectors. Represented as a sequence or workflow of development tasks, the Metaverse can be defined as a pipeline ecosystem to identify each stage and its enabling domains. The literature comprises multiple Metaverse surveys in which some of them are addressing specific domains and others are multidisciplinary. The domain-specific surveys focus on the main enabling technologies of the Metaverse, such as Artificial Intelligence (AI) \cite{huynh2023artificial}, Blockchain \cite{gadekallu2022blockchain, yang2022fusing}, and wireless technologies (5G and 6G) \cite{zawish2022ai}. Other work reviews the market and business advancement in the Metaverse in various countries and the applications offered \cite{ning2021survey}. Existing Metaverse multidisciplinary surveys focus on describing the concepts solely while ignoring the coherence of the enabling technologies and their challenges of deployment \cite{dwivedi2022metaversea, lee2021all}. Besides, there is no standard of structured and well-defined realization guidelines and development workflow or pipeline ecosystem for Metaverse in any of the existing surveys. Finally, the Metaverse offers opportunities for both the academic research and business communities, thus it is important to address and discuss each perspective with regards to the Metaverse pipeline ecosystem and existing challenges. All of the aforementioned problems can slow down and hinder the realization of the Metaverse.

\subsection{Survey Contributions}
The purpose of this survey is to summarize all the efforts in a structured, well-defined, and sequential manner to allow the realization of the Metaverse by relying on experts in miscellaneous fields to engage in the process. 
Therefore, our main objective is to propose a Metaverse pipeline ecosystem that consists of several development stages with the integration of enabling technologies and empowering domains. Similar to the concept of any project development lifecycle, an environment should be developed after passing through several stages or a well-defined pipeline (i.e., workflow). Each part of that pipeline requires a combination of technologies that ensure the addressability of various user concerns and the delivery of the desired experiences. 
In this survey, we present a benchmark of a comprehensive pipeline ecosystem of Metaverse environments combined with a set of enabling technologies. Furthermore, we study what the businesses and academia offer as solutions in the current literature, and present them as test cases with state-of-the-art Metaverse-based architectures and solutions. Therefore, this survey offers for the first time a clear and well-structured realization guidelines and pipeline ecosystem for the Metaverse. In addition, a set of enabling technologies is integrated within each component of the pipeline, providing a multi-class classification of different scientific literature work and business solutions described by functionalities, applications, limitations, and future works. With this pipeline, we address most of the disciplines that are involved in the Metaverse through a holistic view of the ecosystem, as well as the possible technologies that are used or have the potential in augmenting this workflow and addressing existing Metaverse problems. The main objectives of this survey are summarized as follows:
\begin{itemize}
    \item Provide an in-depth, holistic view, multi-disciplinary, academic, and industry-related review of the Metaverse development ecosystem.
    \item Tailor this survey to offer guidelines and benefit a wide range of audience, including field experts, businesses, and Metaverse users. 
    \item Devise a multi-layered pipeline ecosystem composed of nine main stages while integrating enabling technologies, empowering domains, and social enablers within each stage.
    \item Present a multi-class classification of various literature works by reviewing the impact of each enabler on its development. 
    \item Identify new opportunities in the Metaverse within each component by deducing challenges and existing issues that are linked to research directions.
\end{itemize}

We believe that this survey will constitute a major step towards standardizing the pipeline ecosystem and the integration of various enabling and advanced technologies at each stage for realizing the Metaverse. It will also offer opportunities, well-defined guidelines, and taxonomies, for investors, business owners, and researchers in order to find their fitting spot within the booming diverse Metaverse disciplines.

\subsection{Survey Methodology and Organization}
In this survey, we collect the most recent surveys and reviews about the Metaverse in the past few years and study their contributions and limitations. In addition, most of the scientific papers that contribute to the Metaverse since 2018 are studied. Using the Metaverse keyword and the set of enabling technologies, we locate the articles using Google Scholar and Scopus. This survey contains hundreds of scientific articles that are grouped and classified depending on the contributions, disciplines, and use cases applied. Furthermore, we include a list of businesses and study their solutions offered for the Metaverse market as well as the technologies utilized. 
The methodology used in this survey is to identify the key enabling technologies and study their integration with a Metaverse pipeline ecosystem. This pipeline is sequentially ordered and organized as (1) Infrastructure, (2) Environment Digitization, and (3) User Interactions. The Infrastructure is composed of the hardware and equipment, frameworks, and platforms. Environment Digitization consists of avatar modeling, environmental rendering, and sessions. Furthermore, interactions from the user perspective are user-to-user, user-to-objects, and user-to-business in the Metaverse. For each layer and component, we devise a multi-layer classification of the existing literature and business efforts. Furthermore, we study the list of existing issues and challenges behind each component and outline the potential future research directions and solutions.

\subsection{Survey Outline}

\begin{figure}[htp!]
    \centering
    \includegraphics[width=.9\linewidth]{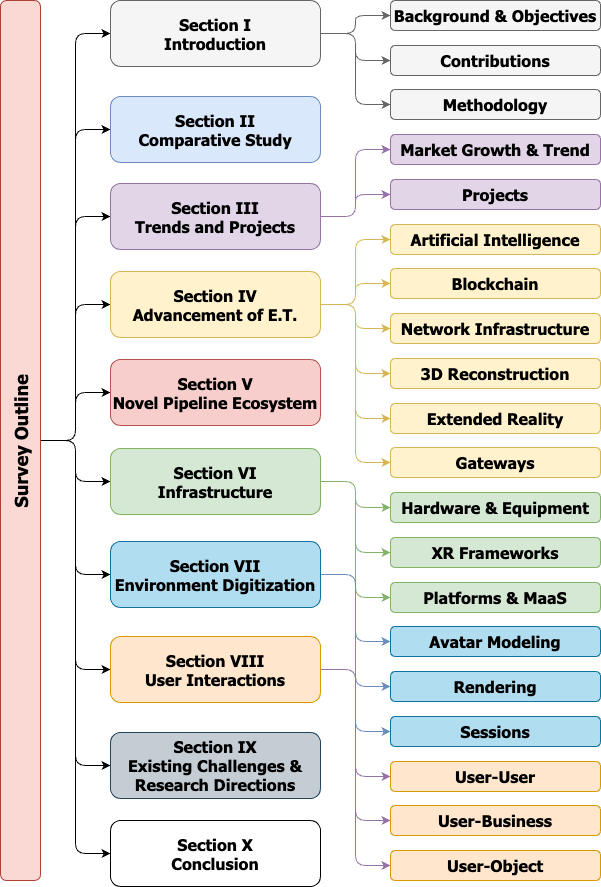}
    \caption{Survey Outline}
    \label{fig:survey_outline}
\end{figure}

The remainder of this survey is organized as follows. Section II presents the comparative studies with the list of existing literature surveys. The Metaverse trends and projects are illustrated and analyzed in Section III. 
Section IV describes the Metaverse foundation through timelines of enabling technologies. Afterward, we present a novel Metaverse pipeline ecosystem in Section V. Then, we elaborate on the main three layers of the Metaverse foundation, i.e. Infrastructure, Environment Digitization, and User Interactions, in Sections VI, VII, and VIII, respectively. Lists of challenges and future directions extracted from the discussions and analysis of each Metaverse component are presented in Section IX. Finally, we conclude the survey in Section X. A detailed organization of our survey is depicted in Figure \ref{fig:survey_outline}.

\section{Comparative Study}


In this section, we provide a review of the current Metaverse surveys. We highlight their points of discussion and analysis in relation with the necessary Metaverse components of the development pipeline from a user or entrepreneur perspective.
The Metaverse gained wide attention from its academic aspect since it became widespread. In terms of surveys, tens of diverse works were published in the last couple of years to summarize the recent advancements in this topic. Such surveys were addressing it from the perspective of various enabling technologies. Therefore, a comparison of surveys that do not address the same technologies cannot be conducted. In this context, we employ several metrics relevant to the Metaverse components and their enabling technologies. We rely on evaluating the literature work by referring to a double-edged metric system that can reveal the direction and focal points of these surveys. To further demonstrate the points of evaluation, we anticipate the following Metaverse components: (1) Hardware and Equipment, (2) Frameworks, Libraries, and Platforms, (3) Avatar and Object Modeling, (4) Environment Rendering, (5) Sessions and User Authentication, (6) User to User Interaction, (7) User to Business Interaction, and (8) User to Objects Interaction. In terms of relevant topics and enabling technologies, we use the following list to distinguish the contribution of the surveys: (A) Artificial intelligence, (B) Blockchain, (C) Networking, (D) Computing, (E) Business, (F) Privacy \& Security, (G) Ethics, and (H) Sociopsychological aspect. The list presents the backbone of the Metaverse in terms of enabling technologies as well as social enablers. From this list, AI and Blockchain are the two pillars for achieving intelligence, security, transparency, and distributed storage. Furthermore, the Computing and Networking enablers form the Metaverse infrastructure to support and host the applications and technologies. In addition, the Business enabler forms a container of companies and products in the Markets supporting and offering solutions for the Metaverse. Moreover, the Privacy \& Security are the crucial factors for the Metaverse success by protecting personal information, digital assets, and online reputation. Besides, social enablers are the Ethical and Sociopsychological implications behind developing and using the Metaverse. We adopt these components in our evaluation to demonstrate the Metaverse technologies and enablers for users and entrepreneurs who are interested in particular topics. 
\begin{figure*}[htp!]
    \centering
    \includegraphics[width=.76\linewidth]{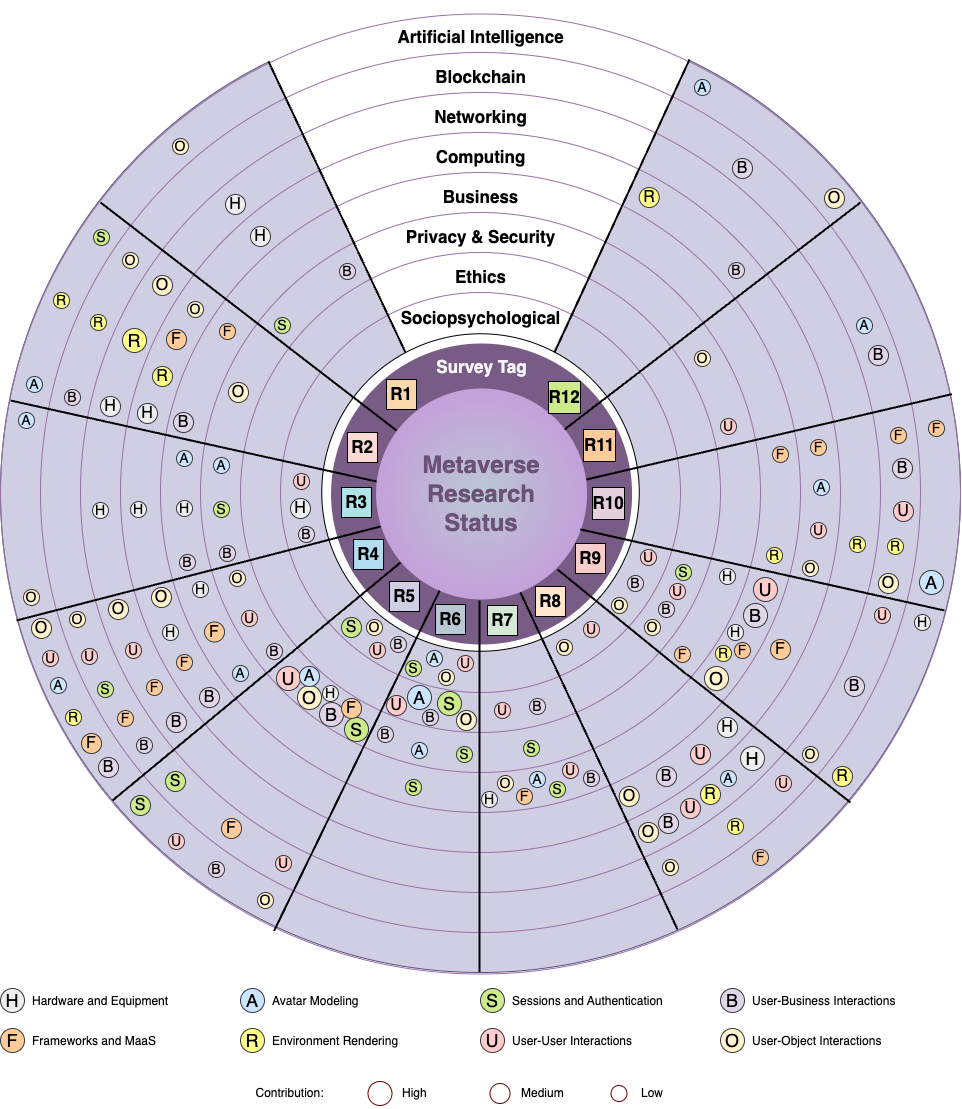}
    
    \begin{tabular}{|c|c|c|c|c|c|c|c|c|c|c|c|c|}
         \hline
         Survey Tag & \boxed{R1} & \boxed{R2} & \boxed{R3} & \boxed{R4} & \boxed{R5} & \boxed{R6} & \boxed{R7} & \boxed{R8} & \boxed{R9} & \boxed{R10} & \boxed{R11} & \boxed{R12} \\
         \hline
         Survey Reference & \cite{tang2022roadmap} 
         & \cite{xu2022full} 
         & \cite{lee2021all} 
         & \cite{huynh2023artificial} 
         & \cite{wang2022survey} 
         & \cite{bibri2022metaverse} 
         & \cite{ning2021survey} 
         & \cite{zawish2022ai} 
         & \cite{dwivedi2022metaversea} 
         & \cite{yang2022fusing} 
         & \cite{gadekallu2022blockchain} 
         & \cite{lee2021creators} \\
         \hline
    \end{tabular}
    \caption{Literature surveys' focus of metaverse: the contribution level of each survey in certain domains}
    \label{fig:galaxy}
\end{figure*}
After identifying the comparison metrics (technologies and social enablers), we carefully study each relevant Metaverse survey in the literature and highlight their focal points and depth. Figure \ref{fig:galaxy} shows in-depth feature extractions of the Metaverse surveys. Generally speaking, surveys address the Metaverse by summarizing and analyzing the literature works of how an enabling technology was able to assist/serve a certain component. To distinguish the surveys' level of contribution, we refer to the depth of discussion by pinpointing with sized circles None (\textit{no circle} .), Low (\textit{small circle} o), Medium (\textit{medium circle} 0), or High (\textit{big circle} O) to each paper regarding each combination of technology and component, as can be noticed in the figure. For instance, the work in \cite{xu2022full} offers a Metaverse discussion regarding its networking aspect, and how Blockchain can aid in its development. From a macro point of view, they decently address the Hardware and equipment component, in addition to the Rendering and User-to-objects interactions in terms of multiple enabling technologies. Furthermore, other components, such as avatar modeling, authentication, and platforms, are slightly discussed. Similarly, \cite{tang2022roadmap} focuses on communications and networking in Metaverse. Precisely, their discussion comprises sensors, DT, edge computing, and Blockchain. Most of the provided information is limited to the authors' opinions. Nevertheless, reflecting back on our evaluation mechanism, their main contribution remains limited. From a Blockchain perspective, the survey in \cite{gadekallu2022blockchain} provides a thorough discussion on Metaverse and the impact of Blockchain on other enabling technologies, including AI and the concept of DT. The discussion provides a deeper technical flavor than other surveys, however, the overall Metaverse ecosystem is not explored in their work. In parallel, a survey on security and privacy in Metaverse was formulated in \cite{wang2022survey}. The authors focus on detailing security threats in different domains and their countermeasures in the Metaverse paradigm. Nonetheless, their mature contribution was limited to the privacy and security aspects only, while the enablers of Metaverse were somehow ignored. A recent survey \cite{lee2021creators} approaches the Metaverse from a computational arts perspective to describe novel artworks in blended virtual-physical realities. The paper neglects however many important fundamentals of the Metaverse, thus, it does not qualify as an instructive survey about its enablers and advanced technologies. The survey in \cite{bibri2022metaverse} examines the practices and ethics of the Metaverse while emphasizing its privacy and sociopsychological aspects of certain components. Nevertheless, their approach has no clear study of the technologies behind Metaverse. 
AI was also addressed in several Metaverse surveys. To name a few, \cite{yang2022fusing}, \cite{zawish2022ai}, and \cite{huynh2023artificial} are reviewing the AI vision in Metaverse, with a prominent flavor. In \cite{yang2022fusing}, the authors tackled how Blockchain and AI fuse with the Metaverse. Their work considers many recent advancements in the literature. Nevertheless, they do not address other important aspects and technologies, as summarized in Figure \ref{fig:galaxy}. Secondly, the survey in \cite{zawish2022ai} focuses on the role of AI in realizing the Metaverse while considering the networking requirements. However, they lack the mature AI discussion for components other than the platform and hardware, in addition to not bearing in mind the privacy and security aspects. The work in \cite{huynh2023artificial} provides an overall review of AI inside Metaverse. Although their AI discussion is covering many aspects of the Metaverse, the information is not deep enough for a reader to comprehend how the Metaverse functions. From a multidisciplinary point of view, there are other diverse Metaverse surveys approaching Metaverse, among which stand out the following works: \cite{lee2021all,dwivedi2022metaversea,ning2021survey}. These three surveys review the state-of-the-art technologies of the Metaverse. Nonetheless, the one thing that should be highlighted about these works is also their discontinued flow of information and the ambiguity to connect the Metaverse components into one complete and structured ecosystem due to addressing each component individually.

In summary, the current surveys in the literature lack a coherent architecture for identifying its components and the role of enabling technologies for each particular component. Hence, they are still limited in terms of contribution and are not consistent with the Metaverse workflow. Therefore, a refined pipeline ecosystem for defining the components of the Metaverse should be devised and discussed in order to ease its realization and bring experts from multidisciplinary backgrounds.

\section{Metaverse Trends \& Projects in Academia and Industry: Statistical Analysis}

In this section, we highlight and discuss the trend surrounding the hype as well as the rise of Metaverse. In the past couple of years, the Metaverse trend has taken an acute upward spiral. Several factors and enabling technologies have contributed to its rise, such as (1) Faster and cheaper computing, (2) better graphics, and (3) faster internet connectivity. However, with all trendy technologies, the future of Metaverse will be decided by the amount of disruption it will cause in mainstream domains, the newer arenas and paths it will pave, as well as the number of use-cases it will facilitate. The ultimate question for any up-and-coming technology remains: Will it improve the Quality of Life (QoL) of its users? 

\subsection{Market Growth \& Trend Increase}
The Metaverse has seen tremendous market growth and trend increase in both industry and academic circles. It is worth noting that several of the industry leaders, who work largely on Metaverse-related projects, rely on academia. Some, even have their own academic research departments internally. A good example of this is Meta's internal research department \cite{meta_research} where it conducts academic research, as well as hosts Ph.D. programs such as \lq Meta Research Ph.D. Fellowship Program\rq.

\begin{figure}[htp!]
\centering
{\includegraphics[width=.8\linewidth]{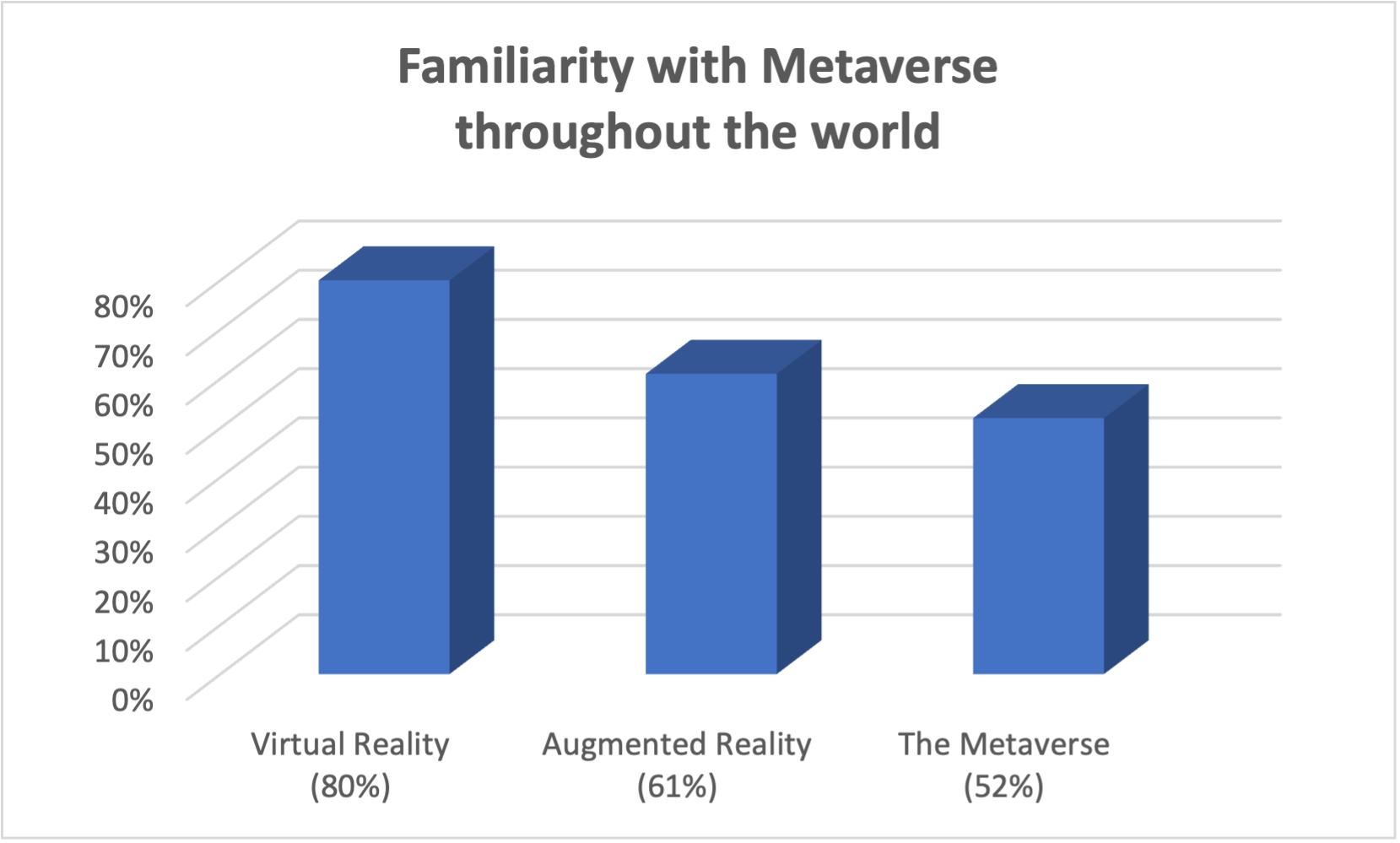}}
\caption{Metaverse Familiarity Throughout the World in 2022 \cite{ipsos_wef}}
\label{fig:metaverse_familiarity}
\end{figure}

\begin{figure}[htp!]
\centering
{\includegraphics[width=.8\linewidth]{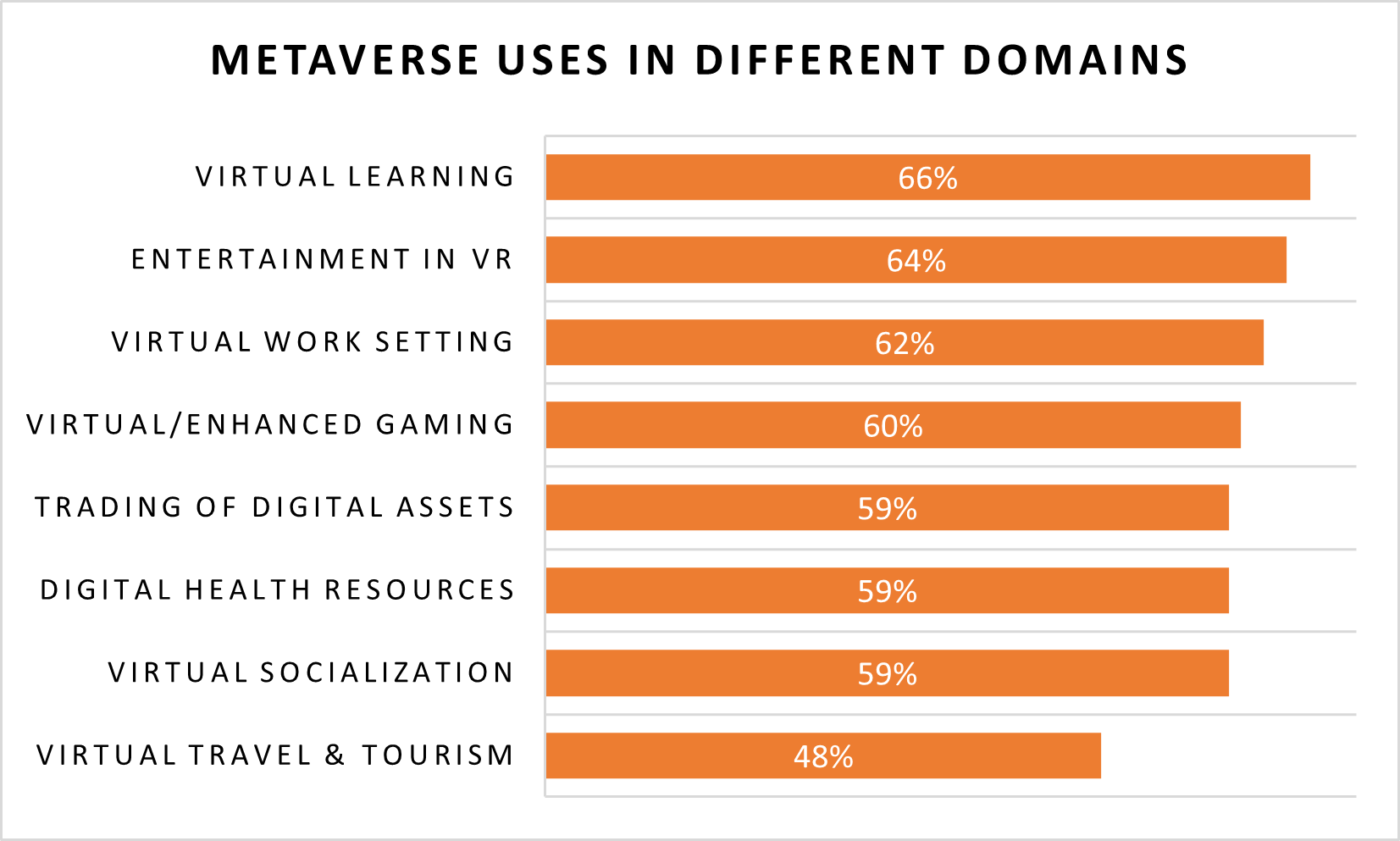}}
\caption{Human Perception of Metaverse Uses in Different Domains, May 2022 \cite{ipsos_wef}}
\label{fig:metaverse_uses_in_different_domains}
\end{figure}

With more people throughout the world getting to know about the Metaverse and getting acquainted with its details and potential, it is expected that the trends regarding the Metaverse will keep on increasing. The marketing research and consulting firm Ipsos conducted research for the World Economic Forum \cite{ipsos_wef} spanning across 29 countries in May 2022 and came up with the following. On average, 80\% of the population knew what virtual reality was, 52\% knew about the Metaverse (Figure \ref{fig:metaverse_familiarity}), and 50\% had positive feelings about dealing with and using extended reality in their everyday lives. The same research asked the participants about their perceptions of how the Metaverse and its applications will impact people's lives, and more than 50\% believed that different Metaverse applications will change people's lives in the next 10 years. Furthermore, the research highlighted the percentage of people's beliefs about different domains in which the Metaverse will play a significant role. Virtual learning had the biggest result while virtual travel and tourism had the lowest perception rate. We show the results from the research for all the domains in Figure \ref{fig:metaverse_uses_in_different_domains}. From the statistics shown in this figure, we can realize that the vision for the Metaverse is multidisciplinary, which brings more complexity and challenges to meet the users' expectations. Specifically, each of these domains requires the use of different technologies to handle, render, and offer the needed experiences.

\subsubsection{Trend Increase in Industry}

\begin{figure}[htp!]
\centering
{\includegraphics[width=.8\linewidth]{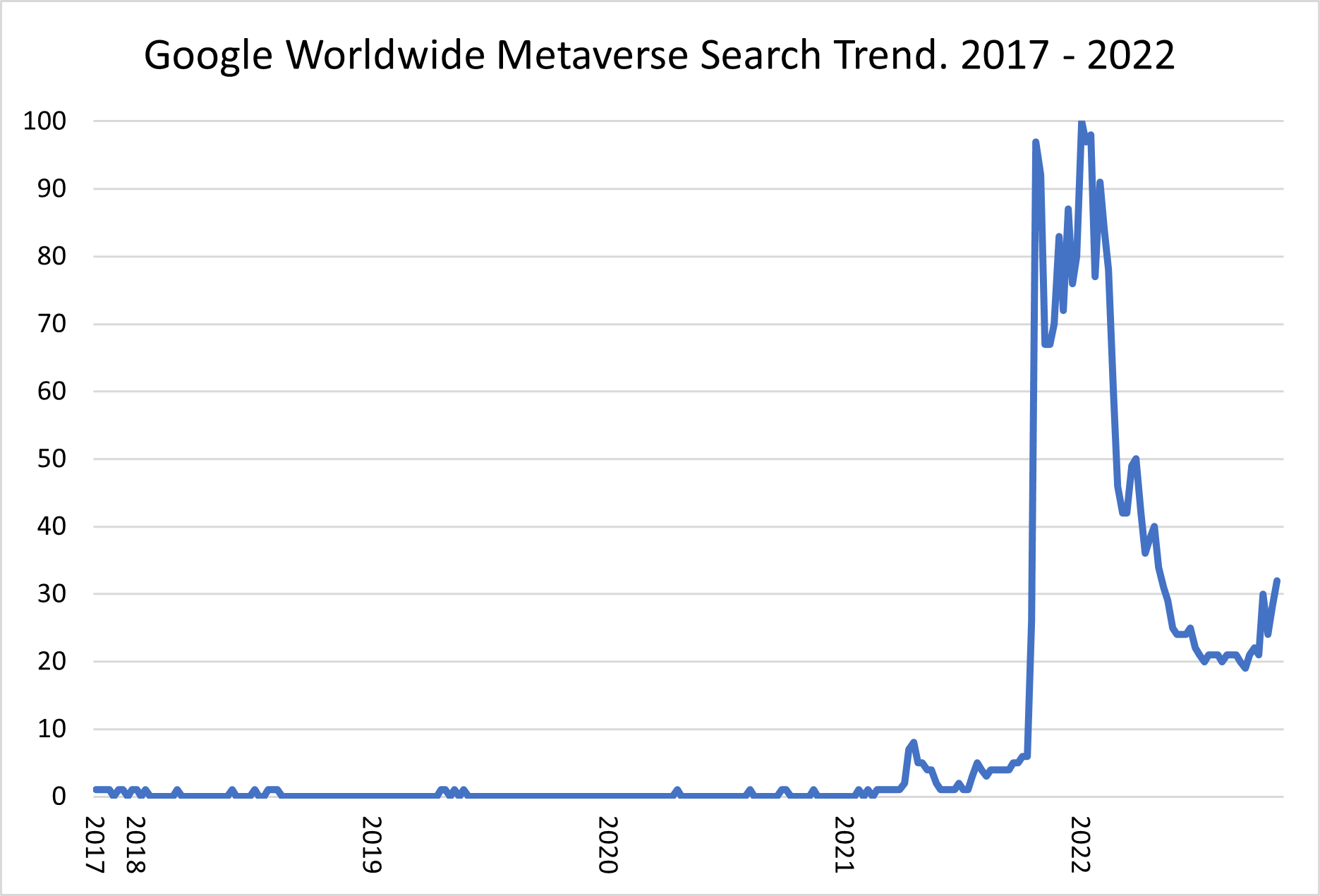}}
\caption{Google trends for the keyword Metaverse search between years 2017 and 2022}
\label{fig:google_trend_metaverse}
\end{figure}

Despite the fact that the notion of Metaverse has been around for 30 years, however, it was only in the last 10 years that it started getting commercial attention. This reached its peak in October 2021, when one of the behemoths of modern social media companies, Facebook, announced that it will rebrand as Meta and focus on Metaverse-related applications. With such a big announcement from a company that has more than 2.8 billion users, people were curious to know what the Metaverse was really about. This can be easily seen in Figure \ref{fig:google_trend_metaverse}, where in the past 5 years, interest and search for Metaverse was almost with a ranking of 0, which suddenly gets boosted to 100 (the highest possible ranking) upon the news of Facebook's new strategy. 

This does not mean that the Metaverse did not exist before, or major corporations did not take it seriously. On the contrary, the notion of Metaverse has been around for a while, especially on the gaming front. Matter of fact, the gaming sector has been the front face of Metaverse for a while and the initial rise for it mainly came from games \cite{huynh2023artificial}. Games and platforms such as Fortnite, Minecraft, and Roblox existed way before the term Metaverse was commercialized. Furthermore, Fortnite was one of the first companies that held a virtual concert in its gaming sphere, where DJ Marshmellow performed a DJ-ing set in early 2019. Other platforms followed where major artists performed virtually in different Metaverse platforms put forward by different companies \cite{concerts}.

Given Facebook's nature as a social platform, their goal was to create social universes through 'Meta Horizons` where users could join using special headsets. These headsets are produced by the company and allow users to hang out with other users, meet new people, play games with each other, and participate in events. Meta's new strategy also attempts to change what a normal day at the office looks like. A more corporate and business aspect of their vision is geared towards a service dubbed `Meta Horizons Workrooms'. Another tech giant, Microsoft, who already has been in the Metaverse loop for a while with its very own Hololens virtual reality gear and its `Microsoft Mesh' platform for virtual reality, purchased a very large gaming company named Activision Blizzard which is the owner and the developer of several successful games such as Call of Duty, World of Warcraft, and Guitar Hero. Microsoft's purchase was a big message to the entire industry that big firms were now spending millions of dollars and betting on Metaverse's future. Microsoft CEO's message was clear regarding the acquisition and the position of one of the biggest software companies, he has been quoted saying \lq Metaverse is essentially about creating games.\rq.

To become a major player in the Metaverse space, an entity needs to be highly technical, engineering-oriented, and have interdisciplinary knowledge of the technological enablers for it. Multiple domain expertise and know-how, such as physics, hardware engineering, software engineering, graphics, and AI are required. To overcome such hurdles and make Metaverse accessible to the general public with no technical knowledge, some companies such as bit.country \footnote{https://bit.country/}, Propel \footnote{https://propel.xyz/maas}, Touchcast \footnote{https://touchcast.com/}, Metaversebooks \footnote{https://metaversebooks.com/} have started to provide services in the form of Metaverse-as-a-Service (MaaS) where users do not need to worry about deployment and technical arrangements to own their Metaverse, rather worry only about content and moderation. These services include Metaverse engine deployment, hosting services for Metaverse needs as well as technical support. We can consider such a service analogous to the cloud services presented by cloud providers where companies do not host or develop their own software or host their own servers anymore. Instead, they subscribe to and utilize the services of these SaaS and IaaS providers. From an economic point of view, a MaaS can save large amounts of money for companies and entities that do not possess the digital expertise to develop their own Metaverses. It can also allow for smaller as well as mid-sized firms to enter the scene in a fast manner.

In addition, with the advent of decentralization and Web 3.0, we are seeing more Metaverse-oriented systems and applications that are working hand in hand with other Web 3.0-based enabling technologies such as blockchains \cite{2022Cao}. For example, major commercial Metaverse platforms such as Decentraland \footnote{https://decentraland.org/}, and The Sandbox \footnote{https://www.sandbox.game/en/} facilitate the purchase of virtual lands, alongside unique virtual items such as paintings and wearable in their platform, offer their own financial tokens as a means for financial transactions and purchases \cite{wang2021non}.

The financial and media company Bloomberg expects the potential market of Metaverse to reach around \$800 billion by 2024, where online gaming, as well as AR and VR hardware production taking around 50\% of the market share. The rest of the market share will be distributed between entertainment, ads, live events, concerts, films, sports, and social media applications. This is a big surge from the market share of Metaverse from 2020 where the total market share of all Metaverse-related applications was around \$400 billion \cite{2022bloomberg}.

\subsubsection{Trend Increase in Academia \& Research}

\begin{figure}[htp!]
\centering
{\includegraphics[width=.8\linewidth]{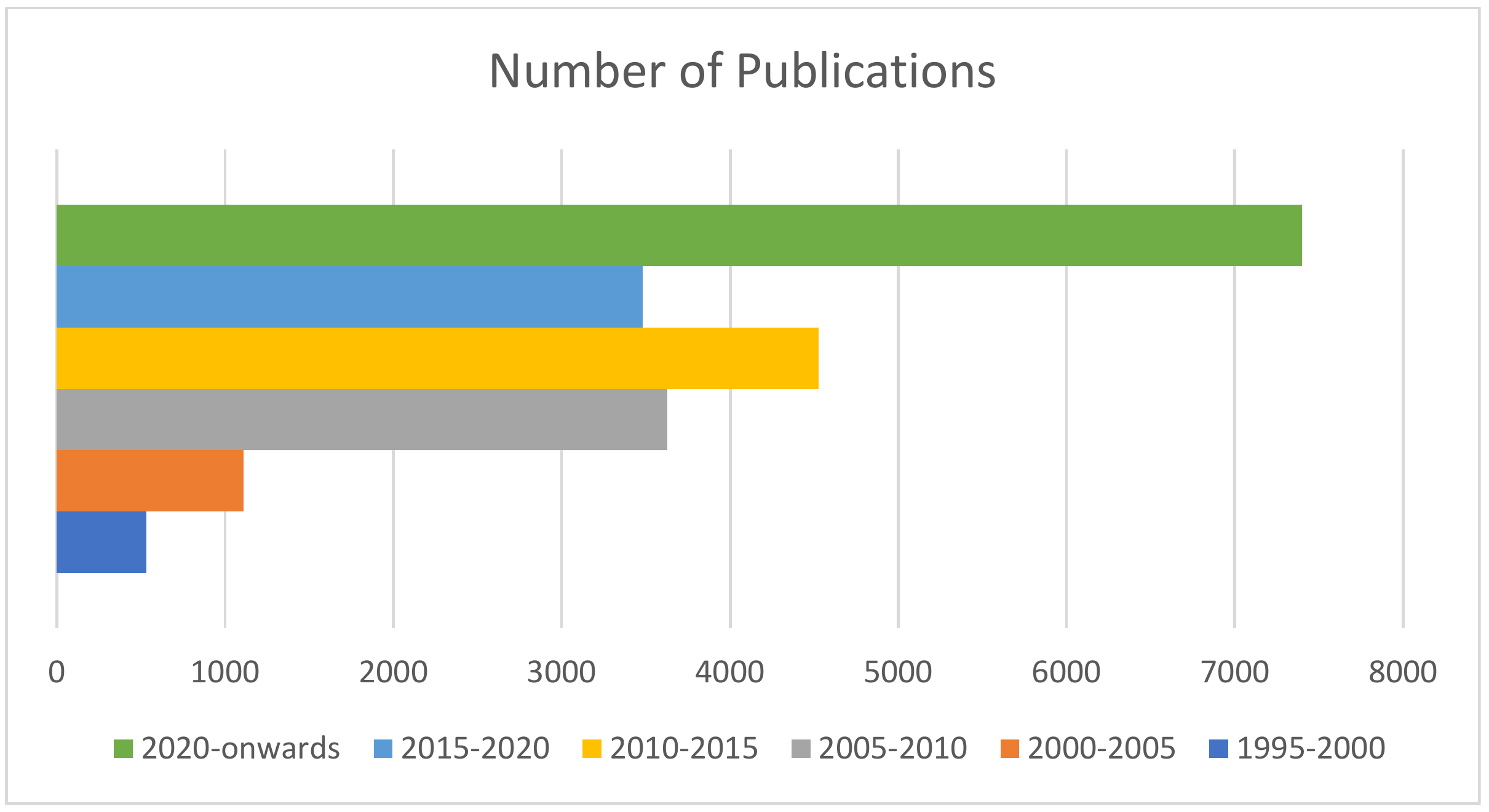}}
\caption{Google Scholar results for Metaverse keyword publications starting 1990}
\label{fig:google_scholar_metaverse}
\end{figure}

The trend for Metaverse in academia has increased remarkably too. A quick search for the keyword `Metaverse' in Google Scholar returns around 22,900 results (November 2022). What is worth noting is that the last couple of years saw a tremendous increase in the publication of academic works pertaining to the Metaverse topic. We searched for academic publications that contain the keyword `Metaverse' on Google Scholar \footnote{https://scholar.google.com/} in 5-year increments starting the year 1990 and showcase our findings in Figure \ref{fig:google_scholar_metaverse}.

We can notice from the graph in Figure \ref{fig:google_scholar_metaverse} that in only the past couple of years, i.e., 2020 and onward, around 7400 academic works were published, while during the 5 years before, this number was only around the half value, which was around 3480 publications. Such a spike in the number of publications highlights the amount of interest the topic of Metaverse has brought up in Academia. Furthermore, while searching for the keywords `virtual reality', which is a major aspect of the Metaverse and considered one of its core components, we were able to locate more than 3,000,000 academic works, with 50,000+ works just published in 2022. As the industry interest in the Metaverse increases, we believe that academia will follow the track and we will witness more publications around the topic.

In parallel, we are also witnessing a rise in academic conferences that target the Metaverse. While in the past, virtual reality or Metaverse-related conferences took place under the umbrella of other conferences, these days we can see the term `Metaverse' being used in the name of the conference, and the entire conference would target Metaverse and its related applications. Some of these conferences are organized by subdivisions of well-known academic bodies such as IEEE (Institute of Electrical and Electronics Engineers) which is organizing the `IEEE-TLT-Metaverse 2023: IEEE Special Issue on Metaverse and the Future of Education' \footnote{https://ieee-edusociety.org/ieee-special-issue-metaverse-and-future-education}, a conference especially dedicated for the topic of education in the realm of the Metaverse, as well as `International Conference on Metaverse Computing, Networking, and Applications' \footnote{https://www.ieee-metacom.org/2023/}, and IEEE Metaverse Congress \footnote{https://engagestandards.ieee.org/IEEE-Metaverse-Congress-Series.html}. Logically, such a conference would produce and allow the publication of several Metaverse-related publications, which will further fuel the research in academia, where researchers will build on the work done by other researchers. 

In addition to academic conferences, the amount of industry, or industry-academia hybrid conferences is also on the rise. Conferences that gather Metaverse industry experts and speakers are happening more frequently across the globe. For instance, the worldwide famous journal Economist is organizing a Metaverse Summit \footnote{https://events.economist.com/metaverse/} as part of its Economist Impact initiative with more than 100 guest speakers. The event is happening both in-person and virtually to secure a large number of attendees. Other events include `Global Metaverse Carnival' \footnote{https://metaverse-club.net/}, `Metaverse Global Congress' \footnote{https://www.sensorsconverge.com/} and `Augmented Enterprise Summit' \footnote{https://augmentedenterprisesummit.com/}. Such conferences are used as a bridging tool to connect different stakeholders including industry experts, entrepreneurs, engineers, researchers, as well as academics around the topics of Metaverse.

\subsection{Metaverse Projects}
Companies, organizations, start-ups, and even cities have started to apply the Metaverse in several different domains by coming up with various projects and applications. In general, we have started seeing projects revolving around Metaverse to be focused in the following several domains: gaming, industrial applications, entertainment, art (museums, shows, galleries), finance (DeFi), e-commerce, smart cities, real-estate, healthcare, manufacturing, and education. 

Current top leading Metaverse projects in the consumer space are usually environments in which they aggregate several domains and applications under one roof. These include the notion of gaming, socializing, decentralized finance, entertainment, and ownership. For example Metaverse platforms like Axie Infinity, Decentraland, and Roblox, make it possible for the user to purchase and own properties and land, and even rent it to others through tokens that run using blockchain technology. Users can also purchase and trade certain memorabilia such as paintings or wearables as NFTs (Non-Fungible Tokens) or even fantasy characters through smart contracts and hold on to them or resell them later for a higher value. Concerts and art exhibitions are also possible in these environments. Such games also include immersive experiences, however, the majority of such games at the moment can be played without any special AR or VR equipment. 

\begin{figure}[htp!]
\centering
{\includegraphics[width=.8\linewidth]{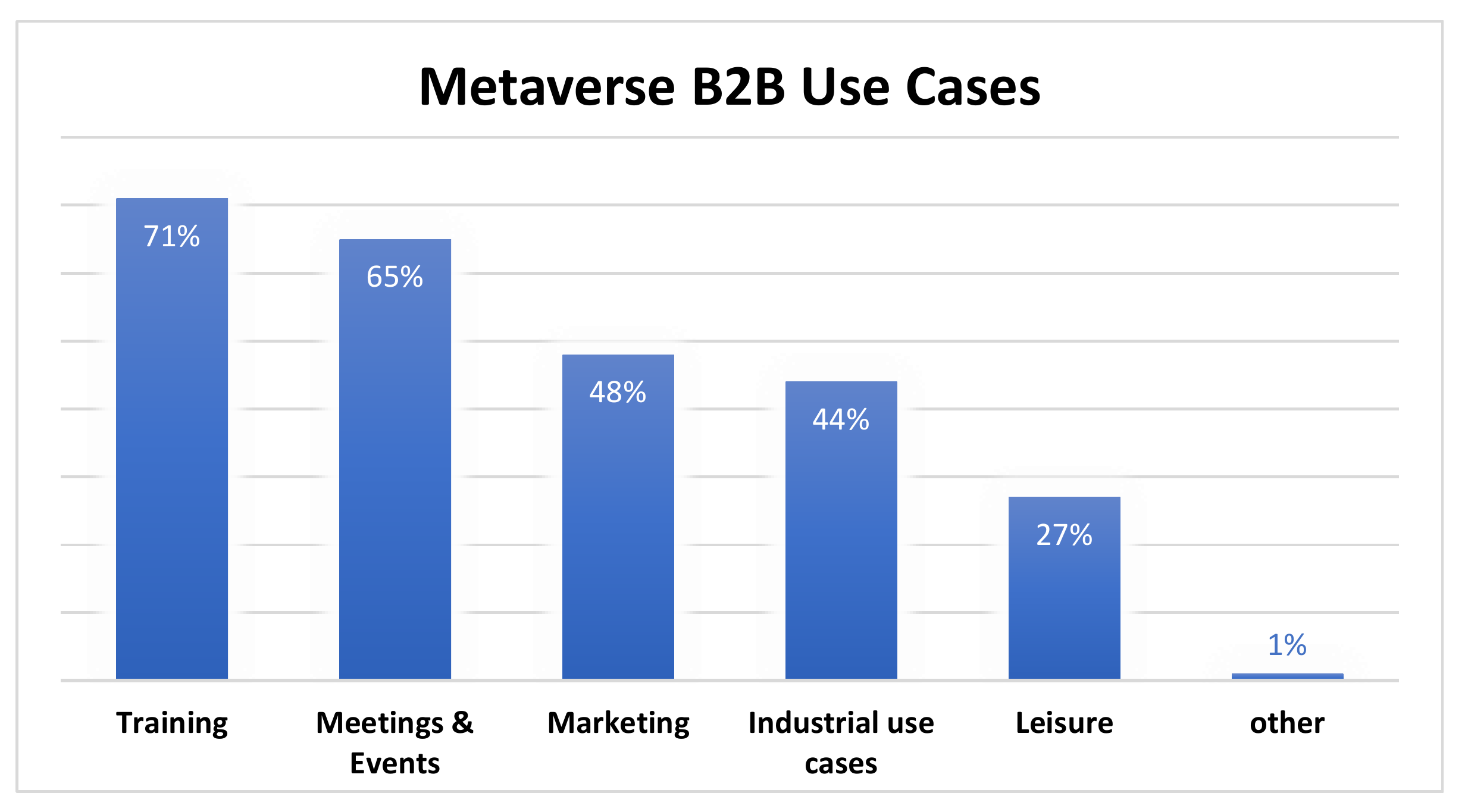}
\caption{Metaverse B2B Use Cases Survey Result by Nokia (extracted from \cite{nokia_b2b})}
\label{fig:B2B_metaverse_use_cases}}
\end{figure}

Metaverse has also found a place in the B2B space where companies are investing a lot. Uses of Metaverse in the B2B spectrum include training, product design, HR-related tasks such as employee hiring and onboarding processes, virtual guidance for hands-on labor work as well as corporate gatherings and meeting events. Professional services company Accenture \footnote{https://www.accenture.com/} has created its own Metaverse for onboarding new employees as well as using it for corporate events and team-building meetings called the N-th Floor. Nokia and Gartner joined forces and conducted a survey \cite{nokia_b2b} about the biggest use cases for B2B-related Metaverse applications that seemed the most compelling, and the domain of training led the poll. The results of their survey are highlighted in Figure \ref{fig:B2B_metaverse_use_cases}. VR-based training is becoming more practical since it can bring so many people together from different parts of geographical locations for training sessions by cutting down traveling time and expenses. The advantage of such training sessions using different kinds of augmented and virtual reality systems and gears is that, in addition to the regular training that would have happened in a 2D environment, participants can engage in hands-on technical and multi-step based training. Such types of training are safer, interactive, and can be more engaging, which makes the training more fun. In addition, training can happen in difficult and technically challenging environments which are hard to recreate, such as rescue and firefighting missions. Companies such as Nokia and Walmart are already relying on Augmented and Virtual reality-based approaches for their staff training. Nokia developed an internal training platform based on augmented reality during the Covid19 pandemic called Nokia Learning Space to train staff on how to deploy Nokia equipment. Participants even got certifications at the end of the training session \cite{nokia_b2b}. On the other hand, Walmart has purchased around 17,000 Oculus go VR headsets to train around 150,000 employees in customer service and management \cite{walmart_17000_oculus}. Other industry leaders are also making big use of augmented and virtual reality. The engineering firm Bosch has developed its own system known as Augmented Reality Platform (CAP) which allows quickly preparing and creating virtual content in augmented reality. Bosch claims that CAP applications save 15\% in average time pertaining to tasks that involve repairs \cite{bosch_microsoft}.

Another major domain where Metaverse is being used in several applications is the health industry. A big use is still taking place in the training aspect of healthcare. 2D images of body parts are converted into 3D floating objects and used for training purposes to prepare better surgeons without risking real lives during a training session. Companies like Enhatch \footnote{https://www.enhatch.com/} have started to provide services known as Intelligent Surgeries where a combination of enabling technologies such as robotics, navigation, and augmented reality provide surgeries to be more cost effecting and faster. Another major event where virtual reality was a core element in healthcare procedure was during the preparation of the complex conjoined twins' separation operation. Doctors and medical team taking part in the operation spent months using virtual reality projections of the twins resulting from MRI and CT scans. Surgeons from different countries, by the use of VR headsets, worked as if they are in the same room. The result was a successful operation that the surgeons attribute to the assistance of the VR \cite{mccallum_2022}.

Cities have also entered the race toward the Metaverse. The city of Seoul in South Korea has been developing a Metaverse platform \cite{seoul_metaverse_2022}. It will be the first municipality government to accomplish this. Dubbed as Metaverse Seoul, the platform will be built in stages over the course of the next couple of years. The procedure will be used for tasks currently conducted by the city including economic policies, as well as civil complaints. The Seoul Fintech Lab will be reproduced in the Metaverse to assist in the economical domain. This will also help the companies to attract foreign investments. The city believes this domain will also be rejuvenated with the help of the Metaverse after it shrunk due to Covid19 pandemic. The educational sector will also find a large place in Seoul's Metaverse plans, as it will include a virtual campus of the Seoul Open University. Immersive courses, lectures, and programs will be conducted. Tourist locations and landmarks will also be converted into virtual sports that people would be able to visit. Tourists will be able to hop on virtual buses and tour the city. Seoul's most attractive concerts and festivals will also be a part of the Metaverse. Parts of public services for the general public and citizens will also be produced in the Metaverse. These will include consultancy, civil complaints, and reservation of public facilities.  Furthermore, a Metaverse version of the mayor’s office will be created and used as a communication platform between the residents and the city's representatives. Metaverse will play a big role in urban management too. VR, AR, and XR technologies will be used to adapt to the disabled as well for their convenience and safety. Finally, the Seoul Metaverse will be big on events, conferences, and working space. It will be used as the main communication channel for events, as well as it will feature remote working environments. The city of Seoul is keen on making the Metaverse project become reality. This shows how the Metaverse can be used by large cities to facilitate the different aspects of their citizens and increase their QoL. Other than Seoul, several other cities such as Dubai \cite{Dubai-metaverse} as well as several Chinese cities and provinces such as Shanghai, Zhejiang, Anhui, and Wuhan have shown interest in also having their virtual versions of their cities in the Metaverse space. Despite the fact that these cities do not have concrete timelines by the time of the writing for their Metaverse ambitions, they are positioning themselves for such advancements in the future.

Projects have also been spun up to govern the Metaverse to make it more interoperable and safe. Such an initiative was proposed during the World Economic Forum's yearly meeting in Davos \cite{world_economic_forum}. 
Dubbed as 'Defining and Building the Metaverse', the initiative will provide assistance in creating an ethical and inclusive Metaverse in which several organizations across the public and private sectors can partake in a plethora of domains such as civil society, academia, business, regulators· The initiative will mainly focus on 2 main areas: The first is the governance of the Metaverse. This part of the initiative will focus on how different technologies can be used to create environments in the Metaverse in a secure, safe, inclusive, and interoperable manner. The second area is the value creation from the Metaverse, as well as identifying the risks and the incentives of those different stakeholders, such as individuals, businesses and society in general will face as the Metaverse progresses. The initiative will also highlight how Metaverse can transform industries, how it can disrupt current value chains, and how it can aid in the creation of new assets and the protection of their rights accordingly. More than 100 entities such as Meta, Microsoft, Lego, Sony, Walmart, Stanford University, and NYU across different industries and academia have committed to partner with World Economic Forum on this initiative. Other initiatives such as the Metaverse Standards Forum \footnote{https://metaverse-standards.org/} are trying to 
unite different stakeholders, companies, and organizations in the realm of the Metaverse to make it more open and interoperable. The Forum itself will not create standards, however, it will provide resources and coordinate and consult the creation standards with the help of other standards organizations working for a better and open Metaverse. 

As stated, art has also been finding a big role in the Metaverse, especially from the perspective of NFTs which are digitally created pieces of art and are being sold through blockchain-enabled technologies. This has led to several so-called mini-museums being created inside well-established Metaverse platforms such as Decentraland. Others have created their own Metaverse platforms that are purely dedicated to becoming digital museums such as Musee Dezentral \footnote{https://musee-dezentral.com/}. This virtual museum is considered to be the first decentralized museum in the world. People can purchase or rent frame space in the museum to exhibit and sell their work. Prominent art broker Sotheby’s also launched its own Metaverse \footnote{https://metaverse.sothebys.com/} which serves as a destination for NFT-based art sales. Furthermore, traditional brick-and-mortar museums are shifting to digital versions to re-grow sales which have been in decline due to Covid19 restrictions from one side, and to keep up with the new emerging trends such as AR and VR from the other side. In fact, many museums are using their own or collaborating with AR and VR applications and offering their art in virtual or augmented reality. For example, the infamous museum Louvre, which hosts Leonardo Da Vinci's Mona Lisa painting, has released an app in which you can enjoy the Mona Lisa painting in 360 degrees \cite{louvre}.

A very important aspect for companies is to drive more and further sales of their products. Promoting user purchases is the ultimate goal for such companies. With the emergence of the Metaverse, more companies are utilizing it to provide immersive shopping experiences and promote their products. With the advent of online selling, and its promotion to live commerce, in which items are showcased via video broadcasts and presented using chat for potential buyers, the Metaverse is playing more of a complimentary role to combine the advantages of live commerce with the immersive nature of experiences. In this regard, the authors of \cite{jeong2022innovative} present a business model on how to incorporate Metaverse with live commerce to provide a better purchasing experience for users. Shoppers can immerse themselves into the Metaverse and can see items in 3-dimensional space from every possible angle. They can also try and test the product virtually on different surfaces and different possible combinations with other products. Such advantages can further assist the purchaser to make an informed decision before buying an item. Based on these, companies are developing newer applications and services for their potential customers to further grow their sales and provide better shopping experiences.



\section{Advancement of Enabling Technologies}

\begin{figure*}
    \centering
    \includegraphics[width=.7\linewidth]{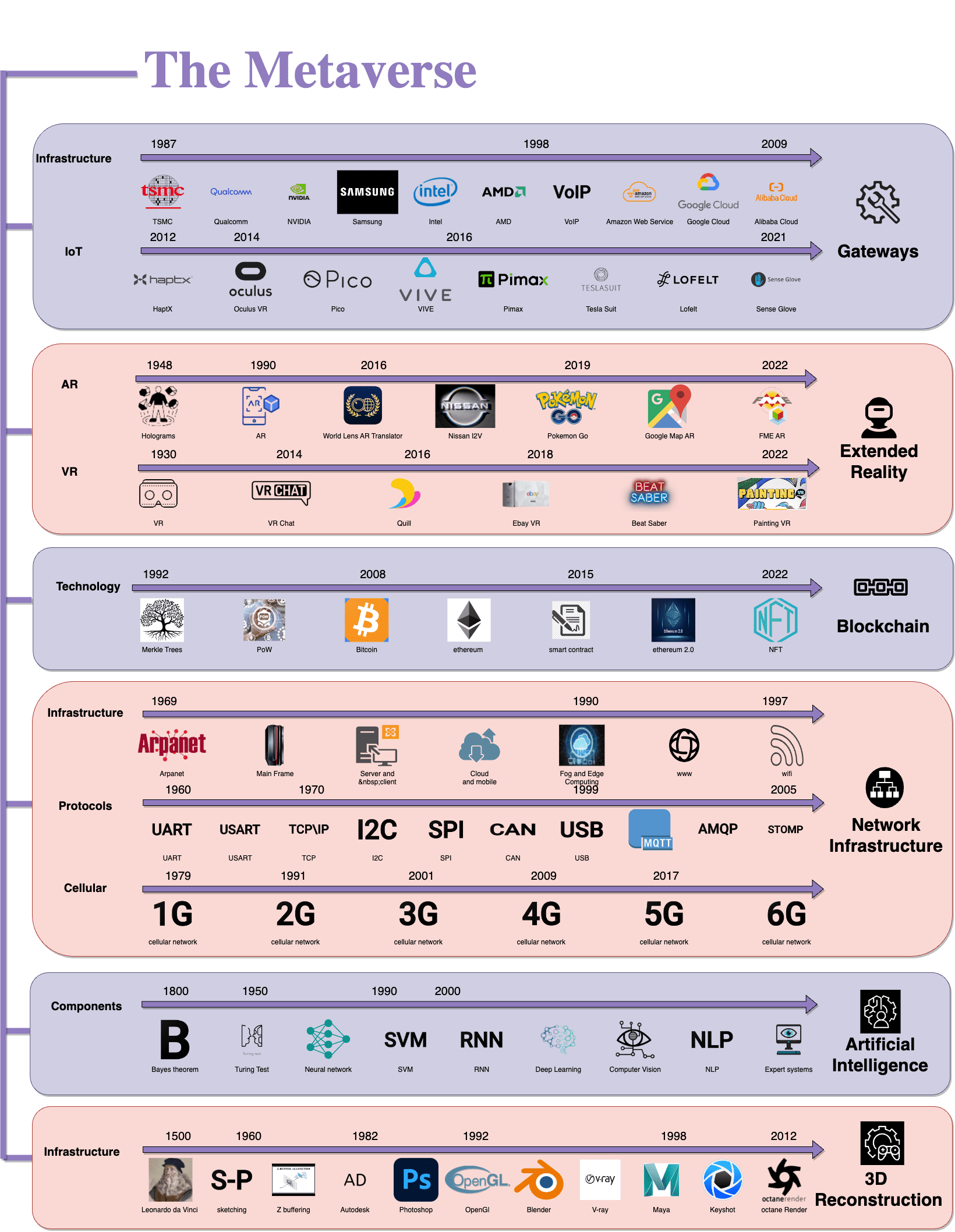}
    \caption{Overview of the Metaverse Enabling Technologies}
    \label{fig:meta_components_1}
\end{figure*}

This section introduces and explores the technological development that gave rise to the Metaverse. The Metaverse development workflow or pipeline employ these enabling technologies to augment the user experience. As presented in Figure \ref{fig:meta_components_1}, the most recent Metaverse development has been made possible by technologies including AI, IoT, AR, VR, Blockchain, Three-Dimensional Modeling, and Edge Computing. It's critical to understand the history of the technology that gave us the right tools and software to build the Metaverse. Each supporting technology's timeline and an overview are presented in this section and illustrated in Figure \ref{fig:meta_components_1}.
For easier tracking, major technologies and sub-components are highlighted in bold.

\subsection{Artificial Intelligence}
At a conference at Dartmouth University in 1956, the term "Artificial Intelligence" was formally used for the first time. Researchers started their contributions of simulating human intelligent activities using machines by learning human behavior \cite{M1}. AI is a synthesis of numerous academic disciplines, leading to a wide range of internal sub-fields, components, and applications. With a more than 70-year history, AI has undergone a prolonged development process that produced numerous smart technologies, which can be employed to construct the Metaverse. AI and its approaches are heavily used in the Metaverse, and its development can be broken down into various parts which are highlighted in the following paragraphs.

\textbf{Machine learning} (ML) \cite{AI1} is  a trendy name we hear everywhere. Businesses are getting competitive and profitable advantages in markets by adopting ML algorithms. Many mathematicians in the 18th century developed and employed statistical and probability theories such as the Bayes theorem, which is still in use today. Turing machine tests were introduced afterward in 1950. Researchers sought to establish whether a machine can think by demonstrating that it can respond to a query and convince a human. The term "perceptron" was initially used in 1957 to describe a method of creating a \textbf{neural network} \cite{AI2} using a rotating resistor. Parallel Computing was originally utilized in 1986. As processing power increased, Parallel Computing aided in the adoption of Neural Network models that can handle large datasets of information generated. In the 1990s, Support-Vector Machines (SVMs) and Recurrent Neural Networks (RNNs) became popular. Afterward, \textbf{Deep Learning} (DL) \cite{AI3} was first presented to explain new techniques in 2006. DL can offer a lot of services to the Metaverse such as Transformers, which help machines to work with natural language. In the 2000s, kernel methods and Unsupervised Learning became widespread. In addition to that, ML algorithms and recent technological breakthroughs have made it possible for \textbf{Image Classification} \cite{AI7} and \textbf{Computer Vision} (CV) \cite{AI4} approaches to solve problems. Google began to assist in testing \textbf{Robot} cars on roadways after MIT initially presented a facial identification framework in 2001.
Modern government and business sectors made extensive and sophisticated use of CV, which plays a significant role in improving humans' ability to interact with the virtual world and gesture recognition in the Metaverse. Moreover, People began to see the value of employing a machine to translate across languages after World War II. \textbf{Natural Language Processing} (NLP) \cite{AI5} has been studied and improved since this period to address more challenging issues like topic discovery and modeling, sentiment analysis, text-to-speech and speech-to-text conversion, and machine translation. In the beginning, NLP used a technique called "Bag-of-Words" to count the occurrences of each word. Following that, the "Word2Vec" and "FastText" algorithms were used. Alongside these breakthroughs, the 2019 publications "XLNet" and "Transformer Models" were popular among scholars. Such techniques aid in the development of speech recognition, allowing users to voice-navigate the Metaverse. This can enhance machine understanding, creative collaboration, and realism in AI storytelling, among other things. With such advancements, \textbf{Expert systems} \cite{AI6}, which mimic human decision-making abilities and seek to resolve complicated problems using knowledge-based reasoning, started to be well used to solve complex problems, especially while using \textbf{Fuzzy logic}, which is a method for representing and changing ambiguous information by evaluating how likely the hypothesis is to be true.
    
To complete a task, AI relies on 5 key elements. (1) \textbf{Learning}, which consists of rote, unsupervised, and supervised learning. (2) \textbf{Reasoning}, often known as logic, can be either deductive or inductive. There are special purpose methods and general purpose approaches for (3) \textbf{problem-solving}. (4) \textbf{Perception}, which provides the AI agent with information about its surroundings, such as sensors and cameras. (5) \textbf{Knowledge representation} transfers the incoming information from sensors to a standard format to process. NLP is an example that translates the input language into a legitimate format to be processed. 
    
As these techniques developed, applications of AI began to emerge. The first robot, "UNIMATE," which replaced humans on an assembly line in the industrial sphere in 1961, was one such application. A chess machine player from IBM named "DEEP BLUE" defeated the world champion in 1997. Apple developed "Siri" in 2011, and today we have "Alexa" by Amazon and many more sophisticated and intelligent AI applications that assisted in creating some Metaverse environments and continue to aid in improving upcoming Metaverses.

\subsection{Blockchain}
Blockchain is a distributed and decentralized digital ledger. It operates entirely in a decentralized manner with no need for a central authority. As a result, Blockchain-based applications and architectural designs benefit from high levels of data availability, security, and privacy \cite{M2}. In 1991, the Blockchain technology concept began when research scientists W. Scott Stornetta and Stuart Haber were developing a workable technique to maintain the backup of digital information \cite{M3}. They described the use of a chain to cryptographically secure blocks in order to preserve the accuracy of earlier data. Afterward, multiple advancements and techniques were proposed under Blockchain. Due to these advances, the Blockchain has emerged as one of the fundamental pillars of the Metaverse, allowing for the use of cryptocurrencies and NFTs to establish a virtual economy that is fully operational and allows for the buying and selling of any virtual good. In this paragraph, various aspects of how Blockchain has evolved and developed are covered.

The secure construction of decentralized blocks was a top priority for academics and developers. To gather information and documents into a single block, they created the \textbf{Merkle Trees} concept in 1992. Later, \textbf{Proof-of-Work} (PoW) \cite{BC1} was adopted to safeguard these blocks against network abuses. The introduction of \textbf{Reusable Proof of Work} (RPoW) and the subsequent \textbf{Peer-to-Peer E-cash} system" known as \textbf{Bitcoin} \cite{M6} proposed by the pseudonym Satoshi Nakamoto marked the 20th century as the "golden period" for the rise of Blockchain. PoW method, which is used to send and verify transactions among decentralized nodes, was made popular by Bitcoin \cite{M6}. Consequently, Blockchain will allow Metaverse enterprises to provide their clients with integrated services that will merge their 2D and 3D digital presences, transforming how clients engage and transact in a secure way. During that period, people began to create decentralized applications, and in 2014, \textbf{Ethereum} \cite{BC3} was published as a cryptocurrency and a decentralized platform that helps users in the Metaverse to exchange assets and funds. Moreover, with such advancement and the use of Blockchain, \textbf{smart contracts} \cite{BC4} were introduced as one of Blockchain's most significant applications in 2015. People began to adopt Blockchain applications during that time and up until 2021, which prompted major corporations like Facebook and Amazon to develop Blockchain services and currencies. To address some limitations, \textbf{Ethereum 2.0} was introduced. In 2021 and 2022, \textbf{NFT} saw the lights \cite{BC5} where owning digital assets became possible and trendy. Users of the Metaverse can sell their NFTs for fiat currency at any time or exchange them for cryptocurrencies to acquire other Metaverse objects. Many software programs, such as "OpenSea" were introduced until reaching the \textbf{Metaverse}, which was motivated by the COVID-19 epidemic. During this period, the culture of working from home started to be extended and became more convenient. Recently in September 2022, Ethereum completed the Blockchain's transition from PoW to \textbf{Proof of Stake}~\footnote{https://ethereum.org/}.

Due to its developments and dependability, Blockchain is used to create and suggest new cryptocurrencies. Blockchain-based smart contracts have recently become widely used to initiate processes and execute them automatically. Blockchain technology has been adopted by banks all around the world to boost productivity and cut expenses. It is currently utilized by several game companies and supply chain management like Walmart to monitor some of their products with the help of IBM's cloud servers. Moreover, it serves nowadays as a main pillar in addressing efficiency and security in the Metaverse.

\subsection{Network infrastructure}

Networks and interconnected components attracted people's interest after computers became mainstream. The year 1969 was the first milestone where ARPANET (Advanced Research Projects Agency Network) connected the first computer to the network \cite{M5}. Onward, \textbf{General-Purpose Mainframe} computers became widely available in the middle of the 20th century, which prompted the development of \textbf{server} computing that offers services to \textbf{client} computers. Early in the 1990s, \textbf{cloud and mobile} \cite{NT6} computing became popular. After that, a \textbf{World Wide Web} (www) proposal was produced in 1989, which sparked the creation of numerous www protocols and search engines. From then on, network infrastructure saw rapid advancements in the realm of cables, ranging from \textbf{Ethernet} to the development of \textbf{WiFi} in 1997. Since the Metaverse is being constructed on top of the internet, it will need the internet and advanced protocols in order to run and offer services to users.
Beginning with MQTT in 1999 and moving on to AMQP and STOM in 2005, meaning \textbf{messaging protocols} have been improved. \textbf{Inter and Intra} System Protocols, such as UART, USART, USB, I2C, SPI, and CAN, have also become popular.
The \textbf{networking and cellular}~\footnote{https://www.rfpage.com/evolution-of-wireless-technologies} infrastructure was also incorporated in the advancement trend by introducing \textbf{1G} in the 1970s to offer voice services. It was later extended to serve voice, data, and web mobile service with low speed under the \textbf{2, 2.5, and 2.75} Generations. Then, \textbf{3G and 3.5G} were released to speed up streaming, video calling, and web surfing. In 2010 we saw the arrival of \textbf{4G}, which allows users to use services at fast speeds, with high-quality HD video conferencing, and with global roaming. We are currently in the 5G era, which provides extraordinarily fast mobile internet, supports the Internet of Things, autonomous driving, smart cities, and intelligent healthcare systems, paving the way for the creation of the 6G network-based Metaverse to further improve the experience. Moreover, the solution to sustaining the Metaverse lies in \textbf{edge and fog computing} which were first introduced by Cisco~\footnote{https://www.cisco.com/} in 2012. In essence, edge and fog computing \cite{NT1} improve response times, save bandwidth, and minimize latency, making it ideal for many Metaverse use cases. Additionally, it places data and applications as near to consumers as feasible, providing them with the necessary local computing capacity while reducing network-based latency and congestion risk.

\subsection{3D reconstruction }
3D reconstruction is the act of recreating the form and appearance of genuine items in three dimensions. It all started in the 15th Century, when the famous artist, Leonardo da Vinci drew the first accurate 3D drawing on paper. James Joseph Sylvester created the matrix mathematical notation \cite{M4} later on in the 18th century. This was a key building block for the creation of computer software and 3D images that enabled the development and expansion of the Metaverse applications, which heavily relies on 3D rendering and visualization. Various 3D reconstruction models and programs have been created over the years and improved. Aspects of software and methods for 3D reconstruction will be covered in this paragraph.

People have been drawn to 3D modeling for ages. With the development of resources and technology, 3D software has become more prevalent. In the early 1960s, Sketchpad, the first 3D sketching program, was proposed and released. Following that, 3D graphics companies began to emerge in 1968 to create new tools for drawing 3D shapes, which led to the emergence of depth buffering or Z buffering by Ed Catmull. Later, other software appeared, including Autodesk 3D, Adobe Photoshop, Redshift, Octane, and many others. Such technology can create a 3D scene from a series of 2D photos collected from various perspectives in seconds, which motivates the idea of having better 3D objects in the Metaverse.
People started by forming the model using the \textbf{Rasterisation} technique, which treated the model as a polygon mesh with embedded information in the vertices. After that, \textbf{Ray Casting} was invented to address various issues and drawbacks with the earlier method. \textbf{Ray Tracing} was later proposed to accurately simulate object shadows. Then, when those technologies were being developed, a number of \textbf{rendering engines} \cite{3D2} were suggested, including Lumion, KeyShot, Corona Renderer, and V-Ray.
The speed at which the Metaverse is being built can be greatly improved due to the advancements in this discipline, where it is now possible to create 3D spaces and objects by simply taking several 2D photographs.

Character animation was limited to fairly rigid movements due to complex resource issues \cite{3D1}. Motion capture, then, implemented intricate facial rigs in games, adding a whole new level of realism to the experience. Additionally, developers began to give video game characters more motions and moves. The development that happened in 2014, brought a considerably more realistic feels to the game by using animations other than the usual walk, run, attack, and jump actions. Today's 3D animations have advanced to the point that, in a game, for instance, the player feels as fluid as they would in a movie. Additionally, major corporations like Meta are working to create 3D objects with more realistic free motions and capabilities. Applications like VR, AR, MR, and the Metaverse are now applicable due to the advancements made in this discipline.


\subsection{Extended Reality}
Extended reality is the term provided for the immersing technology of the combination between \textit{Augmented Reality} and \textit{Virtual Reality}. The extension of the physical world to virtual reality is achieved by the modern world digitalization in our day-to-day needs and wants. With the XR's advanced technology, people can now achieve higher dimensions of immersing their characters in the Metaverse. 
The term virtual reality first appeared in a science fiction story in 1930, in which the writer predicted that a pair of goggles that users can wear would allow them to experience the fictional world through smell, touch, and taste \cite{gigante1993virtual}. To further describe VR, we can refer to it as a 3D environment entirely generated by computer technology \cite{coelho2022authoring}. Users can immerse their presence inside it using attachable devices capable of generating visual and sensible scenes. Some VR apps are in the form of hand painting, in which users can use their imagination to be creative, using the tips of their hands to draw and paint without the need for physical brushing and painting tools \cite{gerry2017paint}. In addition, this technology allows users to interact with their surroundings through a narrative that can provide contextual information about a product or place \cite{pimenta2010defining}. Many other real-world applications are today benefiting from this technology, such as Beat Saber 2018~\footnote{https://beatsaber.fandom.com}, E-bay VR Commerce 2016~\footnote{https://www.ebay.com/}, Quill 2016~\footnote{https://www.quill.org/} and VR Chat 2014~\footnote{https://hello.vrchat.com/}.

In contrast to VR, augmented reality does not create a new reality; instead, it augments the physical reality to add more components to an existing reality and enrich the content we see. AR, since its existence back in 1990, has been seen in several aspects of the real world. One aspect of the usage is in location-aware geospatial computing, in which a system is used to compute and track a location of a moving target. Astronauts are using the technology to assist in repairs to space stations, such as the T2 augmented reality (T2AR) project 2016~\footnote{https://tinyurl.com/T2ARnasa}. Other uses include interior mapping to map and plan venues with FME AR 2020~\footnote{https://www.safe.com/}. Another AR application is sketching, where users can follow virtual lines to draw physical ones on a piece of paper using a smartphone application Sketchar 2017 ~\footnote{https://sketchar.io/}. In addition, AR is also used in creating emojis for users' faces on smartphone devices, as well as many other usages~\footnote{https://www.apple.com/ca/augmented-reality/}. Floating menus and icons have also taken part in augmented reality simulations, where users can interact with buttons and menus to perform some actions \cite{bowman2001design}. Other real-world examples of the usage of AR include the Google maps AR feature 2019~\footnote{https://arvr.google.com/ar/}, Pokémon Go 2016~\footnote{https://pokemongolive.com/} and world lens AR translator 2010~\footnote{https://www.xda-developers.com/}.

A blend of VR and AR is created through a term called \textit{Mixed Reality (MR)} first used in work published dated back to 1994 \cite{milgram1994taxonomy}. The interaction between the physical and virtual worlds is possible, allowing users to introduce both worlds into their comprehension through attachable devices like a headset with a clear lens or a camera. MR's earliest applications are motion picture projection dated back to 1895, holograms in 1948, and volumetric displays in 1945 \cite{johnston2006holographic}. The usages of holograms consist of using virtual 3D images. The visibility of these images depends on some light beams to visualize the properties and details of the items projected.

\subsection{Gateways}
Infrastructure availability and the Internet of Things (IoT) devices are the basis of the creation and utilization of the Metaverse. For us to have connected virtual worlds, a basic infrastructure should assemble the Metaverse. The computation power represented in the chips and processors is vital to run the simulations of the virtual worlds. Companies like Qualcomm, Nvidia, Samsung, Intel, and AMD (1970-1995) are the major suppliers of the chips and processes, providing us with different central processing units (CPU) and graphics processing units (GPU) that are critical to the performance improvement of the virtual world. The advancement and the powerful processing units created are contributing to the creation of a virtual environment. In addition to the processing units, the Voice over Internet Protocol (VoIP) creation in 1973, and the first software using this technology appeared in 1995 \cite{sisalem2013short}. The adoption of VoIP enriches the connected users' communication layer through the Metaverse. Therefore, users can communicate over the voice channel using a microphone and audio device to speak and listen to others while in the Metaverse. 

The reliability of the Metaverse is measured by its security to identify systems that provide access only to legitimate personnel and help protects against adversaries' malicious activities. In addition, access to data storage is crucial to offer scalability and simplicity for individuals to access their digital information across the Metaverse. Furthermore, containerization and orchestration are helping developers and programs to better build and deploy their software over multiple devices and operating systems to offer a service for all individuals. Some of the leading giants provide such services such as VMware~\footnote{https://www.vmware.com/}, AWS~\footnote{https://aws.amazon.com/}, Azure~\footnote{https://azure.microsoft.com/}, Google cloud~\footnote{https://cloud.google.com/} and Alibaba cloud~\footnote{https://www.alibabacloud.com/} (1999-2009). These services contribute to creating and managing the apps on the Metaverse.
In addition to the infrastructural components of the Metaverse, individuals need IoT devices to connect and experience the sensations inside the virtual worlds. Extended reality IoT devices are several, and they are built in a way capable of engaging with all the different sensations of individuals. The user can now interact with the Metaverse through different devices, such as Mobile input techniques, Hand-based input, Head Mounted Displays (HMD), Haptic, Audio, Brain attachable sensors, Holographic, and Smart Glasses. 
More details about these technologies and their usage are provided in Section \ref{sec:hardware}.


The aforementioned technological developments result in significant cooperation in creating more effective and dependable infrastructures, which serve as the main building blocks of the Metaverse creation.

\section{Novel Metaverse Pipeline Ecosystem}

In this survey, we present a unique pipeline ecosystem consisting of a stream of sequential workflows that contributes significantly to this survey and elevates it to a wider and more comprehensive level. 
The proposed pipeline will provide the necessary knowledge that allows the users to fully grasp the Metaverse concept in a clear and direct manner to fit it into their own perspective. 
The novelty of the devised pipeline is twofold. On the one hand, it places the building blocks behind the realization of the Metaverse concept and puts it into reality. On the other hand, it provides valuable guidelines for people from different domains (e.g., Academic, Industrial, and Business) to find themselves within the Metaverse paradigm, and allow their expertise to further enhance its foundation.
\begin{figure*}[h]
    \centering
    \includegraphics[width=0.80\textwidth]{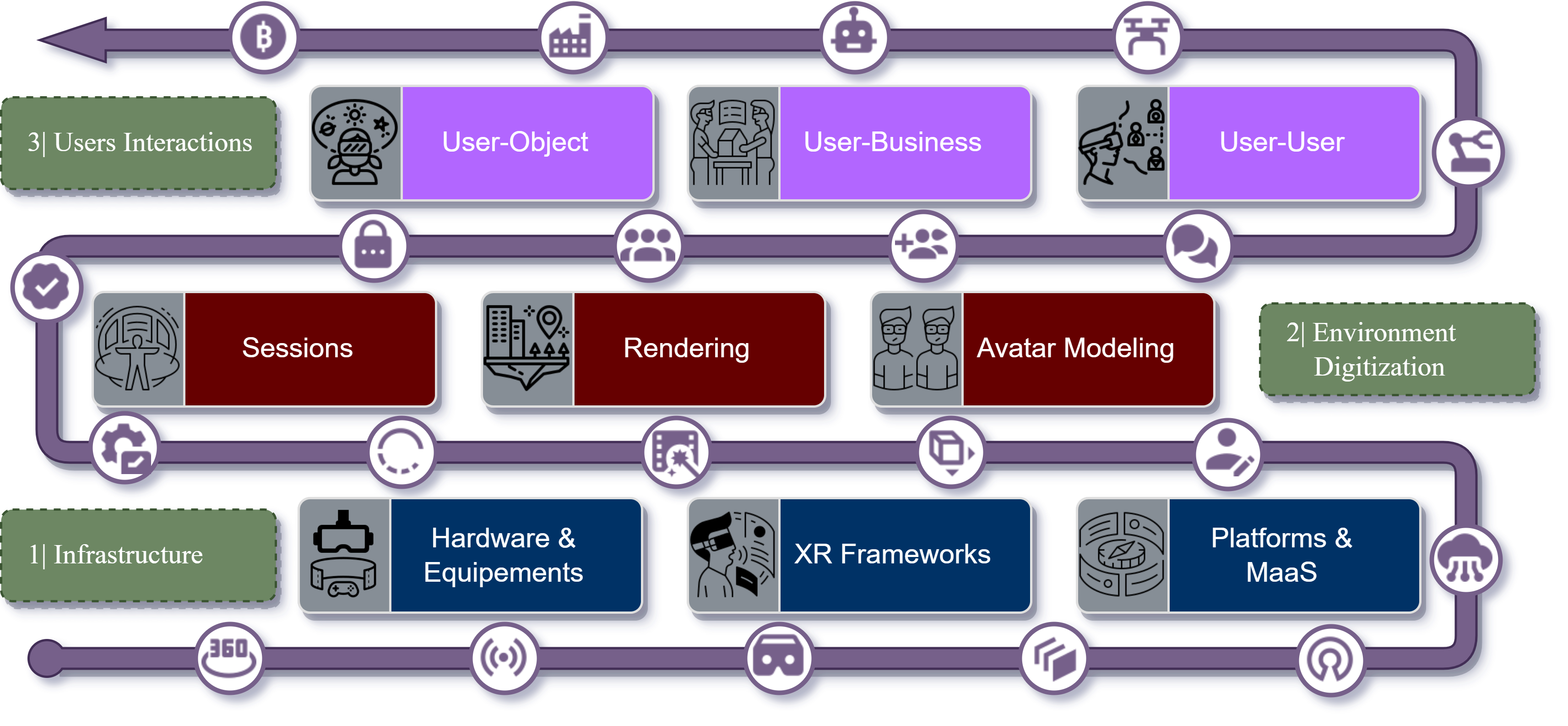}
    \caption{Metaverse Pipeline}
    \label{fig:pipeline_1}
\end{figure*}

\begin{figure*}
    \centering
    \includegraphics[width=\linewidth]{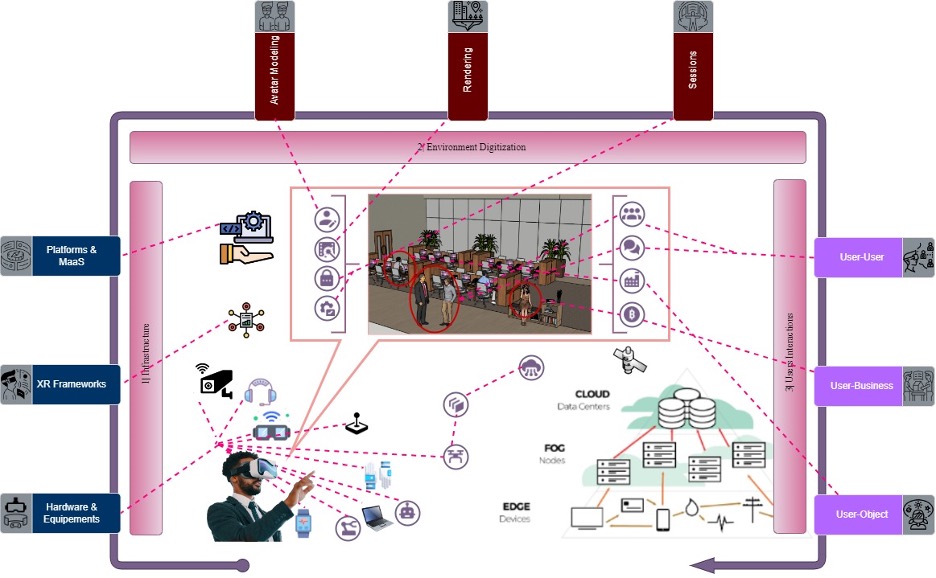}
    \caption{Metaverse Pipeline Layers and Components}
    \label{fig:pipeline_detailed}
\end{figure*}



The pipeline represents the workflow sequencing between the Metaverse's connected entities. This effective sequential design creates a clear order for the Metaverse development process and decomposes it into phases that can simplify its realization. Such pipeline, illustrated in Fig.~\ref{fig:pipeline_1}, comprises three layers/tiers. Starting from the bottom tier of the pipeline, the essential \textbf{Infrastructure} is set to host the Metaverse. The second tier encloses the steps for the Metaverse \textbf{Environment Digitization}. And the top tier defines the \textbf{User Interactions} within the constructed environment. Such a design reflects the system's elegance while also allowing for simpler, but more efficient, system analysis.

Each layer is composed of a set of entities. 
The infrastructure layer is the class responsible for the collaboration between hardware and software. It contains all of the fundamental components required for Metaverse digitization and functioning. This layer is made up of three entities, i.e., \textbf{Hardware}, \textbf{XR Frameworks}, and \textbf{Platforms \& Metaverse-as-a-Service (MaaS)}. Environment Digitization is the second tier in charge of creating Metaverse worlds; it ranges from copying real-world objects to human representation as avatars. 
This layer includes the following entities: \textbf{Avatar Modeling}, \textbf{Rendering}, and \textbf{Sessions}. The aforementioned tiers combined will emerge a Metaverse world that is ready for deployment and usage. Finally, the third tier comprises the interaction components. This layer represents a set of interactions that can happen among actors within the Metaverse. It incorporates three interactive scenarios: \textbf{User-User}, \textbf{User-Business}, and \textbf{User-Objects}.\newline
Figure \ref{fig:pipeline_detailed} presents an overview of each of the components in our multi-layered pipeline ecosystem while connecting them to some of their use cases within the metaverse. 
In the following, we briefly describe each of the aforementioned entities:
\begin{itemize}

\item \textbf{Hardware:} a wide range of hardware and sensors can be used in the Metaverse. The such component includes everything from cameras and microphones to more complex equipment like haptic feedback devices and motion controllers. In precise, the hardware and sensors used depend on the type of Metaverse experience being created. In general, the main objective is to create a realistic and immersive experience for the user. For this purpose, VR and AR are the two main components that tailor the Metaverse experience. VR allows users to be completely immersed in a computer-generated world, while AR provides the ability to promote the digital world inside the real world in a more realistic way. Further, Computer Vision can be infused with the available frameworks to simulate real-world experiences inside the Metaverse or provide the necessary means to enhance the experience's quality. Moreover, Hardware usage may differ based on the intended application usage. For instance, a VR game will use different hardware and sensors than a social VR platform. Such technologies can perceive realism for the users.

\item \textbf{XR Frameworks:} unlike the hardware entity, which provides means to access the Metaverse, frameworks can be defined by the tools used within the Metaverse itself. In this context, XR frameworks play a major role in its development \cite{a1}. They provide a complete skeleton to build interactive solutions where users can directly engage and communicate with their environment. Furthermore, AI technologies have proven their capacities in this domain. For instance, Deep Machine Learning can enhance users’ tracking systems
, improve the rendering quality and reduce the bandwidth cost while increasing latency which is a major concern for users in the Metaverse. 

\item \textbf{Platforms \& MaaS:} is usually capital and consistent with the rest, which is referred to as a collection of technologies, software, and services that enables the realization of Metaverse projects and concepts. The Metaverse is expected to serve as a globally scaled interactive and immersive platform for individuals worldwide~\cite{o1}. The new fully decentralized version of the internet allows you to experience the replicated world in real life and vice versa. Metaverse platforms often stand for the big organizations/companies that work on financing and developing the Metaverse projects to be realized \cite{o2}. These projects include Meta, Decentraland, The Sandbox, AXIE INFINITY, GALA, Enjin Coin, Metahero, and others. All the aforementioned projects serve as Metaverse key enablers, by offering Metaverse-as-a-Service (MaaS) solutions.

\item \textbf{Avatar Modeling:} one of the advantages of the Metaverse is the immersive experience it provides for its users to interact with each other. Thus, Metaverse individuals expect a level of realism when creating their avatars to represent themselves. In this context, Avatars can be customized to look like the person's real-world appearance, or they can be completely different depending on the platform. Multiple methods can be employed to create avatars. These methods include using photogrammetry or AI technologies. Each method has its benefits and drawbacks. For instance,  photogrammetry is a popular method for creating realistic and accurate avatars. However, it can be time-consuming since it requires the average user to capture and provide all the required details manually. Furthermore, 3D software can be expensive and require prior experience to use effectively. In this regard, AI technologies offer flexibility by automatically mapping humans to shaped and skinned avatars while sacrificing some accuracy of the final results.

\item \textbf{Rendering:} is creating and displaying an environment in a three-dimensional representation. It entitles the Metaverse to a high degree of control over the environment. Rendering can be used immersively to visualize the virtual world or to simulate real-world experiences that are difficult to generally achieve such as flying through space or visiting other planets. Examples include rendering a realistic representation of a location for an immersive virtual reality experience or creating a stylized environment for a specific application context. There are a variety of different techniques that can be used for such tasks, which depend on the look and feel of the final product. For realistic environments, lighting and shading techniques can be used to create a believable atmosphere. However, for stylized environments, more abstract approaches may be taken to create a unique look depending on the targeted audience, such as the color brightness and geometric shapes. One of the challenges in this context is the balance between realism and performance. Too much realism can be computationally expensive, while too little realism can ruin the experience. Finding the right balance is usually based on trial and error and depends on the project's specific goals. Another challenge is dealing with the vast amount of data required to create a realistic environment. This data can come from a variety of sources, such as photographs, satellite images, and on-site measurements. Organizing and managing this data can be daunting, but it is necessary to create immersive and realistic environments. 

\item \textbf{Session Management:} Sessions allow users to connect jointly to the Metaverse and explore the virtual worlds. There are a variety of session-based activities that users can participate in, ranging from simple socializing activities to more complex, cooperative, and interactive ones. In this context, session management is a crucial component that requires careful planning and designing. There are many factors to consider, such as security, performance, scalability, and resource management. For instance, managing sessions in a cluster of distributed servers might require specific considerations to allow a large number of users to interact smoothly without causing delays. Moreover, sessions also provide a way for a system to track a user's activity. The system provider can use this information for auditing and security purposes or to improve the user experience by understanding how users interact with the system.


\item \textbf{User-User Interaction:} this type of interaction refers to the different means of communications between users. Users are active members and actors of the system. Everything users do in the Metaverse has a reaction and an effect elsewhere on the ecosystem. It is a fundamental guideline that designers must follow. User-user or peer-peer interaction is one pivotal aspect of the Metaverse experience. People can join the virtual world and interact with each other, similarly to our physical world. They are able to enjoy their daily digital life such as attending a virtual conference with other peers/avatars or simply chatting and gossiping together with friends and strangers. Users can also form parties to discover new areas or team up to overcome gaming challenges. In addition, boundaries and relationships come to life in the Metaverse. For instance, users can apply some restrictions like blacklisting other users or authorizing each other to enter their own properties/zones. 

\item \textbf{User-Business Interaction:} an interaction that resembles the business services consumed by virtual users. the Metaverse is still in its infancy, but it has the potential to grow into a large and thriving market. Numerous businesses are already operating within the Metaverse, and many more are expected to enter the market in the future. Business owners supply their customers with different services such as virtual real estate and online stores where it offers the capacity of selling or purchasing virtual goods. Several factors can contribute to the growth of the Metaverse economy. For instance, the Metaverse is being marketed as a substantial prospective contributor to business, entertainment, and educational advancement. Another factor can be connected to the increasing number of people who can access the Metaverse thanks to its availability, low hardware cost
, and the recent improvement in the communication technologies such as the presence of 6G. This accessibility is expected to increase the Metaverse's popularity, guiding more opportunities to join the market while making the Metaverse more attractive to businesses and individuals alike. 

\item \textbf{User-Object Interaction:} the meaning behind interacting with objects is generally to manipulate things inside the Metaverse. Objects in the Metaverse may represent a digital replica of real-world objects. Hence, these digital objects must abide by the same physical rules as they do in real life. The state of the digital objects should be modified by a user's input, such as being moved, molded, or any other events that may occur in real life to empower an immersive user experience. Digital entanglement of physical-virtual objects is also a critical part of this component. Through the latter, manufacturing, operations, and other physical experiences may take place virtually using digitally entangled IoT and enabling the concept of Digital Twins. Metaverse users can buy/own lands and assets using crypto-currency. Hence, many use cases for user-object interaction can be wrapped up in this component.
\end{itemize}

\begin{figure*}[h]
    \centering
    \includegraphics[width=.9\linewidth]{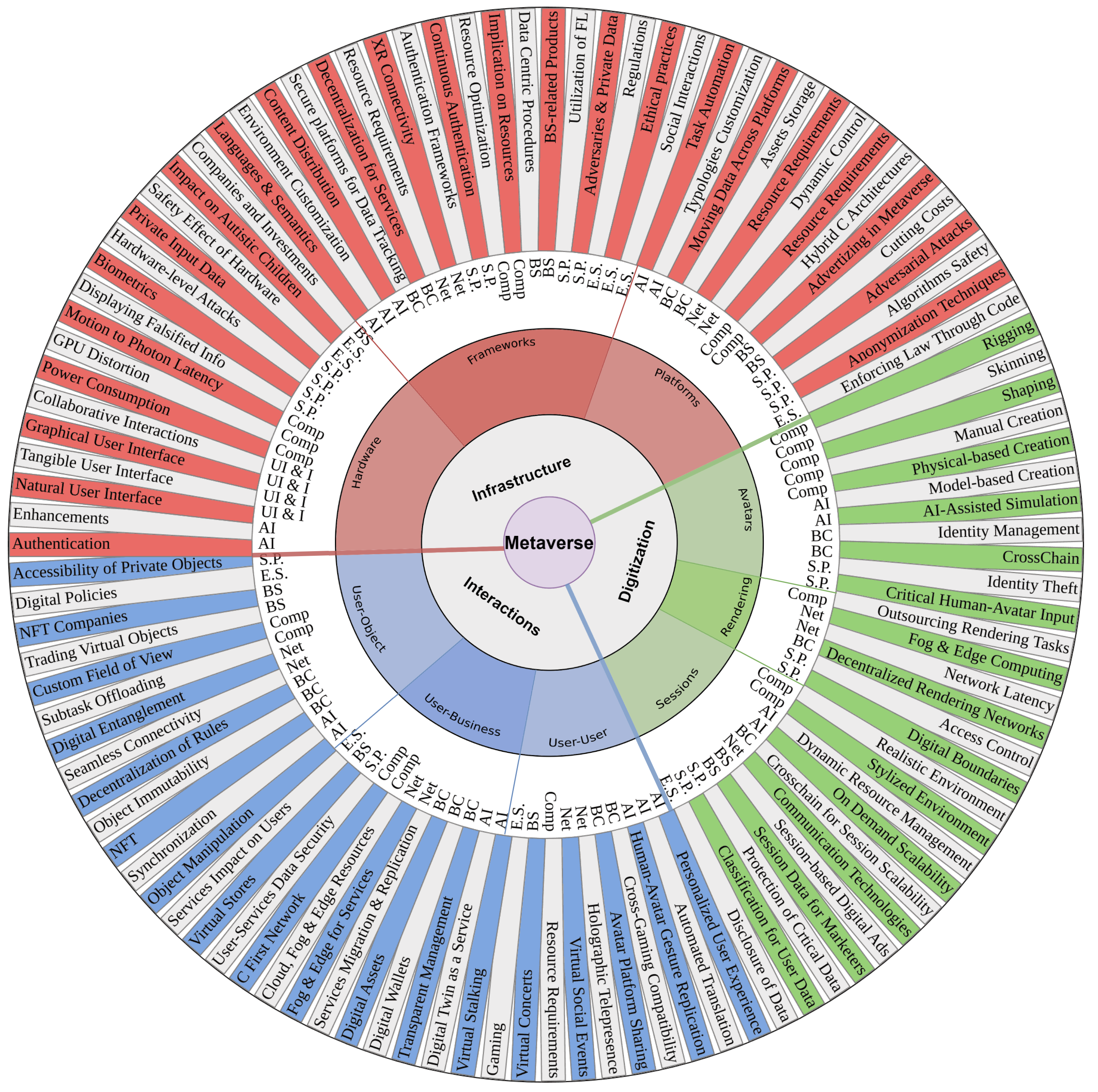}
    \resizebox{\linewidth}{!}{%
    \centering
    \begin{tabular}{|c|c|c|c|c|c|c|c|c|}
         \hline
         Abbreviation & AI & BC & Comp & Net & S.P. & UI \& I & BS & E.S.\\
         \hline
         \hline
         Word & \shortstack{Artificial\\Intelligence}
         & Blockchain
         & \shortstack{Computing\\Infrastructure}
         & \shortstack{Networking\\Infrastructure}
         & \shortstack{Security \&\\ privacy}
         & \shortstack{User-Interaction \&\\ Infrastructure}
         & Business
         & \shortstack{Ethics \&\\ Social-psychology}\\
         \hline
    \end{tabular}
    }
    \caption{Taxonomy: The Metaverse Development Ecosystem}
    \label{fig:taxonomy}
\end{figure*}

In this layered pipeline ecosystem, each of the components is empowered by a set of enabling technologies (i.e., AI, Blockchain, Networking, and Computing) and empowering domains of the Metaverse (i.e., Privacy and Security, Business, Ethics, and Social-psychology) to realize its full potential for serving the Metaverse. A detailed taxonomy is presented in Figure \ref{fig:taxonomy}. The taxonomy shows the components of the Metaverse, branching their relevant topics that are categorized by their appropriate technologies.
Hence, in the following sections, we survey each of the pipeline components against the enabling technologies and empowering domains, including top-of-the-line academic and industrial solutions and efforts.

\section{Metaverse Infrastructure}
\begin{figure}
    \centering
    \includegraphics[scale=0.4]{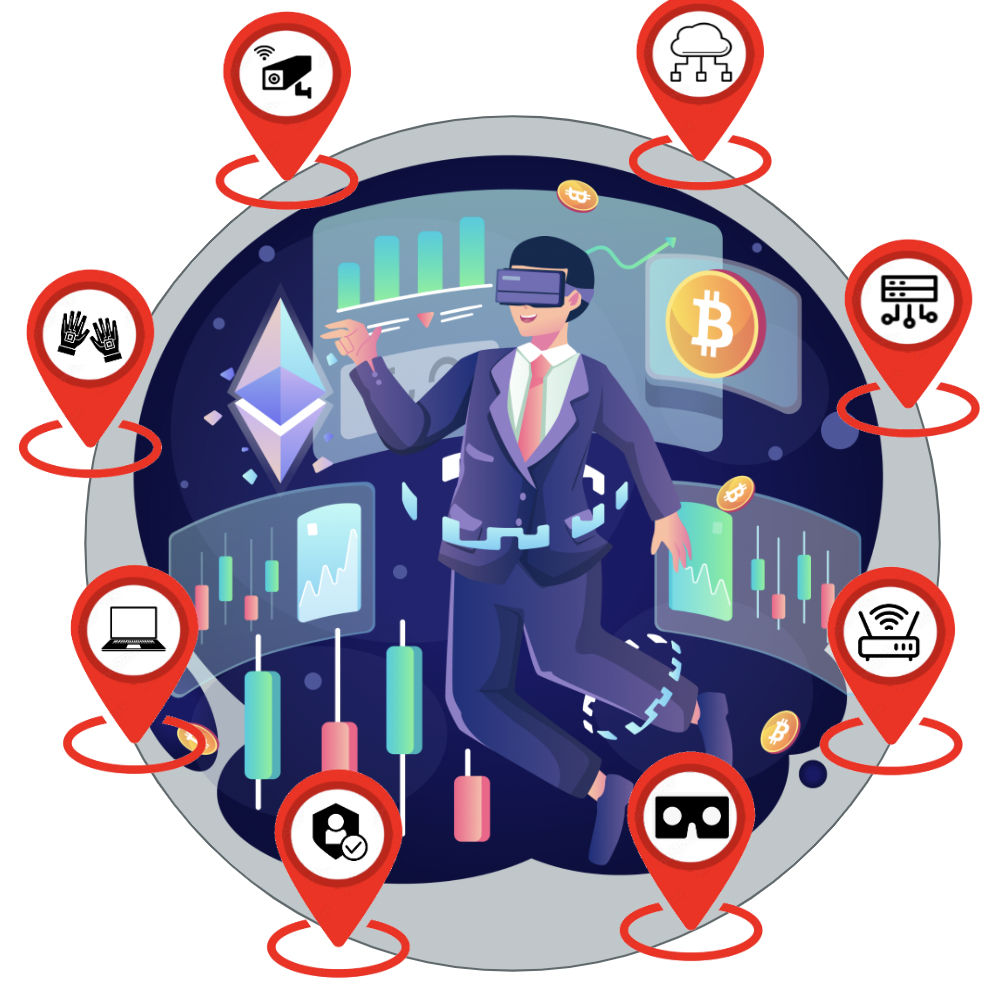}
    \caption{Metaverse Infrastructure}
    \label{fig:meta_infra}
\end{figure}

The hardware and equipment form the Metaverse infrastructure and the backbone for building immersive, secure, efficient, scalable, cost effective, and secure environments. Guaranteeing a well designed and supported environments empowers the development of additional layers in the Metaverse pipeline, including environment digitization and advanced applications for users interactions. Therefore, in this section, we survey the latest advancements for building a powerful and promising Metaverse Infrastructure, which forms the first layer in the proposed multi-layered Metaverse pipeline ecosystem. Our review includes the latest scientific papers and articles targeting one or more of the following infrastructural components: Hardware \& Equipment, XR Frameworks, and Platforms \& MaaS.
In Figure \ref{fig:meta_infra}, we refer to the different supporters of the Metaverse Infrastructure. Examples of such supporters include XR headsets, haptic devices, VR cameras, computing machines, cloud, edge, fog devices, third-party services, and technological platforms.

\subsection{Hardware \& Equipment}\label{sec:hardeq}
\label{sec:hardware}
In this section, we focus on devices and equipment that are built to enable and facilitate the use and immersion into the Metaverse and its related applications. Different experiences in the Metaverse require different types of enabling hardware. For example, for non-immersive experiences, the Metaverse can be explored using traditional day-to-day computing devices (laptops, smartphones), software (web browsers, mobile apps), and traditional sensors (cameras, webcams, and microphones). For immersive experiences, a special type of hardware and equipment is needed that is dedicated and specially built for Metaverse-related applications. These tools provide an entrance into the realm of the Metaverse for the user. \\

\subsubsection{Categories of Metaverse Hardware}
In general, Metaverse-related applications are resource hungry. This is true in the case of networking-related resources such as bandwidth and latency, as well as graphical and computing resources. For example, to have an avatar running inside a Metaverse, at least, rates of 30 frames or more per second are needed to provide an enjoyable level of graphics \cite{kumar2008second}. At the same time, to avoid motion and graphical delays, since these can cause dizziness and sensory disarray to the user, the software running on the device should match the hardware \cite{park2022Metaverse}. In the sequel, we present the various main categories of hardware for serving the Metaverse, including Audiovisual, Hand-based Input, Contact Lenses \& Glasses, and Wristbands. Each category is based on the type of machinery needed and the purpose to support immersiveness. \\

\subsubsection*{Audiovisual}
\begin{figure*}
    \centering
    \begin{subfigure}{0.28\linewidth}
        {\includegraphics[width=\linewidth, height=4.3cm]{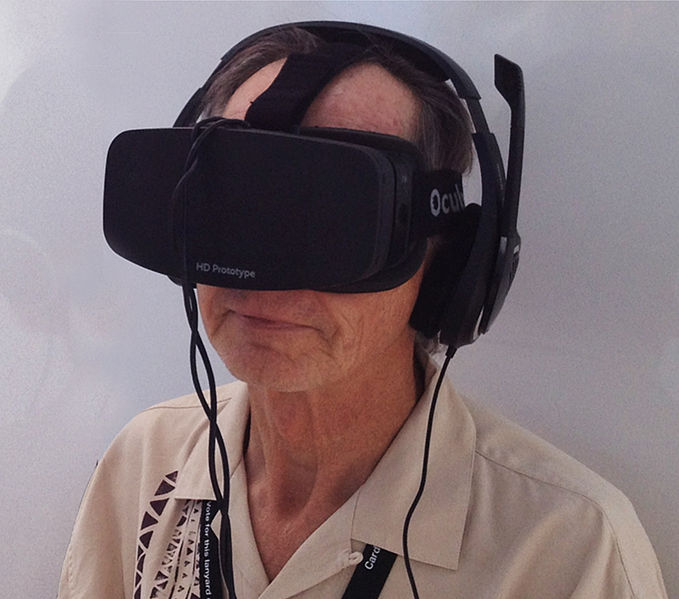}
    \caption{Head Mounted Display \cite{hmdhmd}}
    \label{fig:hmd}}
    \end{subfigure}
    \begin{subfigure}{0.28\linewidth}
        {\includegraphics[width=\linewidth, height=4.3cm]{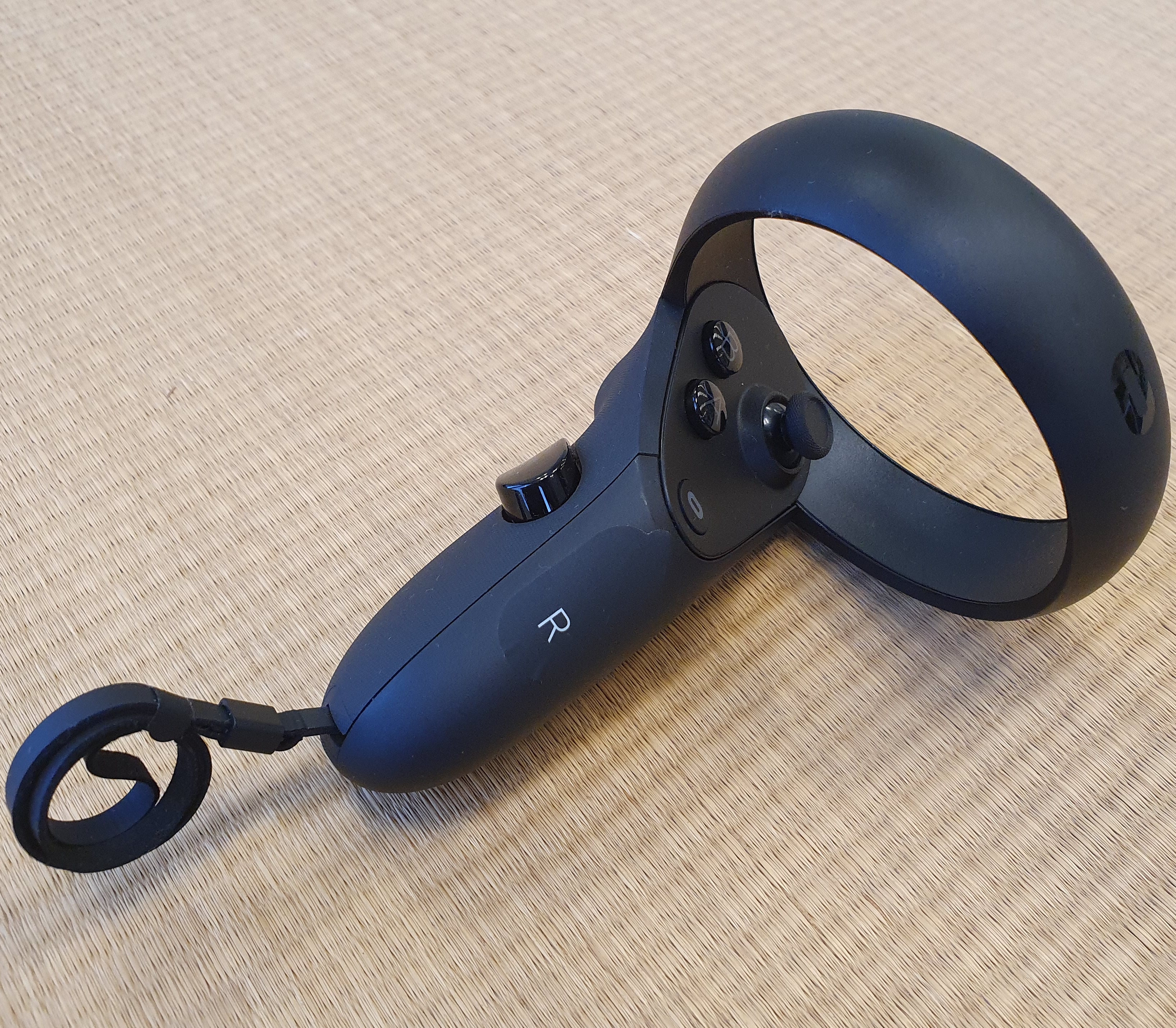}
    \caption{Hand Touch Controller \cite{oculustouch}}
    \label{fig:hand-hold}}
    \end{subfigure}
    \begin{subfigure}{0.28\linewidth}
        {\includegraphics[width=\linewidth, height=4.3cm]{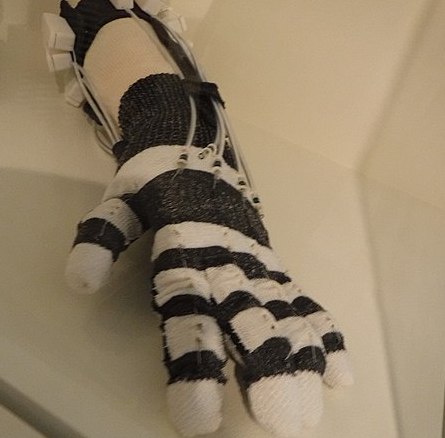}
    \caption{Haptic hand-base controller \cite{exoskin}}
    \label{fig:hand-deploy}}
    \end{subfigure}
    \caption{Examples of different hardware categories for enabling the Metaverse}
\end{figure*}

The primary sensory aspect of the Metaverse is audiovisual. At the moment, the majority of immersive Metaverse-related applications are provided entry using head-mounted displays (HMDs) such as the one in figure \ref{fig:hmd}. Size, field-of-view, resolution, latency, audio quality, and battery life are some of the important factors that make a HMD device perform better than another. Despite the fact that modern HMD models cater to Metaverse-related applications, the notion of HMDs has been proposed a while ago in different augmented reality-related applications such as two head-mounted monocular displays (HMMD) and a held-to-head binocular display (HHBD) in the domain of air-traffic controllers \cite{ruffner2004near}. The way HMDs function is they track the orientation of the movement of the head while it moves, and proportionally adjust the movements of the images and the environment inside the Metaverse. Many works in the literature highlight and discuss the notion and the uses of HMDs. For example, in \cite{LAM2020120035}, the authors focus on the optical engineering aspect of such devices and divide them into three categorical optical solutions, (1) macro-optics, (2) micro-optics, and (3) nano-optics and highlight the characteristics of each. The authors of \cite{ltoh-2021-towards} provide a survey of HMDs in terms of requirements and goals to build interfaces that make the real and virtual worlds submerge. The authors also focus on the challenges such as spatial, temporal, and visual realism that make building such devices and interfaces, not a straightforward process. In \cite{huang-hmd-engineeringeducation}, the authors perform a review of how HMDs in VR can assist and make engineering education better for students and prepare them for engineering careers. In \cite{song2021effectiveness}, the authors study the practical aspects of the effectiveness as well as the advantages and the limitations of conducting training with the help of HMDs in advancing professional skills and safety. Similarly, in \cite{han-2021-measuring} the authors measure the impact of VR in the domain of construction design and perform comparisons between the use of HMDs against regular desktop monitor usage. The authors of \cite{stein2021comparison} also provide a practical work by comparing the eye-tracking latencies between currently commercially available HMDs in the market, while the authors of \cite{pfeil2021distance} research how the distance perception would differ between users wearing a HMD compared to people without a one. \\

By now, it is a known fact that the use of HMDs in immersive VR and AR can sometimes cause dizziness and headaches known as cybersickness, which provides a non-enjoyable experience for users of such head-mounts. In fact, cybersickness is one of the main reasons that drive away users from the use of VR. The notion of cybersickness has been studied and reviewed a lot in the literature. In \cite{caserman2021cybersickness}, \cite{klevckova2021cybersickness}, the authors review and highlight the works done in terms cybersickness caused by the use of modern HMDs and their effects on the user, while the authors of \cite{tian2022review} focus on cybersickness from the perspective of individual susceptibility. Finally, the authors in \cite{ang-2022-uneven} study the effect of various virtual terrains perceived from a HMD, which could lead to cybersickness. \\

\subsubsection*{Hand-based Input}
Another major hardware enablers for the Metaverse are the hand-based input devices \cite{li2021armstrong} \cite{bouzbib2021can}. As the name suggests, these devices are mainly controlled by the hands. Based on the design and the objective of the device, they can be held by the hand such as handles or controllers (figure \ref{fig:hand-hold}), or, they can be based on haptic sensors and installed/worn on the hand (figure \ref{fig:hand-deploy}). Haptic-based input devices add the feeling of the physical presence of certain objects as they can be touched and felt in the virtual space from one side, and allow the user to sense the changes happening in the virtual environment, such as moving objects or accordingly taking physical actions such as pressing a button from the other side. Such devices can facilitate several actions, as well as  interactions. For example, these devices can assist in (1) \textbf{navigation tasks}, such as panning, pointing, teleporting, and virtual walking. Haptic-based input devices also assist in (2) \textbf{interactions} by performing the exact hand-based tasks users do in real life such as sensing and real touch. Another aspect the haptic-based devices aid in is the (3) \textbf{exploration}. This allows the users to touch items in the environment and understand their nature as well as their details, such as the shape of the object. Exploration is mainly done through cues, either tactile or kinesthetic. Another aspect that haptic-based input devices facilitate is the (4) \textbf{manipulation} of objects, such as changing the direction, orientation, or position of a certain object in the virtual reality world. Another aspect is the (5) \textbf{modification} of objects. This is usually referred to as altering properties of objects other than their position or direction \cite{bouzbib2021can} (e.g., physical properties). Furthermore, actuated haptic devices can assist in replicating social scenarios that can elevate one's feelings and emotions by touching and interaction \cite{teyssier2020conveying}, as well as enabling users to feel experiences similar to what they would feel in real life such as tension, force, and resistance. \\

Hand-based haptic and sensory notions have been around for a while \cite{folgheraiter2006new}, \cite{folgheraiter2000blackfingers} \cite{vallino1999haptics} \cite{nojima2002smarttool}, but with the advent of the Metaverse, more haptic-based and HMD VR concepts are being proposed to provide near real-life experiences for users in virtual reality settings and in the Metaverse. For example, \cite{han2018haptic} proposed a novel system called 'Haptic Around' which is a hybrid system that can replicate multiple tactile sensations in VR. The system uses basic tools such as a hot air blower, a fan, and a tool that can create mist as well as lighting to provide a fully immersive experience and mimic sensations such as desert heat or snow cold. Another idea is in \cite{murakami2017altered} where the authors come up with a proposal dubbed as 'Altered touch' in which a small form factor fingertip-based haptic display is developed. The Altered touch can sense force, thermal, and tactile feedback and integrate with augmented reality. The display then is used to modify properties of real objects such as soft, hard, hot, and cold sensations in augmented reality. In \cite{li2021armstrong} the authors propose 'Armstrong' which is an arm-anchored augmented 3D virtual User Interface (UI) that can be seen with the help of a HMD. Users can view the content of the virtual display with the HMD and interact with the content such as scrolling or pinching by using hand-based input devices. Research in this regard is also taking place in the industry in cooperation with universities. Meta has collaborated with Carnegie Mellon University to work on an open-source skin called ReSkin \cite{meta-reskin}. ReSkin can be categorized as a generic tactile-based skin that can mimic sources of contact and assist in haptic AI research. Producing ReSkin is quite inexpensive and Meta is offering with it a much broader open-source ecosystem for the touch processing domain which will include tactile hardware, simulators, data sets, and benchmarks for the AI community to benefit from. \\

\subsubsection*{Contact Lenses \& Glasses}
Other types of input devices such as AR contact lenses and glasses allow bypassing the usage of smartphone cameras or even a HMD to access the Metaverse \cite{morimoto2023optically}. Users need only to wear the lens the way they wear contact lenses. In addition to providing a gateway to the Metaverse, the lens also corrects the vision of its user. Users can control it with simple eye movements. Another tool for augmented reality outside HMDs and hand-based sensors is the use of smart glasses. A major advantage for these Metaverse enablers is that they shift away from big and large form-factor head-mount devices and provide similar services that the large-sized devices provided in much smaller and compact factors \cite{wu2022metaverse}. In that sense, they are a pleasant upgrade for what users actually now already use in their daily life such as contact lenses and glasses.\\

\subsubsection*{Wristband}
Furthermore, a newer category of hardware has emerged recently which is the usage of wristbands. These are small devices, worn similarly to watches, and can detect and process slight finger movements and accordingly convert them into other types of inputs and commands. Such wristbands are being integrated more into AR systems where users can type without any actual physical keyboards, by just tapping on any surface, or play games without any physical controllers by just moving their fingers \cite{stokel2021facebook}. \\

\subsubsection{A Review of Business Driven Hardware}
Depending on the type of experience, (i.e., AR, VR, MR), different types of gears would provide different levels of accommodation. For example, for immersive experiences such as going to concerts or museums, VR headsets are needed, while for online shopping and e-commerce AR experiences, different types of equipment might be more adequate. It is noteworthy that even software and web companies such as Microsoft and Meta have jumped on the wagon of producing Metaverse-related hardware. As time passes, and more companies enter the Metaverse race, the challenge is no more producing any Metaverse gear, but rather producing affordable and lightweight Metaverse gear to mass people to purchase and use. Furthermore, in addition to producing hardware, companies are also trying to provide Metaverse-related environments and platforms for developers so that apps, games, and different kinds of software can be written to derive a Metaverse App community which will indirectly increase the usage of the produced hardware. To this end, we showcase in Table \ref{tab:Metaverse_hardware} some of the current Metaverse-related gear and equipment that is being worked on. For each type of device, we mention the company behind it, the product's name, its type, what subdomain of extended reality it serves, and the price. 
In the remainder of this section, we review the existing market products provided by major companies interested in developing hardware for the Metaverse. We highlight the name of the company in bold, and italicize the product name.\\

\newcolumntype{B}{ >{\centering\arraybackslash} m{0.15\linewidth} }
\newcolumntype{C}{ >{\centering\arraybackslash} m{0.20\linewidth} }
\newcolumntype{D}{ >{\centering\arraybackslash} m{0.17\linewidth} }
\newcolumntype{E}{ >{\centering\arraybackslash} m{0.12\linewidth} }
\newcolumntype{F}{ >{\centering\arraybackslash} m{0.25\linewidth} }

\begin{table*} 
\centering
\caption{Metaverse Related Gears \& Equipment} 
\label{tab:Metaverse_hardware}
{%
\begin{tabularx}{\textwidth}{|B|C|D|E|F|} \hline 
\small
\textbf{Brand} & \textbf{Name} & \textbf{Device Type} & \textbf{Domain} & \textbf{Price} \\
\hline 
Meta & Meta Quest Pro & Headset/Controllers & VR & Starting \$1,700.00 USD  \\ \hline 
Lenovo & ThinkReality VRX & Headset/Controllers & VR & To be announced later  \\ \hline 
Meta & Meta Quest 2& Headset/Controllers & VR & Starting \$400.00 USD  \\ \hline 
Sony & PS VR 2 & Headset/Controllers & VR & Starting \$549.00 USD    \\ \hline 
Valve & Valve Index & Headset/Controllers & VR & Starting ~\$730.00 USD    \\ \hline 
Microsoft & Hololens 2 & Headset & AR/MR  & \$3,500.00 USD  \\ \hline 
Microsoft & Hololens 2 Industrial & Headset & AR/MR  & \$4,950.00 USD  \\ \hline 
Microsoft & Trimble XR Hololens 2 & Headset & AR/MR  & \$5,199.00 USD  \\ \hline 
HP & Reverb G2 & Headset/Controllers & VR & \$599.00 USD  \\ \hline 
Magic Leap & Magic Leap 2 & Headset/Smartglasses & AR & Starting \$3,299.00 USD  \\ \hline 
HTC & Vive Flow & Headset & VR & Starting ~\$500.00 USD  \\ \hline 
HTC & Vive Pro & Headset/Controllers & VR & Starting ~\$1370.00 USD  \\ \hline 
HTC & Vive Focus 3 & Headset & VR & Starting ~\$1300.00 USD  \\ \hline 
HTC & Vive Cosmos & Headset & VR & Starting ~\$500.00 USD  \\ \hline 
HaptX & HaptX Gloves G1 & Haptics/Gloves & Hand-based & Starting \$5,495.00 USD   \\ \hline 
Google & Glass 2 & smartglasses & AR & \$999.00 USD  \\ \hline 
Epson & Moverio 2 & smartglasses & AR & \$699.00 USD  \\ \hline 
Vuzix & M400 & smartglasses & AR &  Starting\$1800.00 USD  \\ \hline 
Vuzix & BLADE 2 & smartglasses & AR & Starting \$1300.00 USD  \\ \hline 
Mojo Vision & Mojo Lens & Smart Contact Lenses & AR & To be announced later \\ \hline 
TAP & TAP XR & Wristband & AR/VR & \$249.00 USD \\ \hline 

\end{tabularx} }
\end{table*} 

\subsubsection*{HMD Devices}
After purchasing the Metaverse headsets producing company \textbf{Oculus}, \textbf{Meta} revamped the headset line and currently offers the \textit{Meta Quest} \footnote{https://www.meta.com/quest/} series of head-mount devices in two forms. The \textit{Meta Quest Pro} as well as the \textit{Meta Quest 2 series}. The difference lies in the hardware specifications between the devices. The Quest Pro version uses thinner lenses than the Quest 2 version as well as it has better resolution. However, both provide an immersive experience and entrance to the Metaverse world. The devices can be used in end-user applications such as games and creative apps, as well as enterprise-level applications such as doing meetings and other work tasks in the Metaverse.

\textbf{Lenovo} has been releasing Metaverse-related equipment and gear since 2018. Their latest release is called \textit{ThinkReality VRX} \footnote{https://www.lenovo.com/thinkrealityvrx} and offers solutions for immersive collaborations and training  on the enterprise level. An interesting notion about the device is that it supports an open ecosystem for developers to write customized software for the ThinkReality VRX headsets through its XR developer program called \textit{Snapdragon Spaces} using their \textit{openXR} based SDK (Software development kit). 

Even though \textbf{Microsoft} is a software company at heart, they also have been manufacturing and offering hardware for a while such as laptops, mouses, keyboards, headphones, and webcams. They have also been manufacturing Metaverse-related equipment through their \textit{Hololens} \footnote{https://www.microsoft.com/hololens} line of headsets. Currently, their Hololens comes in three different editions that caters to different target audiences. The base edition which is the \textit{Hololens 2} can be used in regular environments, while the \textit{Hololens 2 Industrial edition} and the \textit{Trimble XR10} with HoloLens 2 editions are geared more towards industrial settings. Microsoft also offers a development edition that developers can use it to build MR-based applications for the device. 

\textbf{HP} has been producing computing hardware since its inception in 1939, that's why it is no shocker that they are also producing Metaverse-related equipment. Their product is called \textit{HP Reverb G2} \footnote{https://www.hp.com/us-en/vr/reverb-g2-vr-headset.html} and is a HMD with hand-controllers which can be utilized for VR-related applications in the domains of Architecture and engineering, healthcare, education and training. 

\textbf{Magic Leap} was founded in 2010 as a dedicated visual gear-producing company. They produce head-mount base equipment geared towards enterprise and industrial Metaverse-related applications called \textit{Magic Leap 2} \footnote{https://www.magicleap.com/magic-leap-2}. It has a much compact and simpler form factor than traditional head-mount equipment in the sense that it can be worn as eyewear, although it is bigger than a regular eyeglass size. They also provide tools for programmers so they can develop software for the device. 

Phone maker \textbf{HTC} is also producing Metaverse-related headsets. Their series is named \textit{Vive} \footnote{https://www.vive.com/} and contains sub-ranges of products with varying hardware capabilities. \textit{Valve} also has its own Metaverse called \textit{Viverse} \footnote{https://www.viverse.com/} which is a full-fledged Metaverse where users can join and partake in events as well as purchase digital artwork. 

\subsubsection*{Gloves, Glasses, Contact Lenses, \& Wristbands}

The company \textbf{HaptX} \footnote{https://haptx.com/} provides haptic gloves to achieve touching and haptic sensory-related tasks in the Metaverse. Their \textit{G1} gloves line provides micro-fluid-based technology geared toward industrial and mission-critical applications. HaptX also provides a library of tools in the form of an SDK where developers can write applications for Unreal and Unity engines for the Metaverse. 

Moreover, \textbf{Google} produces AR-related gear through their \textit{Google Glass 2} \footnote{https://www.google.com/glass/start/} product which is geared towards enterprises. Unlike traditional HMDs, Glass 2 is lightweight and compact which uses a transparent display and allows for hands-free work, which can prove to be useful in some industrial tasks. 

Similar to Google's Glass 2, \textbf{Epson} and \textbf{Vuzix} provide lines of AR enablers through eyeglasses. Epsons's \textit{Moverio} \footnote{https://tinyurl.com/nh49sy4d} provides AR gateways and usability in different industries, while Vuzix \footnote{https://www.vuzix.com/} has several sub-range models which integrate with various \textit{MDM} (mobile device management) software to provide ease of use. 

Apart from HMDs and smartglasses, some companies such as \textbf{Mojo Vision} \footnote{https://www.mojo.vision/} are producing smart contact lenses that can be worn on the eye and provide AR experiences for the user. This is similar to wearing a traditional contact lens. In addition to enabling AR experiences for the wearer, The \textit{Mojo lenses} provide also prescriptive features for the eye in case the user had to wear contact lenses anyhow. 

Finally, the wristband from \textbf{TAP} dubbed as \textit{TAP XR} \footnote{https://www.tapwithus.com/product/tap-xr/} provides a solution to the Metaverse which eliminates the need for physical or even virtual keyboards and controllers by enabling the user on tapping any surface or even pointing to a direction and accepting that as an input command. Furthermore, \textbf{Meta} developed the \textit{Metaverse Wristband} \cite{stokel2021facebook}, which tracks the communication between the nerves and the brain once worn. The purpose of this device is to understand and adapt to user movements.\\

\subsubsection*{Gaming Consoles}
\textbf{Sony}'s Play Station game consoles have been major sellers in the electronic gaming industry. To this end, it is no surprise that Sony has also joined the Metaverse hardware race and now offers its own VR headsets known as \textit{PS VR 2} \footnote{https://www.playstation.com/ps-vr2/}. The VR headsets which also come with controllers are geared more towards gaming and integrate with the games in the PlayStation game consoles. Game developers can develop their games and add immersive experiences for the players. 

The company \textbf{Valve} has been developing some of the most played games in the industry since its formation such as Half-Life, Dota, and Counter-Strike. They also own and operate one of the biggest game portals in the world known as \textbf{Steam}. Similar to Sony, they have also been producing their own VR headset that accommodates their games to provide more immersive VR experiences to their users. Their model is called \textit{Valve Index} \footnote{https://store.steampowered.com/valveindex} and presents an entire VR kit that comes equipped with a headset and hand controllers. 

\subsubsection{Hardware Enabling Technologies}
We study in this section the list of enabling technologies that support Hardware development and growth in the market of the virtual world as part of the Metaverse pipeline.

\textbf{AI.} With the rise of AI applications, its subdomains such as machine and deep learning (ML \& DL) have found their way to also being used inside hardware equipment pertaining to Metaverse-related applications such as HMDs. It is worth mentioning that several of the points discussed and highlighted in this section are used in conjunction with each other. For example, when Federated Learning (FL) is used with Metaverse-related hardware, privacy is also applied alongside computing factors. Thus, we can say that many of these criteria are intermingled and work together to provide common goals to the users. For instance, these devices are in need of access control and authentication. Works such as \cite{iris-dl} and \cite{benchmark-iris} use DL techniques for authentication and access control purposes. In \cite{iris-dl}, the authors make use of DL to provide better authentication mechanisms for the distorted iris regions in HMDs, while in \cite{benchmark-iris} the authors also make use of different DL techniques for iris recognition. The authors of \cite{doughty2021surgeonassist} leverage the advantages of Neural Networks (NN) and propose a lightweight framework, dubbed as SurgeonAssist-Net, which provides virtual assistance for a predefined set of surgical tasks. The proposed framework can be used on commercially available HMDs such as Microsoft's Hololens. The authors of \cite{holosound}, also make use of NNs in HMDs, however, this time, in the domain of speech and sound feedback, especially for the deaf and people with listening difficulties. Their approach relies on Convolutional Neural Nets and is called 'HoloSound' which utilizes deep learning for classifying and visualizing the identity of the sound and its location alongside providing speech transcription for the users.\\ \\

\textbf{User interface and interaction.}
Given the wearable nature of these devices, the way they are designed and the way the users will interact with them make all the difference in these devices' usability and adaptability. For example, hardware might be compact in size and have small form factors. To achieve this, the internal components of these devices should be smaller. This is becoming easier to be achieved because semiconductors are also becoming smaller. Furthermore,  smaller micro-electromechanical systems (MEMS) and compact batteries which have longer-lasting capabilities are also a requirement for designing more user-friendly Metaverse gear \cite{george2021Metaverse}. The effects of user interfaces with HMDs are highlighted in \cite{jin2022comparison}, where the authors discuss the impact of various interfaces in the realm of interactive narratives on AR-related hardware such as HMDs. There are several types of interfaces that can be used in AR, such as Natural User Interfaces (NUI), Tangible User
Interface (TUI), and Graphical User Interfaces (GUI). Providing users with enjoyable interfaces for AR/VR hardware has been the goal of several research works. For example, the authors of \cite{sym12010053} propose an asymmetric interface for HMD (direct interaction) and non-HMD (multi-viewpoint interaction) users that provide a better experience with respect to their virtual reality environment. The authors of \cite{sym11040476} propose a novel collaboration-based interaction method between users wearing HMDs in virtual reality environments. This is achieved through a communication interface that allows the users to exchange information and feedback with their hands and feet. \\

\textbf{Computing \& Power.} Given HMDs aim to provide more immersive experiences, they render and contain multiple views per instance as opposed to 2D imaging. Accordingly, this requires more computing power from the HMD. The authors of \cite{power-evaluation} compare the power consumption of HMDs that offer 3D rendered views to traditional 2D views and provide statistical analysis as to which modules of the HMD consume more power and provide a couple of power-saving mechanisms. Since HMDs are getting smaller in size, their components are also becoming smaller in size. Several attempts to ease the burden on the internal components of the HMDs (i.e., CPU, GPU) in terms of computing and rendering are being moved to software-based techniques. For example, the authors of \cite{oled} use foveated rendering which is a method that reduces the rendering computation of the GPU inside a HMD. Besides, the authors of \cite{Real-time-correction} try to overcome the problem of optical distortions on the level of GPU in real-time using a software-based approach where a distortion map is built to fix the distortion and fix the quality of the images. Finally, in \cite{temporal-real-time}, the authors convert the process of asynchronous time warping (ATW) that reduces the motion-to-photon latency in HMDs into a programmatic one by proposing a field-programmable gate array (FPGA) based approach which handles the temporal quality compensation without the need for any ATWs.  \\

\textbf{Security.} Just like any network-oriented component, HMDs, and other Metaverse-related hardware and equipment are vulnerable to several security risks. These real-time devices create and capture large amounts of input that can be sensitive. For example, hackers can change the perceived reality and inject poisonous data that can have a detrimental effect on the usability of the AR or VR application, such as the driver's focus on a road. Other attacks can include displaying falsified information and causing cybersickness \cite{beyond-reality}. Another important security aspect in Metaverse-related hardware is the issue of authentication. Recently, biometric authentication has come to the surface given its little memory load \cite{review-multimodal}.
The authors of \cite{luo2020oculock} propose a novel method of authentication for HMDs called 'Oculock' which is a human visual system (HVS) based approach. The proposal works through an electro-oculography (EOG) driven human visual system sensing framework and fast authenticating scheme in which different physiological and behavioral features, which are extracted to assist in the authentication. In \cite{LookUnlock}, the authors present a different authentication approach called 'LookUnlock', which uses passwords made from spatial and virtual targets from the environment. This approach can circumvent shoulder-surfing attacks and provide an additional layer of security for HMD users. Finally, the authors of \cite{fusing-iris} use iris and periocular recognition to authenticate users through the HMDs, which capture the eye image. By using score-level fusion for the iris and the periocular regions, the authentication performance is improved. \\

\textbf{Privacy.} As these wearables become more ubiquitous and easily affordable commercially, users are more concerned about their privacy during their usage. On the hardware level, unlike other attacks such as the one targeting the application layer, the concern is to keep the low-level code that makes the hardware run safe, secure, and easily available for patches in case of a compromise. Attackers can use hardware-level attacks to take over a device, and later work their way into other layers of the hardware and siphon information. Furthermore, the amount of data captured by these devices makes them interesting platforms to perform different types of learning (i.e., ML, DL). However, these models would entail access to private data. Several works have tried to tackle this issue by proposing the usage of FL \cite{fl-survey} with HMDs. For example, in \cite{chao2022privacy}, the authors propose a new framework that preserves the user's privacy and uses FL in online viewport prediction in 360-degree streaming. Unlike other learning scenarios, collected and captured data does not leave the HMD towards 3rd party storage for analysis, rather the learning takes place on the HMD. In \cite{wang2021output}, the authors also use FL coupled with RL to solve the multi-user AR output strategy problem. In their approach, the RL model gets trained on individual AR devices to produce an output strategy, which is aggregated on federated servers.\\

\textbf{Ethics and Sociopsychology.} Various literature efforts have studied the ethical implications of this hardware as they directly interact and intersect with different social, behavioral, and cognitive aspects of humans in general. For example, the authors of \cite{asking-ethical} supply an overview of various ethical principles in research especially pertaining to child development. The authors of \cite{SOUTHGATE201919}, also discuss and explore the ethical as well as safety effects of using existing hardware in low-income high schools, especially in information technology and science classes. Moreover, immersive experiences through various AR/VR glasses and HMDs have had social and psychological impacts. For example, the authors of \cite{bradley2018autism} provide a systematic review regarding the use of HMDs in the education of autistic individuals. Moreover, the authors of \cite{fieldtrip}, highlight the benefits of performing a virtual field trip using HMDs, against showing parts of the trip in regular 2D videos. The reviews reveal that using immersive experiences via the HMDs has more positive effects in terms of self-efficacy and interest near the students even after a couple of weeks from the experiment. The authors of \cite{sports} examine the effects of VR in the domain of sport psychology practice. Their work discusses some of the practicalities of using VR technologies through HMDs and provides recommendations to stakeholders that work in the domain of sports psychology as to how they can use it in their practice. \\

\subsection{XR Frameworks}
XR Framework is the second component in the Metaverse Infrastructure layer of the pipeline. An extended reality (XR) framework is a software development kit (SDK), toolkit, or engine used to create a decentralized Metaverse for augmented reality and virtual reality devices. It allows individuals to easily create and develop applications for the Metaverse by combining various open-source tools and engines. Individuals and business owners can use an XR framework to create 2D/3D environments, manage and coordinate interactions with XR devices, and establish networking connections using client-server technology, among other things. It helps to alleviate the burden of building these tools from scratch and can provide a range of benefits for those looking to develop applications for the Metaverse.
{\color{white}
\footnote{https://www.nvidia.com/en-us/design-visualization/technologies/holodeck/ \label{tbn:holo}}   
\footnote{https://www.meetup.com/colorado-vr-ar/ \label{tbn:covar}}
\footnote{https://unity.com/ \label{tbn:unity3d}}                                  
\footnote{https://www.microsoft.com/en-us/mesh \label{tbn:microsoft}}
\footnote{https://www.spatial.com/ \label{tbn:spatial}}
\footnote{https://www.infinitecanvas.gg/ \label{tbn:infinitecanvas}} 
\footnote{https://create.roblox.com/docs/reference/engine \label{tbn:roblox}}
\footnote{https://developer.oculus.com/ \label{tbn:oculus}}
\footnote{https://www.blender.org/ \label{tbn:blender}} 
\footnote{https://store.steampowered.com/steamvr \label{tbn:steamvr}}
\footnote{https://docs-multiplayer.unity3d.com/ \label{tbn:unitymlapi}}
\footnote{https://www.photonengine.com/ \label{tbn:photon}} 
\footnote{https://www.openmined.org/ \label{tbn:openminded}} 
\footnote{https://aframe.io/ \label{tbn:aframe}}
}

\begin{table*}
\centering
\caption{Emerging XR Tools, Engines, and Frameworks in the Market}
\label{tab:xr_frameworks_ex}
\resizebox{\textwidth}{!}{%
\begin{tabular}{|l|l|}
\hline
\textbf{Tool/Engine/Framework}                      & \textbf{Usage}                                                                                                                           \\ \hline
NVIDIA   Holodeck \footref{tbn:holo}                               & Photorealistic Collaborative Design in VR                                                                                              \\ \hline
CoVAR \footref{tbn:covar}                                              & Remote collaboration tool for AR/VR applications                                                                                       \\ \hline
Unity 3D   engine \footref{tbn:unity3d}                            & Creating 2d/3d   materials, rendering, physics,…                                                                                         \\ \hline
Microsoft Mesh \footref{tbn:microsoft}                                     & Tools for spatial rendering, holoportation technology and avatar creation                                                              \\ \hline
Spatial 3D   technologies \footref{tbn:spatial}                          & 3D modelling and data interop tools                                                                                                       \\ \hline
Infinite   Canvas \footref{tbn:infinitecanvas}                                 & Tools for development, engineering and supporting for gaming                                                                           \\ \hline
Roblox   rendering 3D engine \footref{tbn:roblox}                       & 3D development engine   for texturing, environment rendering                                                                             \\ \hline
Meta AR/VR   developer tools \footref{tbn:oculus}                       & Supports designing,   building, and supporting AR and VR applications                                                                     \\ \hline
Blender \footref{tbn:blender}                                           & Provide necessary   tools for creating visual effects, animated films, 3d models, and motion graphics                                   \\ \hline
SteamVR  \footref{tbn:steamvr}                                               & Decouples input logic   from individual controllers and VR headsets for user social interaction                                           \\ \hline
Unity Multiplayer Networking (MLAPI) \footref{tbn:unitymlapi}                   & Client-server   connections and maintain connectivity for the clients                                                                    \\ \hline
Photon Engine \footref{tbn:photon}                                         & Support multiplayer   game development in cross platforms                                                                                \\ \hline
OpenMinded \footref{tbn:openminded}                                        & Open source framework for tools, resources, and libraries related to privacy                                                              \\ \hline
A-Frame \footref{tbn:aframe}                                            & An open source web framework for tools, strategies, and engines related to business implementations                                       \\ \hline
\end{tabular}%
}
\end{table*}

\begin{table*}
\centering
\caption{Comprehensive Overview of XR Tools, Engines and Frameworks Available in the Literature Works}
\label{tab:xr_frameworks_prop}
\resizebox{\textwidth}{!}{%
\begin{tabular}{|l|l|}
\hline
\textbf{Tool/Engine/Framework}                      & \textbf{Usage}                                                                                                                            \\ \hline
Kim, et al.   2020 \cite{kim2020xr}                 & Create a virtual   environment in collaboration between its different entities using various XR   devices                                \\ \hline
Chua, et al.   2022 \cite{chua2022resource}         & Allocate resources   efficiently for vehicle users in the virtual world                                                                  \\ \hline
Xu, et al.   2022 \cite{xu2022metaverse}            & Encryption model for   the addresses of the users, their devices and services based on Blockchain   technology                           \\ \hline
Kang, et al.   2022 \cite{kang2022blockchain}       & Privacy-preserving   connection for the connected devices based on Blockchain                                                            \\ \hline
Zhu, et al.   2022 \cite{zhu2022metaaid}            & MetaAID: enrich language and semantic technologies in creating digital avatars and twins                                                \\ \hline
Yang, et al.   2022 \cite{yang2022secure}           & Protect the security   and privacy of the user data                                                                                      \\ \hline
Chu, et al.   2022 \cite{chu2022metaslicing}        & Optimize resources   utilization and improve the users' quality of service                                                               \\ \hline
Steinau, et   al. 2019 \cite{steinau2019dalec}      & Systematically   estimating and comparing data-centric procedures for business lifecycle   processes                                     \\ \hline
Canhoto, et   al. 2020 \cite{canhoto2020artificial} & Map AI components in   identifying the value of artificial intelligence and machine learning for   businesses                            \\ \hline
Lee, et al.   2021 \cite{lee2021e3xr}               & E3XR: Analyze the   design of an XR system based on the ethics and learning theory of the human   aspects                                \\ \hline
Gong, et al.   2021 \cite{gong2021framework}        & Improve future XR   systems' usability and user acceptance                                                                               \\ \hline
Rompapas, et   al. 2021 \cite{rompapas2021project}  & Project Esky:   Collaboration in creating high-fidelity XR experiences among different XR   devices                                       \\ \hline
Kern, et al.   2021 \cite{kern2021off}              & OTSS: 2D/3D Drawing   and sketching with XR devices                                                                                      \\ \hline
Cannavo, et al.	2020	\cite{cannavo2020blockchain}&	Facilitate decentralized ownership and management of virtual assets using Blockchain technology	                                         \\ \hline
Bouachir, et al.	2022	\cite{bouachir2022ai}	&	Usage of AI for Metaverse intelligent agents to create, explore and distribute content over the Metaverse	                            \\ \hline
Yu, et al.	2021	\cite{yu2021developing}	        &	Usage of AI technology by recommending a personalized learning path or feedback on the user's performance	                            \\ \hline
De, et al.	2019	\cite{de2019security}	        &	Security controls that can be used for XR frameworks to enrich security aspects of XR applications	                                     \\ \hline
Chang, et al.	2022	\cite{chang2022metaear}  	&	Continuous authentication approach to authenticate users continuously while in the Metaverse	                                        \\ \hline
Cai, et al.	2022	\cite{cai2022compute}          	&	Studies the acceleration of the Metaverse development on data storage, processing and streaming	                                         \\ \hline
Han, et al.	2022	\cite{han2022comic}	            &	Propose tools to overcome the challenges of the limited power source and networking bandwidth speed	                                     \\ \hline
Dagher, et al.	2018	\cite{dagher2018ancile}	    &	Address  privacy and security issues in the healthcare industry	                                                                          \\ \hline
Nair, et al.	2022	\cite{nair2022exploring}	&	Address the privacy risks associated with the Metaverse applications               	                                                   \\ \hline
Bibri, et al.	2022 \cite{bibri2022metaverse}	&	Provide practices to maintain an ethical and private Metaverse interactions	                                                             \\ \hline
\end{tabular}%
}
\end{table*}

An XR framework typically includes hardware and software components that work together to create and manage immersive virtual experiences. When creating a system architecture to develop an XR framework, several development phases need to be considered to have a complete framework infrastructure that could address the different needs of developers and artists in creating the Metaverse applications. For instance, individuals should consider what technologies the framework will address, where the application will work and using which technology, what hardware devices are required to run, and how we could visualize and interact with it. The technologies concerning the development of such an application need to address audio, video, text chat, client-server communication, graphical editors, and digital avatar creations. In addition to these hardware and software components, an XR framework may also include guidelines and standards for developers to follow when creating immersive virtual experiences within the framework. These guidelines may cover user experience, performance, security, and other issues. We have listed the XR tools, engines, and frameworks in Table \ref{tab:xr_frameworks_ex}. The table includes a list of popular and widely-used technologies and a brief description of their features and capabilities. Additionally, in Table \ref{tab:xr_frameworks_prop}, we have included literature proposed references for further research and development in the field of XR. \linebreak


\textbf{Communication and Networking.} In terms of networking and communication, frameworks should maintain a smooth communication layer between the user interaction with the Metaverse and the users' interaction with themselves in the virtual environment. As such, XR devices request intensive resources, and 5G may face performance limitations. Therefore developers should be able to identify the network challenges and deliver an AR/VR capability in the networks. Some infrastructure technologies can be considered for the matter, such as caching, multi-casting, traffic engineering, quality of service (QoS) optimizations, etc. Other solutions, such as Unity Multiple Networking (MLAPI) \footref{tbn:unity3d}, and Photon Engine \footref{tbn:photon} can serve as client-server connections and maintain connectivity for the clients with the usage of Remote Procedure Calls (RPC). Such tools can enrich the networking layers, and developers widely use them. Furthermore, protocols should adapt for updating user positions with low latency or receiving audio/visual sequences with minimal delays, for that matter. A work in \cite{chua2022resource} addresses the issue of transmission latency and provides a potential solution for the system to allocate resources efficiently for vehicle users in the virtual world.\\

\textbf{AI.} AI has been constantly used in application development, and Metaverse creation using XR frameworks can benefit from the digital intelligence it provides. AI capabilities can enhance realism and immersion, facilitating natural language interaction between users and enabling intelligent agents for an immersive and interactive experience. Examples of potential AI applications for XR frameworks include chatbots, natural language processing, avatar creation, efficient resource optimization, and image and object detection. AI will be working behind the scenes in creating customized avatars to connect people from different cultures by minimizing the language barriers between them and helping with user interaction and object visualization in the virtual world. Specialized models in each field will need to be trained to adapt to the problem in every possible aspect, where intelligence is critical for developing the Metaverse. For instance, the work in \cite{zhu2022metaaid} proposes a framework named metaAID that enriches language and semantic technologies in creating digital avatars and twins. It enables the Metaverse application's content creation while addressing the users' preferences and needs. The framework uses AI in the creation of intelligent agents. In addition, the proposed framework allows humans to customize and personalize the surrounding environment based on their preferences. A different work in \cite{bouachir2022ai} discusses the usage of AI to build Metaverse intelligent agents to create, explore and distribute content over the Metaverse. While another work in \cite{yu2021developing} discusses using an intelligent virtual reality interactive system to learn how to brew pour-over coffee. The system is based on the ADDIE model, a widely used instructional design model for Analysis, Design, Development, Implementation, and Evaluation. The system leverages AI technology to allow the VR environment to adapt to the users by recommending a personalized learning path or feedback on the user's performance.\\

\textbf{Blockchain.} Blockchain technology can facilitate new immersive experiences and interactions in an XR framework. Specifically, Metaverse communication and enabling transparent and secure tracking of user activity could be developed based on Blockchain technology, in what capability and features it could provide for users' privacy and security. One work that considers Blockchain in XR frameworks \cite{xu2022metaverse} in which it proposes an encryption model for the addresses of the users, their devices and services based on Blockchain technology, which adds anonymity over the connectivity of physical and virtual entities across the users and devices used with the connection. The system leverages the distributed and secure nature of Blockchain to create and exchange virtual assets. Furthermore, the work in \cite{cannavo2020blockchain} proposes the usage of Blockchain and XR technologies to enable decentralized marketplaces for trading virtual goods and services within the XR environment, facilitate decentralized ownership and management of virtual assets, and enable the invention and distribution of immersive content. \\

\textbf{Security.} Introducing security techniques into any XR framework can help users verify and authenticate themselves when they need access to the Metaverse or while meeting other avatars. Authentication and recognition of the users must be a top priority in any XR framework development to ensure that this data is protected from unauthorized access. Some tools that can be used to protect against such malicious activity are using authentication factors, including face and voice recognition and two-factor authentication using physical and behavioral biometrics. A work in \cite{yang2022secure} introduces an XR framework where two-factor authentication can be used based on chameleon signature and biometric-based authentication to protect the security and privacy of the user. The paper discusses developing a secure authentication framework to ensure the traceability of avatars in the Metaverse. The authors proposed a framework that combines Blockchain technology with biometric authentication to enable the traceability of avatars within the Metaverse. Furthermore, the framework uses Blockchain to store users' biometric data and facilitate the creation of unique identities for their avatars. This would allow the avatars to be traced back to their real-world counterparts. Another work that discusses the security aspect of an XR framework is presented in \cite{de2019security}, where a study is done on XR security and the system's privacy. The authors identified several security controls that can be used for supporting XR frameworks, including data encryption, access controls, firewalls, intrusion detection systems, secure transmission protocols, and secure boot and firmware update procedures. In addition, introducing Blockchain and AI technologies can boost the security of XR systems. Furthermore, the work in \cite{chang2022metaear} uses a continuous authentication approach to authenticate users continuously while in the Metaverse. The system uses an acoustic channel for authentication based on the ear shape of the users and the sounds they emit while interacting in the XR environment. Moreover, The system uses ear shape recognition and audio processing techniques to verify users in the background continuously. \\

\textbf{Computing.} Edge and fog computing can boost the development process of XR framework. Such technologies are being used in XR frameworks to enable real-time processing and interaction with virtual environments and objects and to support the expansion of intelligent XR frameworks. A work in \cite{cai2022compute}  shows how the acceleration of the Metaverse development will result in excessive demands on data storage, processing and streaming. The demands of such Metaverse applications will increase the merging of the cloud into the network and integration of the edge and fog computing. Furthermore, the work in \cite{han2022comic} proposes a framework that provides various tools to overcome the challenges of the limited power source and networking bandwidth speed through the usage of multi-use motion prediction, encode/decode architecture for creating collaborative content, facilitate rendering through cooperation in local and remote tasks. In addition, to address the high demand for resources and computing power, a work in \cite{chu2022metaslicing} proposes admission control algorithms powered by AI to optimize resource utilization and improve the users' quality of service.  \\

\textbf{Business.} An XR framework for business can be used in marketing and sales to demonstrate the effectiveness of a product and engage further with customers. Furthermore, the XR framework can be used for training and education to create a simulated environment where users can practice operations and procedures safely. In addition, collaboration and teamwork will be gained through the immersive experience. Some literary works propose a framework for systematically estimating and comparing data-centric procedures for business lifecycle processes aiming at improving the efficiency and effectiveness of the business management process \cite{steinau2019dalec}. The system consists of a set of evaluation methods and criteria to be flexible and adaptable to different management processes. Another framework is proposed to map AI components in identifying the value of AI and machine learning for ML \cite{canhoto2020artificial}. The framework consists of identifying business objectives using AI and ML, assessing the risks and challenges, evaluating value creation and destruction, and developing strategies to mitigate potential risks. In addition, Unity and Unreal engine offer visual edits, networking communication tools, and physics engines that can help developers create immersive and interactive customer experiences to enrich business applications. Furthermore, A-Frame \cite{aframe} is an open-source web framework that can be used for XR framework development, which offers a wide range of tools for developers to enrich their business products. Such strategies, tools, and engines can be leveraged to integrate with the development of an XR business to facilitate business growth. \\
    
\textbf{Privacy.} The individual should be able to assess how the information is collected and for what reason, avoid any unnecessary data collection or user privacy-preserving tools when necessary, and integrate them into their framework development. In \cite{kang2022blockchain}, the authors propose an XR framework based on Blockchain technology, which uses FL to provide a privacy-preserving connection for the connected devices so that it uses model weights instead of private row data with a third party—enabling decentralized machine learning for the industrial Metaverse. Such techniques can be used when developing an XR framework to enrich its privacy and security capabilities. A work in \cite{dagher2018ancile} is proposed, aiming at analyzing how the framework could address persistent privacy and security issues in the healthcare industry, in addition to the interaction with the various needs of patients, providers, and third parties. Furthermore, a different work in \cite{nair2022exploring} discusses the privacy risks associated with the Metaverse regarding how adversaries can gain access to personal data from a popular Metaverse application like VRChat. The Metaverse hosts a massive amount of personal data that can be collected and processed. Such data may include location, biometric, behavioural, and financial data. The authors highlight effective measures to protect user privacy in this emerging virtual space. In addition, OpenMinded \footref{tbn:openminded} is an open-source framework that offers a variety of tools, resources, and libraries to create privacy-preserving XR applications that aim to protect the privacy of users.  \\

\textbf{Ethics \& Sociopsychology.} The Metaverse ethics regulation can be made at a framework level by integrating some regulations into the development process. Governments can take specific regulations for engines, tools, and frameworks to ensure ethical procedures are called upon when designing such systems. For instance, a framework introduced in \cite{lee2021e3xr} named E3XR aims to analyze the design of an XR system based on the ethics and learning theory of the human aspects. In addition, a study in \cite{bibri2022metaverse} sheds light on paying more attention to privacy and focuses on the ethical practices of the Metaverse and how it can affect surveillance in terms of the data, location and capitalism. Such practices and best ethical practices must exist in any XR framework to maintain a high level of privacy for the user by having moral rules to follow. Furthermore, several non-profit organizations and institutions \cite{stanfordvrVHIL, ieetHome, effElectronicFrontier} are already studying how emerging technologies such as XR affect the social and ethical considerations of humans. Moreover, they propose a wide range of resources, legal guides and policies to enhance the ethical considerations of XR frameworks.

Some efforts are integrated into an XR framework's design and development process. A framework is developed in \cite{gong2021framework} to improve future XR systems' usability and user acceptance. Furthermore, several institutions in \cite{stanfordvrVHIL, ieetHome, effElectronicFrontier} are also studying the sociopsychological impacts of emerging technologies on the users, such as how the technology can affect the user level of social presence, empathy and social interaction.

\subsection{Platforms \& MaaS}
On top of the previous two discussed components, the Hardware and XR Frameworks, enterprises can start implementing platforms that offer the base or infrastructure for Metaverse applications, including Metaverse-as-a-Service (MaaS) solutions. They are virtual platforms and ready-to-use solutions that combine AR, VR, and many other technologies to produce virtual experiences.
Platforms and MaaS form the third component in the Metaverse pipeline. This component comprise development and scripting tools, hosting and networking tools, as well as tools for monetization and commerce. MaaS companies may also offer supplementary services such as analytics, user interaction, and community management. This can enable new types of social interaction, entertainment, and commerce, as well as provide new chances for invention and experimentation by motivating companies and people to develop and operate their own virtual worlds without investing in the infrastructure. Decentraland and other Metaverse platforms facilitate the construction, trading, and exploration of virtual environments. The growing gaming and entertainment sectors have played a significant role in the evolution of Illuvium, Roblox, and Sandbox. Blocks of virtual real estate can be found on the Metaverse platform Bloktopia, and social networks can be found on ZEPETO. Major corporations are working to develop such platforms with the aid of MaaS so people can utilize the Metaverse for their employment, education, and social relationships. Metaverse platforms are also now widely used in the domains of education and medicine.

Several factors need to be taken into account when MaaS businesses offer their services and when deploying such enormous platforms to be efficient, trustworthy, safe, and most significantly, to target user requests since every platform must have specific targeted users. To achieve this, people must first specify the platform's intended users, the technology that must be utilized to fulfill the demands, the kind of computing power required to manage it, and the security and ethics measures that must be taken. \\

\begin{table*}[htp!]
    \centering
    \caption{List of platforms and industrial solutions}
    \begin{tabular}{|l|l|l|}
    \hline
        \textbf{Founder} & \textbf{Tool/Engine/Framwork} &  \textbf{Usage}\\
        \hline
        Verizon & BlueJeans & Video Platform \\
        \hline
        Nicolas Julia and Adrien Montfort & Sorare & Fantasy Football Games\\
        \hline
        Animoca Brands & The Sandbox & Own and Monetize Assets \\
        \hline
        David Baszucki and Erik Cassel & Roblox & Multiplayer Online \\
        \hline
        Ariel Meilich and Esteban Ordano & Decentraland & 3D Virtual World  \\
        \hline
        Kieran and Aaron Warwick & Illuvium & Play-To-Earn Game\\
        \hline
        Robert Gryn & Metahero & Gaming, Profile Pictures, and Social Media\\
        \hline
        Michael Wagner & Star Atlas & Conquer and Gather Game\\
        \hline
        Ross Tavakoli & Bloktopia & Study and Meet New People\\
        \hline
    \end{tabular}
    \label{tab:maas_industry}
\end{table*}

\textbf{AI.} 
Metaverse platforms employ AI algorithms to personalize the Metaverse experience for users by providing customized information, recommendations, and experiences, as well as by recommending virtual items for purchase based on their particular preferences and conduct. In \cite{artificialneww}, the author argues that NLP and data sharing technologies together with data visualization and sentiment analysis can enhance customer satisfaction and expectations in the Metaverse.  In addition, AI can be used to automate repetitive tasks and streamline processes, such as content production, moderation, and analytics, to boost productivity \cite{artificial1}. Moreover, Metaverse platforms may face numerous risks and frauds in the present day; therefore, Metaverse platforms and MaaS use AI to detect and prevent security threats and fraud in the Metaverse by, for example, detecting and blocking dangerous bots and identifying and warning of suspicious behavior. AI plays a crucial role in customizing Metaverse environments and typologies, for instance by adjusting the difficulty of a game or the layout of a virtual environment to the user's preferences. However, as with any other system, the use of technology can be advantageous in many ways but dangerous if certain factors are ignored. Implementing AI into Metaverse platforms and MaaS can be challenging, sophisticated, and knowledge-intensive. If AI is utilized to create lifelike avatars and objects, MaaS and platform owners must verify that these items are not exploited for malicious purposes, such as phishing and impersonation.\\

\textbf{Blockchain.}
Blockchain technology is predominantly employed by MaaS companies and Metaverse platforms to enable new forms of ownership, monetization, and governance. Blockchain facilitates the storage and transfer of avatars and assets between platforms in the Metaverse. Utilizing Blockchain technology lends credibility and trust to such transactions. By connecting the Metaverse, \cite{blockch1} is pursuing the ability to move or teleport avatars from one platform to another. As stated earlier, Blockchain technology adds a layer of security to platforms, allowing for the secure storage of assets on such platforms. MetaRepo is a new solution proposed in \cite{Blockch2} that enables users to easily and securely store and utilize assets in the Metaverse. Using Blockchain and its designed structure, they intended to provide users with a novel way for interacting with other Metaverse universes without the need for additional verification and security checks. Moreover, Blockchain enables the implementation of smart contracts for the execution of functions and the automatic management of the transfer of digital assets \cite{blockcha3}. 

\textbf{Networking.} 
The network plays a significant part in evaluating the experience quality of every Metaverse platform and MaaS services. The emergence and development of the 6G network infrastructure enables Metaverse systems to benefit from ultra-reliable and low-latency communications and the support for huge numbers of connected devices; Metaverse is now widely recognized as the Internet fuel of the next-generation Internet \cite{cheng2022will}. The vast amount of user interactions with numerous streams must be gathered and handled in real-time. Specifically, Metaverse platforms must take into account end-to-end latency requirements, while certain Metaverse systems apply 7 to 15 ms end-to-end delay limits \cite{networkma2} and it must be less than 1 ms in critical applications such as a surgery. The authors of \cite{networkma3} suggest a model for optimizing Metaverse applications end-to-end and offer dynamic control.  \\

\textbf{Computing.} 
Power is an essential part to take into consideration when deploying services by MaaS and Metaverse platforms on the servers. Intel states that Metaverse requires a 1,000-fold increase in processing power\footnote{https://www.thehindu.com/sci-tech/technology/intel-says-Metaverse-needs-a-1000-times-computing-power-boost/article37977295.ece}. With the development of computing and supercomputers, it is now feasible to have a fully functional Metaverse platform. However, standard and conventional computing power cannot support the adoption of the Metaverse, as Metaverse platforms require supercomputers while measuring processing speed in terms of floating-point operations per second (FLOPS) rather than the conventional Gigahertz (GHz). Companies that host such platforms must guarantee the availability of power and cooling systems. This deployment has a significant influence on hardware and energy consumption \cite{computing4} since the Metaverse requires extensive simulations and rendering tasks. Emulations in a number of Metaverse platforms, such as Second Life, are centralized and cloud-based \cite{computing1}. In \cite{computing2}, the authors proposed a new concept for Fog and Edge hybrid computing architecture to use the computing capacity of edge devices in order to perform intensive computational tasks. In addition, \cite{computing3} emphasizes the usage of mobile edge computing (MEC) by delivering resources close to endpoints in the Metaverse by comparing it to centralized mobile cloud computing. \\

\textbf{Business.}
Companies responsible for Metaverse platforms and MaaS must limit the cost of their services by balancing the expected input with the cost incurred by users. On Metaverse, it is possible to develop new company concepts, allowing users to integrate and profit from innovative and new types of businesses. In addition, modern firms are utilizing the Metaverse to improve their products and minimize costs. The authors of \cite{bb1} explain why businesses should advertise in the Metaverse. The Nike-Roblox case study was examined to demonstrate the importance of advertising and communication in the Metaverse, as well as how this affects their business, in particular the interactions with regional and local markets. Moreover, Disney is constrained by its users and customers, thus it is partnering with the Zepeto Metaverse to extend its consumer base by employing data mining techniques \cite{bb2}. Aside from that, large corporations are pursuing the usage of Metaverse to lower costs by utilizing tele-learning and creating virtual prototypes, which are less expensive to create virtually than physically. The authors of \cite{bb3} targeted the mobility in the Metaverse while taking scenarios of meta-empowered advanced driver-assistance systems to illustrate how Metaverse will shape the future of the existing architectures. Furthermore, company cars can use the Metaverse to simulate real-life car crashes to improve their cars while spending a lot less money on such simulations.   \\


\textbf{Security \& Privacy.} 
Security and privacy issues related to data generated from the Metaverse are of major concern for users. Having access to such data raises many concerns regarding the manner in which these platforms utilize the data \cite{privacymar4}. AI models can be developed to anticipate unreported information about users, such as their political and sex tolerance. Multiple levels of synthetic data dependability must be considered by platform owners. By eliminating data biases and ensuring data privacy, authenticating synthetic data can improve the fairness and safety of AI algorithms \cite{privacymario3}. Such sensitive information must be secured and kept private. Implementing "Know Your Customer" (KYC) \cite{privacyma1} in which businesses check the identities of their clients and identify potential hazards for money laundering or financing terrorism is a wonderful way to improve customer experience \cite{privacyma2}. Implementing anonymization techniques to secure users and their data is an effective strategy to empower privacy. Furthermore, MaaS cooperation raises doubts regarding their trust. Therefore, Metaverse platforms should implement security and trust safeguards through the service providers. In addition, before implementing a service or platform, owners must examine the security measures that a Metaverse platform enforces to protect the safety of users and the quality of their experience, as a large number of attacks are aimed at these platforms. For instance, adversaries can compromise wireless access to the Metaverse by altering the inputs to the learning algorithms used for user authentication \cite{securityma1}.   \\

\textbf{Ethics \& Sociopsychology.} 
Ethical considerations are essential for maintaining a safe and clean environment for user interactions. The users of these platforms might experience misbehavior, spam, harassment, and conflicts with other users \cite{sociomar1}. The software code of the Metaverse can constrain the shape of the Metaverse, similar to the physical laws of nature. Online surroundings and user behavior are influenced by code \cite{socioma2}. Businesses and developers can select which features will be included on the web platform \cite{socioma3}. In addition, the personal boundary function is significant since it was developed on the Horizon platform after an avatar was sexually attacked by a group \cite{fernandez2022life}. In addition, the platform must be founded on the principle of complete democratization in order to be open and fair. The authors of \cite{socioma5} emphasized the need for moderators to resolve community issues in the platform. In addition, they proposed an incentive structure to encourage beneficial user behavior. Users' inappropriate actions must be dealt with instantly. Misconduct on the platforms must be strictly regulated and punished. 

\section{Metaverse Environment Digitization}

\begin{figure}
    \centering
    \includegraphics[width=\linewidth]{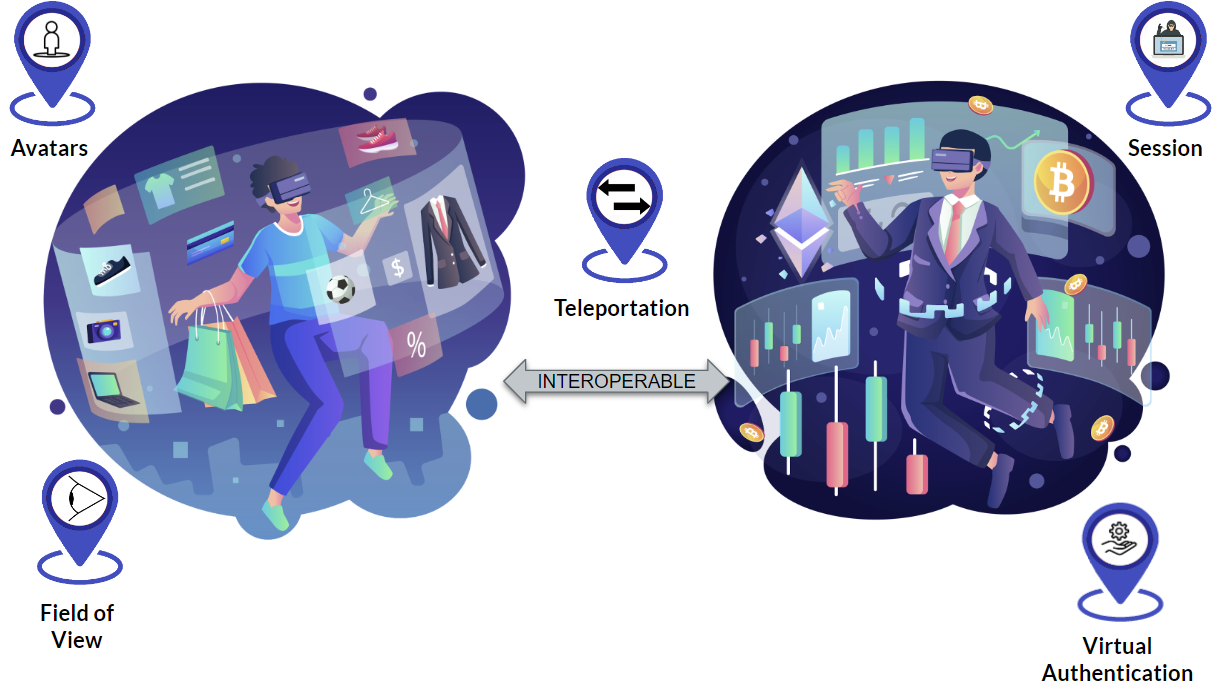}
    \caption{Environment Digitization}
    \label{fig:taxonomy_environment_digitization}
\end{figure}

Building on top of the infrastructural developments, the next step is to start forming the virtual components and representation in the Metaverse, which we refer to as Environment Digitization. The Metaverse Environment Digitization layer, the second layer of the pipeline, is responsible for the creation and management of the virtual world's pieces. It ranges from copying real-world objects to human representation as avatars. The main components of this layer are Avatar Modeling, Rendering, and Sessions. Below we detail each one of them. The first entry point to users in the Metaverse is their avatar representation. Afterwards, rendering the avatar representation as well as objects and surroundings is the role of rendering engines. Consequently, sessions are a result of data generated from rendered environment, that requires careful data handling, including storage and query operations. To this end, we study in this section existing advancement and state-of-the art research and industrial achievements in each of the components forming the Metaverse environment digitization. In Figure \ref{fig:taxonomy_environment_digitization}, we present an overview of some of the characteristics of environment digitization as part of existing research focus.

\subsection{Avatar Modeling}
\begin{figure}[htp!]
    \centering
    \includegraphics[width=0.8\linewidth]{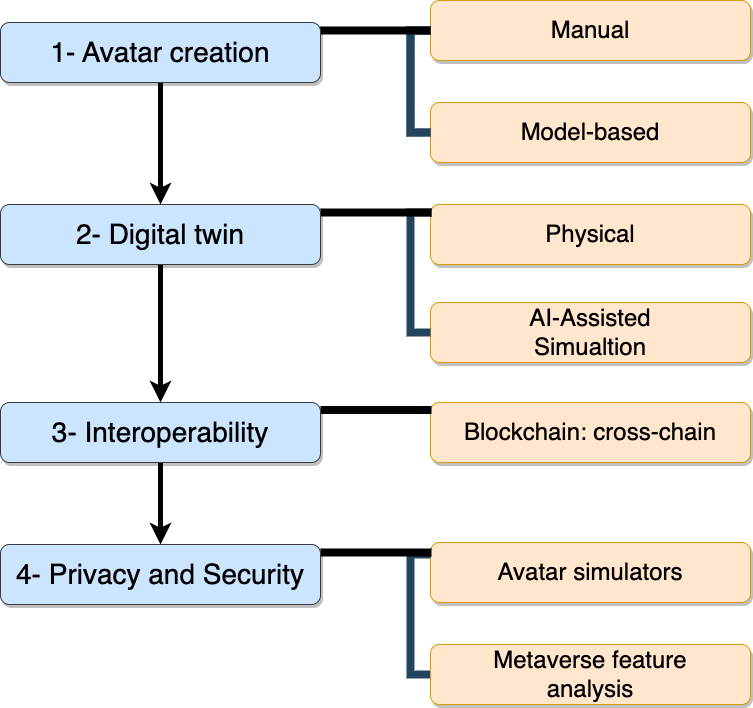}
    \caption{Avatars in the Metaverse}
    \label{fig:ar_avatars}
\end{figure}

This section discusses the various technologies involved in creating Metaverse avatars as part of the Environment Digitization layer of the pipeline. Such a procedure is complex by nature due to an expected level of realism. Metaverse allows its users to represent themselves as avatars while giving them new abilities to control and interact with their environment. Moreover, new types of motions, including leg and hand contraction, are newly introduced by the Metaverse in which avatars should be able to replicate one-to-one physical users' motion. Thus, we instigate a unique sub-pipeline, presented in Fig. \ref{fig:ar_avatars}, dedicated to the available Metaverse's avatar methodologies while Table \ref{tab:av-ref} aggregates the references used in this section for each category. The pipeline considers the following: 1) Avatar Creation, 2) DT, 3) Interoperability, and finally, 4) Privacy and Security.

\begin{table}[htp!]
\centering
\caption{Avatar Modeling References}
\label{tab:av-ref}
\begin{tabular}{|l|l|}
\hline
\textbf{Category}    & \textbf{References}                                                                              \\ \hline
Avatar Creation      & \begin{tabular}[c]{@{}l@{}}\cite{ ar1,ar2,ar3,ar4,ar5,ar6}\\ \cite{ar7,ar8,ar9,ar10}\end{tabular} \\ \hline
Digital Twin (DT)         & \cite{af0, af1, af3, af4, af5}                                                              \\ \hline
Interoperability     & \cite{ap1, ap2, ap3, ap4}                                                                        \\ \hline
Privacy and Security & \cite{as1,dwivedi2022metaversea}                                                                                       \\ \hline
\end{tabular}
\end{table}

\textbf{Avatar creation.}
Avatar creation is a significant aspect of the Metaverse as it gives a capacity for a high level of realism which affects the user experience. Users expect a level of self-representation and the capability of portraying themselves as they do in real life. In contrast, some others want an unorthodox creation of their avatars, given the necessary tools to complete this process. This domain includes three fundamentals: Rigging, Skinning, and Shaping. 

\textit{Rigging:} is the process of creating a skeletal structure for a 3D model. The skeleton is typically made up of bones connected by joints and can be used to control the model in the virtual environment. 

\textit{Skinning:} is creating a digital skin of avatars that can be deformed to match the motion of a real-world counterpart. This allows the avatar to mimic the appearance and movement of the user convincingly. Once the rigging is in place, the skinning process uses the skeleton as the foundation for the skinned mesh. The skinned mesh is typically created by starting with a basic shape followed by adding vertices to match the skeleton.

\textit{Shaping:} is the process of modifying the avatar's look to match its motion. When rigging and skinning take place, Shaping is when skinned mesh uses the underlying skeleton as a guide to enable deformations. 

Currently, avatar creation can be done by either a manual \cite{ar1} or a model-based procedure. Manual creation is the most commonly used technique and can be categorized into three types: Photogrammetry \cite{ar2,ar3,ar4}, 3D Scanning, and Character Creation Toolkits. The final results of these processes are a complete avatar that includes the three principles mentioned above. 

\textit{Photogrammetry:} can be used to create realistic and accurate 3D models of users' shapes. It is done by taking a series of photographs from different angles and then using them as input to reconstruct a 3D model from these photos. The major limitation in Photogrammetry is the high dependency on the quality of the given images. Such can be affected by many factors, including the type of camera used, angle, lighting, and distance from the subject. Time can also be a factor, as Photogrammetry can take a Long duration to get a suitable model, depending on the complexity of the image.

\textit{3D Scanning:} is a process of collecting digital data on the shape and appearance of a person in order to build its complementing avatar. This data can be collected using a device, such as a laser scanner, to record measurements and take them as input for the digital 3D model creation. Similar to Photogrammetry, 3D scanning is time-consuming, requires some level of expertise and is expensive for the typical Metaverse consumer. 

\textit{Character Creation Toolkit:} can be taken from two perspectives, either in the form of consumers or designers. Consumer kits are business models that build avatars for users in return for monetary value. In this case, users are given a minimum set of tools to hand-pick their avatars based on predefined assets \footnote{https://xr-marketplace.com/en/makeavatar} \footnote{https://readyplayer.me/} \footnote{https://maketafi.com/}. These toolkits usually include various customization options, such as face shape (eyes, nose, mouth, jaw), body style (skinned, fat, tall, short) and colours (hair, body, eyes). However, this technique has a major constraint in terms of realism as it limits the possibilities of replicating the user's intentions. As for designers, creating avatars for the Metaverse requires a considerable amount of manual labor and a high level of expertise in avatar design domains \footnote{https://docs.unity3d.com/Manual/AvatarCreationandSetup.html} \footnote{https://www.unrealengine.com/en-US/metahuman}. 

The main challenges in manual creation are the time, required knowledge, and material, which might not be convenient when targeting typical Metaverse consumers. As such, the Model-Based techniques address these issues. These techniques employ AI-Based technologies to create a highly realistic one-to-one avatar from a single image. A machine learning algorithm analyzes a collection of images to learn what features are common, including faces and shapes, and then creates a model that can be used to generate new avatars from a given image. This approach has several advantages:
\begin{itemize}
    \item It can be used to create realistic avatars that look like the person they are meant to represent. 
    \item It is possible to create remarkably expressive avatars showing a wide range of emotions.
    \item It takes less time and effort than the manual approach.
\end{itemize}

Different AI technologies are employed to address the aforementioned principles. For instance, most studies consider Linear blend skinning (LBS) \cite{a0}, a technique used to deform a character's mesh based on the latter bone structure movement. In LBS, a character's mesh is represented as a set of weighted points (vertices) that are connected by lines (edges). Each vertex is assigned a weight which determines how much it will be affected by the movement of the bone structure. The bone structure is represented by a set of bones connected in hierarchical structures. Each bone has a position and orientation in 3D space. When a bone moves, all the vertices connected to it will also move according to its weight. In \cite{ar5}, the authors trained a model based on a generic 3D human body template. It uses a combination of linear blend skinning and pose-dependent blend shapes to represent the shape and pose of a human body. The model is designed to capture the joint angles of a human body and is composed of a set of parameters that can be used to animate the Metaverse’s avatars. The model is created by fitting 3D scans to a template, which is then used to generate a set of parameters that can be used to generate an avatar. The model consists of a set of joints and a set of joint angles determined by fitting real people's body scans to a template. 
The work in \cite{ar6} trains a facial shape and expression model from 4D scans. The model is based on a deep convolutional neural network and is trained on a large dataset of 4D scans. The model can accurately map 3D coordinates onto facial features such as eyes, nose, and mouth. In addition, it is also capable of predicting facial expressions, including emotions and expressions of surprise, based on the scan data. In \cite{ar7}, the authors propose a modeling scheme for capturing hands. Their scheme uses a combination of 3D scans of various hand positions to train their model to capture the geometry of hands. Different works followed a similar architecture but with incremental advances in accuracy and efficiency. For instance, in \cite{ar9}, the authors extended the SMPL model and combined it with FLAME \cite{ar6} and MANO \cite{ar7} to generate a mesh geometry with higher fidelity from a single image. Conversely, the focus in \cite{ar8}  was on avatar clothing. The authors used 3D scans of clothed humans to train a model capable of generating a pose-aware deformable clothed avatar. 
The authors in \cite{ar10} propose an auto-rigging approach by matching morphable models to 3D scans. Once the rigged model is identified, the authors apply the skeleton and then the skin to the model. The results are similar to the handmade 3D meshes.  Additionally, using morphable models gives the capability of reshaping the results to cope with the intended shape.

\textbf{Digital Twin.} 
The DT represents the physical entity as a digital asset in the Metaverse context. According to \cite{af0}, DT can be either one-one direct control from the physical entity or a model-based simulation trained from the interactions of the physical entities. Regarding direct control, the state of the arts focuses on sensor-based technologies and pose estimation based on image stream input. Section \ref{sec:hardeq} depicts some of the sensor-based avatar motions. Regarding pose estimation technologies, the challenge can be summarized as predicting the 3D keypoints from an image while projecting it on the avatar. \cite{af1} provides a survey on the existing approaches of pose estimation-based avatar controls while detailing the advantages and disadvantages of each work. For instance, in \cite{af3}, the authors provide a scheme able to predict the 3d keypoints of an image even when it is incomplete or truncated. \cite{af4} aims to solve another problem where keypoints that do not have equal weighted visibility should not have a similar effect during the training procedure. In \cite{af5}, rather than motions based on a single image, the authors build a context in which prediction is based on a fusion of images taken from multiple views. By contrast, model-based motion does not require the existence of physical entities once the predictive model is trained. Such a technology can be used to simulate virtual interaction between the Metaverse's avatars as predictive models of the physical world or to improve the in-verse interaction between users' avatars and their environment. 

\textbf{Interoperability.}
No matter where Metaverse users are located or what technology they use, virtual applications should allow free movement between the different worlds. Interoperability is essential to complete this task as it allows users to keep their avatars while efficiently transferring them between digital worlds. For instance, in \cite{ap1}, the author emphasizes the interoperability usage of seamlessly transferring avatars and getting access to any environment without compromising or changing the existing credentials. However, interconnecting multiple virtual worlds will suffer from different standards in the contexts of identity management, currency, modeling, and communication protocols. 
Blockchain technologies can address all of these issues. In this context, blockchain-based enabling technology supports the interoperability of user avatars into any Metaverses bypassing the hardware and software limitations. However, placing all available technologies under one blockchain network is far-fetched. As such, the usage of cross-chain technologies is crucial for enabling interoperability in the Metaverse. For instance, \cite{ap2, ap3, ap4} discusses the benefits of blockchain interoperability, including improved scalability, interoperability of avatars, and improved security. Additionally, it enables the development of more complex transactions, as well as the ability to access a wider range of environments.

\textbf{Security \& Ethics.}
Technologies and advancements in the Metaverses avatar allow for intricate and deep representation of the physical users on their avatar counterparts. For instance, recent technologies can accurately project users' expressions on their avatars by either their context (spoken or written) or from a stream of image input. As such, these technologies open opportunities for various security risks, fraud, identity theft, and psychological manipulation. For instance, a user can deploy an ML agent to analyze the avatar's facial expressions and interactions to predict the physical user's personality. As such, an outcome of interaction can be predicted beforehand. Additionally, since an avatar is the identity of its users in the Metaverse, stealing an avatar is equivalent to stealing the original user's identity, which can be employed for impersonation and manipulation schemes. Thus, avatar authentication has a major impact on the Metaverse. In this case, avatar interactions, such as voice, text, and behaviors, can be used for identification. However, it is challenging without having a custom scheme or authority of control over the created avatars. In this case, adversaries can create a model of the physical entities capable of accurately simulating the physical users \cite{as1,dwivedi2022metaversea}. Consequently, such authority requires additional personal information that breaches users' privacy.

\subsection{Rendering}
The second component of Environment Digitization layer in the pipeline is rendering. Rendering is a crucial component of computer graphics that transforms 3D models into images and animations, thereby playing a key role in developing immersive Metaverse environments. This customization enhances the user experience by providing a personalized space. Various techniques and machine learning models can be used to create multi-dimensional models, adjusting the lighting and shadows on objects and surfaces. Despite its importance, rendering can be computationally expensive, and there is a trade-off between the texture quality and the models' performance. The primary goal of rendering is to achieve real-time realism and stylization for an immersive experience. The techniques aim to create vivid views, showcasing the features of visual context and enabling customization of the 3D environments to meet the user's preferences \cite{davis2009avatars}. \\

Modeling, animating, and composing differ from rendering in that they focus on creating geometric shapes or objects in a 3D scene or moving specific frames to create animations and compositions, in which are created by combining multiple layers of images. To create 3D models and images, algorithms, mathematical equations, and AI tools are used to consider material, lighting, brightness, contrast, and other visual elements. The output of these models can be used in VR/AR games, movies, and applications. \\
Various tools, techniques, and software are used to create high-quality images and animations to improve the visual aspect of the Metaverse environment. One such technique is generating high-quality environment maps for use in front-facing AR applications, as demonstrated in \cite{zhao2023multi}. Another approach, proposed in \cite{monroy2018dynamic}, uses real-time AR technology for environmental acquisition and rendering on mobile devices. Other tools for modeling and environment simulation include Unity3D \cite{unityUnity},  Auto-CAD \cite{ autodeskAutodeskEmpowers}, Blender \cite{blenderBlenderorgHome}, ZBrush \cite{pixologicZBrushAllinonedigital}, and others.
The development of computer graphics has seen significant growth with the use of rendering techniques, particularly in the movement from 2D to 3D environments and in the Metaverse. Advances in hardware, such as the CPU and GPU, have enabled high-performance rendering and the creation of high-quality models, images, and videos. In \cite{zhao2022litar}, a technique for two-field lighting reconstruction is proposed for generating high-quality environmental maps for a limited field of view (FoV) captured from a camera. Such methods can address the challenges of mobile devices' limited sensing capability and mobility-induced noise. Popular rendering techniques for environments include ray tracing, rasterization, and global illumination. The Metaverse allowed individuals with disabilities to freely move within and explore its virtual environment \cite{hughes2019disruptive}. Responsive subtitling is also provided to offer a personalized experience for each user. In \cite{roxas2018occlusion}, a method to improve occlusions in the blending-based method by predicting visibility and utilizing modular image information is proposed. In addition to the visual aspect, the acoustic aspect is also essential for providing an immersive experience for users. For example, \cite{jot2021rendering} proposes a positional audio playback with extended coverage and resolution in rendering, using a flexible configuration for loudspeakers and encoding and rendering audio scenes in Ambisonic format.

\textbf{Realistic and Stylized Environments.}
Rendering in the Metaverse can be divided into two main approaches: realistic and stylized environments. \textit{Realistic environments} aim to simulate the physical structure of objects and environments realistically, including the behavior of light, lighting, shading, and texture. Creating a realistic environment involves data capturing, preprocessing, modeling, and rendering. Realistic environments are typically used in video games, real-world simulations, surgeries, and architectural simulations. In \cite{xu2022rendering}, a rendering-aware learning scheme for efficient loss computation and photo-realistic virtual object rendering is proposed.
\textit{Stylized environments} aim to create a unique environment for each simulation to provide an immersive experience for the user. Such environments do not have to follow realistic physics, allowing for custom color palettes, shading, and ambient environments. The creation of stylized environments involves specialized software for 3D modeling, with artists creating initial shapes and appearances tailored to a custom-made scene. Stylized techniques are used in sci-fi-related games, dream-like simulations, and fantasy environments.

\textbf{AI.} 
Intelligence is a major player in rendering, especially for simulating lights reflections, study the physics of surfaces, improve image quality, generate virtual content, in addition to applying styles and completing missing parts of images. First, Ray tracing is a rendering technique that uses AI for simulating light reflection in the Metaverse as a result of interacting with objects. Examples of existing solutions applying Ray tracing techniques using AI are the Nvidia's RTX platform and Microsoft DirectX Ray Tracing (DXR) \cite{velho2020immersive}.
In addition, Deep Learning Super Sampling (DLSS) is a promising application for rendering that uses AI and has a potential for improving rendering engines in the Metaverse. DLSS uses AI to upscale low resolution images, which can reduce the computation required for rendering scenes in the Metaverse with realistic and high quality results. The Nvidia's RTX platform utilizes DLSS \cite{burgess2020rtx}. Moreover, AI has been adopted to perform style transfer between images, thus it can be applied to produced effects and style objects with reduce computation \cite{xiang2022panoramic}. Furthermore, neural rendering is another application of AI for improving rendering engines supporting the Metaverse. In neural rendering, AI is used for completing missing parts with realistic results using incomplete or low quality images \cite{thies2019deferred}.

\textbf{Networking and Computing.}
The 3D simulation is more complex when considering the Metaverse. Different 3D objects that will change depending on the situation will need to be rendered. It will need many resources to render a detailed 3D space. Right now, none of the products on the mass market can perform that\footnote{https://uxplanet.org/metaverse-design-guide-part-3-5e2f4fc6cd61}. The main idea behind resource management revolves around resource allocation, workload balance, task scheduling, and QoS to achieve performance improvements \cite{mijuskovic2021resource}.
One solution to the complexity problem is outsourcing rendering tasks to external servers while sending them back as a video stream to the user's device. Adopting this strategy can be beneficial to reduce the costs for the consumers. Also, a delay problem in rendering while using wireless VR devices persists because of the intensive computation and communications. The following paper \cite{si2022resource} proposed a mobile augmented reality MAR-based connection model for the Metaverse and proposes a communication resources allocation algorithm based on outer approximation (OA) to achieve the best utility. Such a solution reduces communication and computation costs. Furthermore, fog servers will also help to solve the problem of latency and high delay because these servers can be located closer to the user to minimize network latency. \\

\textbf{Blockchain.}Rendering tasks are highly dependent on both central and graphical processing units. The server that handles the rendering tasks will be unable to maintain a sustainable rendering speed due to the increasing demand. One solution can integrate a decentralized network of computers to achieve cumulative rendering power; one such network is the Render Network. It is a distributed GPU power empowered by the Blockchain; it provides a platform for sharing GPU power across users. By relying on such a Platform, users can join the network and offer their idle GPU resources in exchange for monetary profit. Render calls the clients needing GPU power "Creators" and the providers available on the network "Node Operators." When creators need something rendered, they send their files to the network, and a job is created. The Node Operators are then given these tasks, and those who render are given Render Tokens (RNDR)\footnote{https://www.gemini.com/cryptopedia/render-network-3d-rendering-software-render-token-rndr-token}. The key advantage for creators is the low overall cost of generating their extremely complex files because they do not need to buy and maintain the computers and to optimize the resources. Moreover, in the concept of Blockchain, the stakeholders need to access and hold assets in different virtual worlds. Data interoperability across these virtual worlds is limited due to the different environments in which they are built. It is possible to exchange data on two or more blockchains located in distinct virtual worlds using a cross-chain protocol. Users can migrate more easily between these virtual worlds because of a cross-chain protocol in the Blockchain's interoperability. This protocol allows the exchange of possessions like avatars, NFTs, and payment between virtual worlds, and it will provide the groundwork for widespread Metaverse adoption \cite{gadekallu2022blockchain}. \\

\textbf{Security and Privacy.}
The main objective of the Metaverse is to provide a shared environment that simulates the physical world, allowing for personalized digital assets to be rendered publicly in front of multiple users. Nevertheless, Such a shared environment makes it a potential target for security and privacy attacks and increases the risk of malicious activities such as hacking, fraud, and abuse. Thus, it is essential to have strong security measures in place to prevent such incidents and guarantee a safe and secure experience for all users.

Access control checks during asset rendering are a traditional way of ensuring security and privacy in a virtual environment. In this context, integrating an access control mechanism limits the visibility and control of information and assets to authorized entities only. Furthermore, complex access control also includes authorization levels that benefit specific types where the shared environment is a critical aspect of the application. Such an environment includes virtual teaching space \cite{apr4}, social apps, and other entertainment platforms. Typical access control integration is done through Access Control List \cite{apr1}. However, this mechanism incurs serious drawbacks, such as a lack of delegation capabilities and forcing a central authority which might incur performance and scalability bottlenecks.

An improved version of access control is proposed in \cite{apr2}. The authors present a spatial-based mechanism that benefits from simulating the location of entire worlds in the Metaverse applications to limit access based on boundaries. Such boundaries include access limitation to either a specific place inside a region or the whole region. The main advantage of this approach is shifting access from a basic access check to conditional access based on object location, where the access or denial is given by going through a sequence of boundaries where the object is located. However, considering only the object's location produces issues when virtual environment places are rendered in different altitudes \cite{apr4}. In \cite{apr3}, boundaries are accessed through ID cards, keys that can open areas within the Metaverses. Such keys can be shared with others allowing access delegation with multiple parties. Such a methodology is indicated by object-capability security, as such an object is unique and can be interacted with in a specific way. In addition to previous approaches, the authors presented a visual implementation of their work by simulating their mechanism in a real virtual environment. Furthermore, patents regarding data privacy have been registered. For instance, \cite {apr5} propose a blocklisting mechanism that allows users to ignore others. The main advantage is the avoidance mechanism that depends on the location of blocked users. Rather than denying access to a user in a specific location, the system will notify the primary user of the existence probability of the blocked user in a location. Moreover, in \cite{apr6}, another patent is deployed to protect the Metaverse universe from malicious users and the damage incurred from their existence. Such damage includes damaging virtual world properties or affecting account reputation. Thus, a rollback mechanism is proposed in an attempt to restore the virtual world to a previous state prior to the damage incurred.

\subsection{Sessions}
The last component of the Environment Digitization in the Metaverse pipeline is Sessions. This section focuses on Metaverse sessions' data allocation, collection, security, and usability. Sessions indicate the user's engagement in the Metaverse and the location where the user's sensitive data is handled. For that, session management should be treated carefully. Sessions are created while launching the application prior to logging in. They manage data such as object state that represents the environment object position, status, and the user's personal data. As a result, controlling Metaverse session data is critical for various reasons, including customer experience, data and security regulation, and marketing effectiveness. Depending on the session activities, we can differentiate between two kinds of sessions. A private session belongs to a specific user. It takes on a more personal tone.
In contrast, sessions can be public between multiple users where they can join the same environment and interact with each other and with the environment's objects. In such cases, if the state of an object has been manipulated by a user, the object state should be edited for all the other users. Thus, sessions play a significant role in the real-time synchronization between users.

\textbf{AI.} Session structure development and optimization have been studied in the literature using AI from different aspects. By applying AI for decision-making, the communication, computing, and caching resources can be collaboratively optimized~\cite{o4}. Moreover, AI can play a major role in session data access management in general. Through resource optimization and data learning, AI can influence the user experience and increase user engagement. Dealing with dynamic data sizes, such as session data, usually requires a combination of resource allocation, including cloud, fog, and edge computing technologies~\cite{o3}. In terms of session optimization, AI plays a significant role by applying data categorization and then distributing the session to multiple destinations/servers based on the classification results. Thus, observing and analyzing the data traffic of sessions is essential. Depending on the data priority level, AI contributed to resource allocation, which is beneficial in the case of stream processing scenarios~\cite{o5}. Another problem in the session data is scalability. Sessions are dynamic regarding the number of joiners, which is positively correlated to session resource availability. AI can contribute to resource allocation through on-demand-based scalability~\cite{o6}.

\textbf{Blockchain.} Critical aspects in the Metaverse are related to the session data storing, privacy, and integrity. The considerable amount of personal data generated from sessions raises a lot of privacy and security concerns~\cite{nair2022exploring}. Blockchain might be a suitable solution for session data threats due to its unique characteristics such as data integrity, transparency, and decentralization~\cite{gadekallu2022blockchain, kang2022blockchain}. Moreover, blockchain can facilitate communication between sessions from different spaces and platforms. Therefore, applying cross-chain for better data flow and transmission enables the communication between various types of Blockchain and offers better scalability~\cite{o9}.

\textbf{Networking \& Communication.} The Metaverse relies heavily on session networking and communication. The immersive experience requires stable connectivity which is prone to numerous unexpected communication issues. The rising complexity and volume of session data for new Metaverse applications, in particular, pose a tremendous barrier to data sharing security~\cite{o10}. In order to ensure the quality of user experience in session communication, ultra-high capacity and reliability for the wireless system are needed, which the existing 5G system cannot provide. A revolution in networking and communication is required to attain the Metaverse paradise. For instance, the 6G wireless technology appears to be a potential answer due to its ubiquitous connectivity, ultra-low latency, ultra-high capacity and dependability, and tight security~\cite{tang2022roadmap}.

\textbf{Business.} User data are beneficial due to the immersive nature of the technology and the amount of time it will be utilized. Metaverse platforms capture considerably more sensitive information about consumers~\cite{barrera2023marketing}. This level of detail will be tremendously beneficial to brands. Through session data, business owners analyze users' behavior to drive targeted materials in front of people in the Metaverse. For example, billboards along a virtual street or a non-playable figure standing on the sidewalk enjoying the product enhance product positioning and marketers on a grand scale. Moreover, businesses may investigate intimate and intrusive aspects of their consumers' lives~\cite {o13}. In addition, similar to social media, business owners may share data with marketers, who could subsequently display related product advertisements~\cite{o14}. Furthermore, this data might be utilized to feed applications' algorithms to maintain users on their platform for a longer time~\cite{o15}.

\textbf{Privacy \& Security.} Privacy and security of session data should be protected. Despite the benefits of session data and the improvements that may be brought to the Metaverse, several forms of data in the Metaverse must be protected based on the user's activity and type. In the case where the user is a business, the size and type of data to deal with are substantially larger and different from those of a regular user, who just has his personal data~\cite{o16,o17}. Metaverse content, personal data, analytical data, and qualitative data are all examples of Metaverse data that should be secured~\cite{privacymar4,falchuk2018social}. For instance, some basic solutions can be applied such as data classification and terms continuous updates. Additionally, more novel and reliable methods can be applied such as the use of federated learning, which will work as a promising solution for data privacy~\cite{o20}. 



\textbf{Sociopsychological \& Ethics.} Since session data are very sensitive. Thus companies should be using all the information gained from sessions ethically~\cite{o22, bibri2022metaverse}. From a Sociopsychological perspective, the disclosure of sensitive session data, like the visited places and Metaverse's activities, can negatively affect the user's psychological attitude and may affect the user's trust in the Metaverse. It can impact the user mentally and emotionally~\cite{o24}, since users trust the data collectors to secure their data while respecting their privacy.

\begin{table}[h]
\centering
\caption{Sessions References}
\label{tab:se-ref}
\begin{tabular}{|l|l|}
\hline
\textbf{Category}    & \textbf{References}                                                                              \\ \hline
AI      & \cite{o3, o4, o5, o6} \\ \hline
Blockchain         & \cite{nair2022exploring, gadekallu2022blockchain, kang2022blockchain, o9}                                                              \\ \hline
Networking \& Communication    & \cite{o10, tang2022roadmap}                                                                        \\ \hline
Business & \cite{barrera2023marketing,o13,o14,o15}                                                                                       \\ \hline
Privacy \& Security & \cite{o16,o17,privacymar4,falchuk2018social,o20}                                                                                       \\ \hline
Sociopsychological \& Ethics & \cite{o22, bibri2022metaverse, o24}                                                                                       \\ \hline
\end{tabular}
\end{table}

\section{Metaverse User Interactions}



\begin{figure}
    \centering
    \includegraphics[width=0.9\linewidth]{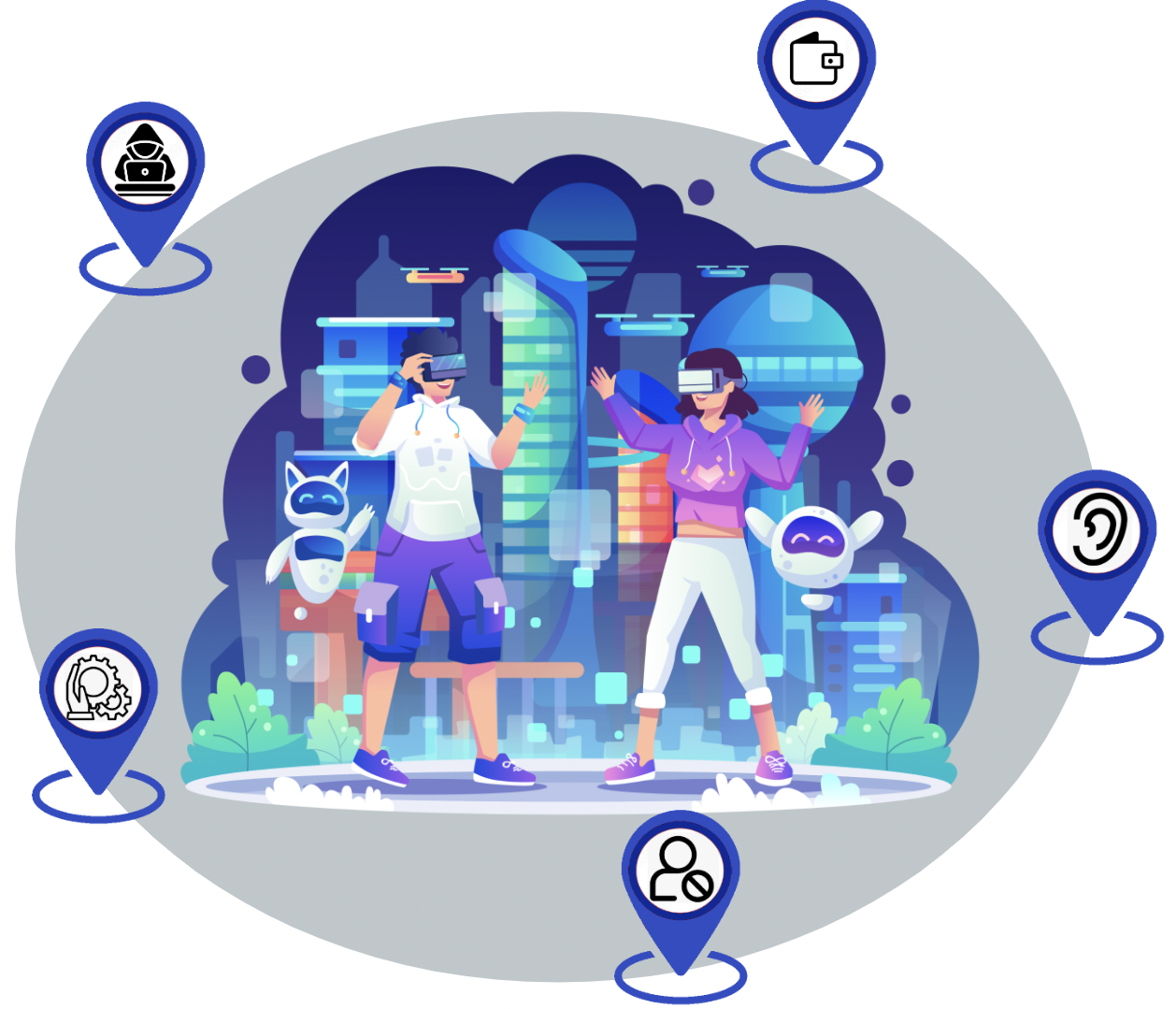}
    \caption{Metaverse User Interactions}
    \label{fig:metaverse_interactions}
\end{figure}

Using the Metaverse infrastructure and building effective rendering engines, the last layer of the Metaverse pipeline is building user applications to empower immersiveness and engagement. These applications present different types of interactions involving the user. In this section, we complete our proposed architecture with the last layer, i.e., the Interaction layer. The main types of interactions in the Metaverse from a user perspective are categorized as (1) user-to-user interactions, (2) user-to-business interactions, and (3) user-to-object interactions. In Figure \ref{fig:metaverse_interactions}, we present an overview of existing applications characteristics presented in the literature, including digital entanglement, digital stalking, business transactions management, as well as security measures.

\subsection{User-User Interactions}
As part of the user interactions layer in the Metaverse pipeline, we present the User-User interactions. A user-to-user interaction mainly refers to the interaction that involves a user - represented by his/her avatar - with one or numerous others. These interactions can take the form of various virtual social encounters. Below, we name a few types:
\begin{itemize}
    \item \textit{Socializing}: Users in the Metaverse can socialize with each other in many ways, including chatting, forming groups, and attending virtual events and parties. Such a category is the main key to enabling most of the Metaverse applications.

    \item \textit{Gaming}: Gaming is a popular form of user-to-user interaction in the Metaverse. As Virtual reality games are already approved by the users in the gaming world, it is very likely that these games will be integrated into the Metaverse to allow a smooth experience for Metaverse gamers. A variety of games will be offered through the Metaverse, which will vary from First-person shooter (FPS) games to massively multiplayer online role-playing games (MMORPGs)

    \item \textit{Virtual events}: Events and concerts can take place virtually without diminishing the user experience. Avatars can interact at virtual concerts, conferences, or exhibitions in the same manner they do physically. This can entail many advantages such as reducing costs and increasing user accessibility, in addition to allowing a customized experience for each user individually.
\end{itemize}

\begin{figure}
    \centering
    \includegraphics[width=0.4\linewidth]{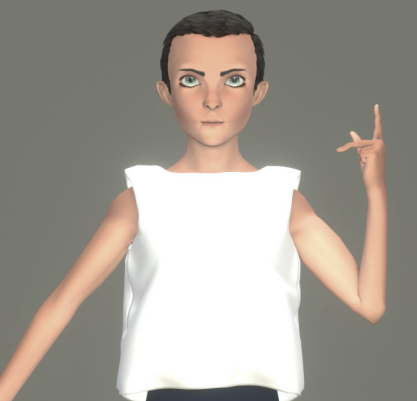}
    \caption{Avatar mimicking a letter \cite{batnasan2022arsl21l}}
    \label{fig:interaction_avatar_letter}
\end{figure}

\textbf{AI.} AI is one of the main technological pillars that can enhance the user experience of user-user interaction. First of all, it enables a sophisticated personalization experience where avatars can be recommended for attending virtual events or training sessions according to their preferences. In addition, it can ease up communications between linguistically diverse users. The authors in \cite{syafrony2022universal} expect that the AI technologies utilized in nowadays applications for automated translation (e.g., Facebook) can be further adopted seamlessly in the Metaverse concept to facilitate communication. Some recent research in the literature took this aspect a step further when they addressed the problem of socializing with people with hearing loss. The authors in \cite{batnasan2022arsl21l} proposed that avatars can be trained to mimic the gesture of the speaking person using an AI system (Figure \ref{fig:interaction_avatar_letter}). \\

\begin{figure}
    \centering
    \includegraphics[width=0.9\linewidth]{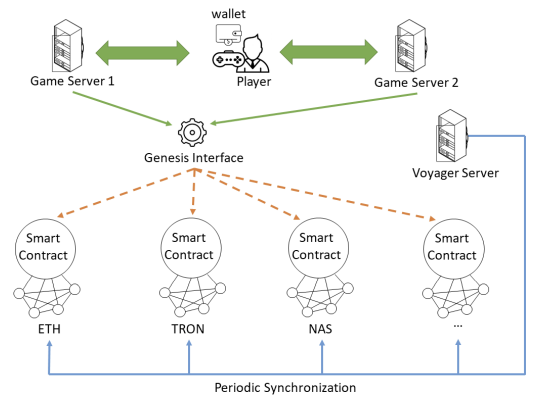}
    \caption{Interoperable Blockchain Gaming Framework Across Multiple Games and Chains \cite{cai2019demo}}
    \label{fig:interaction_gaming_blockchain}
\end{figure}

\textbf{Blockchain.} Furthermore, Blockchain can enforce the Metaverse with a diverse of advantages in such a component. Many works emphasized the importance of integrating Blockchain within the Metaverse to enable powerful features such as data security, privacy, and interoperability in multiple user interaction scenarios \cite{huynh2023artificial}. To demonstrate, Blockchain can be helpful in offering cross-game compatibility, where users can interact with other users while preserving their progress across games \cite{besanccon2019towards}. For instance, the work in \cite{cai2019demo} devised a framework that offers interoperability using Blockchain across multiple games and chains. The authors claim that their framework (Figure \ref{fig:interaction_gaming_blockchain}) is capable of facilitating next-generation Blockchain games. They emphasized deploying avatars on smart contracts to enable their growth during the game's progress. In addition to utilizing the Genesis interface to allow games to share the avatar while facilitating an asynchronous gaming experience. \\

\textbf{Networking.} The need for enabling connectivity for interactions rises as the users connecting to the networks are distributed in a highly dynamic, distributed, and ultra-massive network \cite{zawish2022ai}. Most MMORPGs and virtual events will require a solid networking infrastructure where users can feel the presence of others regardless of where they connect from. State-of-the-art emphasized the role of these networking technologies to enable an immersive experience for the users using Beyond 5G and 6G. For instance, Holographic Telepresence was addressed in several works where researchers studied how a 6G of 0.1ms, backed by terabits of bandwidth per second, is needed to allow such an experience \cite{chowdhury20206g}. \\

\textbf{Computing.} Moreover, any social event in the Metaverse requires computing resources depending on the level of realism the app seeks. According to \cite{hill_2022}, the majority of the VR games' minimum requirements are a processor Intel i5-4590 (or AMD Ryzen 5 1500X), 8GB RAM, and an Nvidia GeForce GTX 1060 (or AMD Radeon RX 400 Series) in order to provide an adequate session for the users where they can interact within it.  \\

\begin{figure}
    \centering
    \includegraphics[width=.8\linewidth]{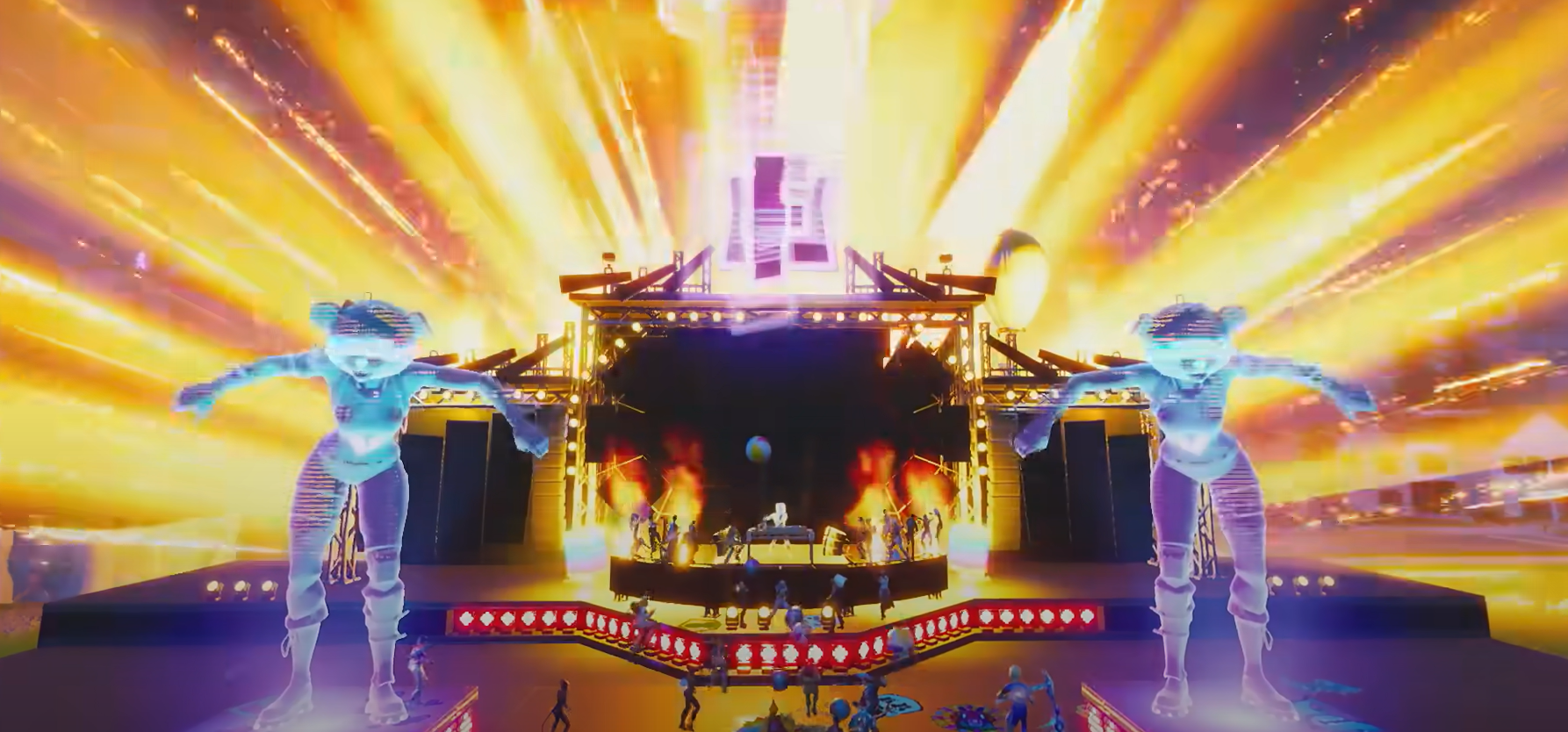}
    \caption{Marshmello's Virtual Concert inside Fortnite}
    \label{fig:interaction_concert}
\end{figure}

\textbf{Business.} Virtual events are gaining a lot of attention lately and companies are racing toward attracting viewers to boost their visibility. Many artists, singers, and performers are collaborating with VR companies to advertise online. To emphasize, one of the biggest events that took place virtually in 2019 was a concert inside Fortnite featuring the artist \textit{Marshmello} \cite{webster_2019}. Such a virtual event gathered 10.7M players to see that artist. The official recap on Youtube acquired 62M views as of February 2023. It is worth mentioning that the event had colorful effects and holograms that are not possible yet in our physical world (Figure \ref{fig:interaction_concert}). Moreover, the famous Canadian singer \textit{Justin Bieber} collaborated with the virtual entertainment company `Wave' to perform his first live show as an avatar \cite{brew_2022}.  \\

\textbf{Ethics \& Social.} Many threats that already exist in our physical world are subject to carrying their burden into the digital one. Threats that may result from interactions among people can be augmented in the Metaverse. Virtual stalking, for instance, is considered a violation of the user's personal life. Stalkers with mental health disorders might track the victim's avatar to spy and collect data for mal-intension purposes. The effect of such kinds of crimes and harassment may have an amplified effect as well on the physical world even when they take effect in the Metaverse \cite{wang2022survey}.
\newline While a lot of research was done on the previously mentioned enabling technologies, there is an urge to study more the relation between the user-to-user interactions with the social enablers, including users' privacy and security, ethical, and psychosociological aspects as the current literature are limited. We address some of these issues in Section IX.  \\

\subsection{User-Business Interactions}
The second type of interactions we study as part of the Metaverse pipeline is the User-Business Interactions. In this section, we describe the set of services offered by businesses to users represented as digital avatars in the Metaverse. For instance, a user can request to ride a car, purchase land or a house, buy clothes, ask for a real-time language translator, or attend universities and get trained. In this current literature, there are no precedented attempts to categorize the User to Business interactions based on the types of applications. Therefore, we conducted a literature review to categorize the groups of applications tailored for offering services to avatars in the Metaverse based on the purpose of the applications and underlying technology used. The resulting categories include (1) Acquisition, (2) Employment, (3) Education, and (4) Entertainment. In the sequel, we present the set of services provided in the Metaverse for each of these categories. Afterward, we describe a common set of enabling technologies that are shared across the different applications offered by businesses. 

\begin{itemize}
    \item \textit{Acquisition}: Purchase a property, such as land, building, house, apartment, car, clothes, etc., and convert it to a digital asset.
    \item \textit{Employment}: Attend the virtual twin of the companies and perform the tasks required. From one perspective, the avatar is part of the business providing services to other avatars in the Metaverse. From another perspective, the avatar works for a company or opens a business that is handled by a service provider. In this case, the job of the service provider is to offer the infrastructure needed to manage a proper work environment such as handling tasks and organizing meetings, delivering supplies, and facilitating coordination and communications among employees and customers.
    \item \textit{Education and training}: Another form of user-business interaction will be offered for education and training purposes. Avatars will be able to attend virtual classes and training sessions. Furthermore, they can work cooperatively to resolve assignments and discuss projects. 
    \item \textit{Entertainment}: attend a sporting event, concert, casino, cinema, and theater virtually from the comfort and safety of their own homes.
\end{itemize}
Business applications simplify and relieve virtual users beyond worries about the underlying complexity, trust, privacy, and security. Services offered by businesses to avatars are created, maintained, and optimized through a set of enabling technologies that are customized in favor of the Metaverse environment.

\textbf{AI.} AI is at the core of most of the services provided by businesses in the Metaverse. Through the data generated from various user interactions in the Metaverse, the volume of generated data is enormous and can be employed to reinforce knowledge and enhance intelligence by personalizing experiences and services to improve user satisfaction. Such data is the target of many business owners to increase their market value by investing in the Metaverse and developing applications pointing to the richness of AI solutions. Digital Twin (DT) is one of the driving forces behind building applications in various Metaverse environments \cite{lv2022building}. In this context, DT is used as a service offered by businesses. For instance, the authors in \cite{lopez2022university} and \cite{duan2021metaverse} provide an example of how to build a DT of a university inside the Metaverse. More precisely, the authors described existing efforts in the business sector to create virtual universities for improving learning quality and overcoming spatial and temporal limitations to perceive knowledge. Bright opportunities exist to improve students' immersiveness in learning through learning by doing, either through virtual simulations or a deeper view of their heritage and culture by visiting historical places and sites and traveling through time. Furthermore, the authors in \cite{cimino2019review} present a DT for building manufacturers and automated supply chains, which offer services for customers digitally and deliver the products to their homes.
In terms of using AI for growing business and serving users, the possibilities are endless and include building virtual robots inside the Metaverse to serve virtual users, either by offering chatbots, real-time translation, tour guides, event management, and tasks automation on-demand \cite{huynh2023artificial}. \\

\textbf{Blockchain.} Multiple businesses in the Metaverse attempt to use Blockchain as part of their services for transparent management and increased value by gaining the trust of users \cite{ryskeldiev2018distributed}. As part of the AI ecosystem, it is important to use trustworthy data, which can be guaranteed through the use of Blockchain \cite{jeon2022blockchain}. An important aspect to offer a business to users is to allow them to pay for it. Through Blockchain, users can use the digital wallet of cryptocurrencies (e.g., Bitcoin, Ethereum, Dogecoin, etc) to pay for services and transfer funds safely and securely \cite{vidal2022new}. The Blockchain in the Metaverse is also used for holding the ownership of digital assets such as lands, houses, or other digital objects. The work in \cite{christodoulou2022nfts} studies the importance of using Blockchain in the Metaverse for businesses while focusing on the importance of cyberattacks detection and improvement of intra and inter-organizational communications among manufacturers through the Blockchain \cite{mourtzis2023blockchain}. \\
Moreover, businesses are more likely to target a variety of interconnected Metaverse environments, or parallel Metaverse, to offer their services. For the user's convenience, the same subscriptions, digital wallets, ownerships, and benefits should be shared across multiple environments. To this end, it is important to utilize the cross-chain concept to facilitate teleportation and maintain high standards of security and transparency. For instance, the authors in \cite{kang2022blockchain}, propose the use of cross-chain-empowered federated learning solutions to improve the security and privacy in the industrial Metaverse. \\

\textbf{Networking.}
The increasing volume of data generated through the IoT devices and users' interactions with Business services raises the need for producing faster network connectivity for real-time application response \cite{cai2022compute}. For instance, real-time digital asset rendering and physical world synchronization for supporting DT require higher network resources \cite{tang2022roadmap}. Furthermore, data aggregation for processing and conducting intelligent inferences requires networking support. Furthermore, it is vital to maintain seamless transitions or teleportation of the same service across multiple Metaverse Environments when the user moves \cite{cheng2022will}. With the development of 5G, these network requirements are slightly relieved, due to the high internet speed and scalability offered compared to previous generations \cite{njoku2022role}. Since a high data rate for rendering is required, where computation is more likely not performed on the headsets or locally, the use of fog and edge computing is becoming handy. Supporting the immersiveness, immediacy, consistency, and reliable API requests between the users and applications is possible through the 5G \cite{fu2022survey}. \\

\textbf{Computing.} Computing requirement is tremendous in the Metaverse due to the need for data processing, AI solutions, XR, and content creation in general. Suitable computing infrastructure that must be adopted by businesses include cloud computing, fog and edge computing, and computing first network (CFN) \cite{fu2022survey}. While cloud computing offers massive computing power allowing fast processing of large volume of data and handling a vast number of API and service calls, networking delays is the main issue behind relying solely on cloud computing to support the Metaverse \cite{ning2021survey}. Consequently, edge and fog computing can complement the cloud computing power by offering additional computing resources near the user, eliminating the issue of high delays for virtual rendering and real-time response \cite{xu2022full}. However, fog and edge computing have limited computing power to support the Metaverse, especially when the number of users increases while counting more on interoperability across different environments. Finally, a business can resort to CFN, which improves computing and networking availability by leveraging more servers from within the edge computing layer but at different locations, thus empowering distributed edge computing \cite{krol2019compute}. \\

\textbf{Privacy and Security.}
Businesses must deal with malicious users attempting to steal users' data and attack services and applications in the Metaverse \cite{wang2022survey}. Privacy and security risks are connected to each of the enabling technologies supporting the Metaverse. Through attacks on authentication mechanisms, malicious users might attempt to impersonate an avatar and steal the access credentials through simple techniques such as email phishing or by imitating voice, behavior, and appearance \cite{falchuk2018social}. As a result of the huge volume of data from users' interactions which are mainly transferred for analysis and AI model updates, such data can be accessed by the attacker through the network for stealing valuable information or tampering. 

\textbf{Business.} Business in the virtual world has a different feel following all the advancements of technologies, more importantly, AR, MR, and AI \cite{nalbant2023development}. Marketing and branding using AI with the help of generated data from interactions and the existence of intelligent virtual assistants. To this end, more opportunities exist for businesses to scale and grow by offering more flexibility for the users to check the products in AR mode. While the user is in a virtual shopping store, the computing machines in the background hold the history of data related to all interactions and time spent looking for certain categories, allowing the development of more robust marketing. All these factors lead businesses to invest millions of dollars in developing Metaverse applications to widen their market and benefit from the hype of virtual technology and intelligence \cite{barrera2023marketing}. \\

\textbf{Ethical \& Social.} Ethical and social aspects are of immense importance in the Metaverse where businesses can develop and serve applications without precautions of the impacts on human lives \cite{kshetri2022policy}. A major issue is related to data collection, where concerns are raised about the amount of data collected, the sensitivity of such data, and the privacy implications \cite{fernandez2022life}. Furthermore, aligned with the objectives of Metaverse to offer immersiveness and near-real-life experience, some applications offered by businesses can have a negative impact on users' lives (e.g gaming industry). Furthermore, lack of awareness and cyberbullying are major concerns that can increase with the circumstances and freedom of behavior provided in the virtual world \cite{qasem2022effect}.\\
From another perspective, when utilized for the correct purposes, the Metaverse is a great opportunity for expanding businesses and providing opportunities for users to develop economies and earn living by making money as they work \cite{kshetri2022policy}.

The Metaverse offers a golden opportunity for businesses to grow and increase their profit and market value by embracing the new advancement in virtual technologies and intelligence. Despite the growing trend towards building more applications for serving users in the Metaverse, it is challenging to account for the concerns related to trusting AI, securing user data and transactions, mitigating security issues and vulnerability, and respecting ethical guidelines and social implications on human lives.

\subsection{User-Object Interactions}
Finally, another type of interactions as part of the User Interactions layer of the Metaverse pipeline is the User-Object Interactions. The user-to-object interactions represent any sort of interaction where the avatar is in contact with virtual models or a digital environment in the Metaverse. These interactions can take various forms in the Metaverse such as owning objects, using objects, and having DTs of IoT devices. Furthermore, object rendering and manipulation can be affected by the rendering features used and the types of applications, such as having non-static environments, a custom field of view or viewport, and offering advanced texturing and sensations. Users are assumed to be able to manipulate any of these digital objects as they do in the physical world in order to provide immersion. Below, we discuss these categories and highlight the research efforts toward enabling this type of interaction.
\begin{itemize}
    \item \textit{Owning objects}: This is where users can exclusively possess and control items in the Metaverse. Items may include real-estates, cars, pictures, or cosmetics.
    \item \textit{Using objects}: This category refers to the ability of the user being able to manipulate the item state by triggering a certain action or to use it for other purposes, such as user relocating. 
    \item \textit{DT IoT devices}: Some devices integrated within the Metaverse can represent a real-world IoT device that could be controlled in a similar manner. Digital smartwatches are an example where a user can track his progress while completing a task (e.g., gaming, health-related, or work-related tasks) inside the Metaverse. Strong interaction occurs when such devices exhibit real-time manipulation on both physical and digital ends.
    \item \textit{Custom field of view}: Personalized features and objects based on the user preferences, such as manipulating the surrounding object colors and shapes for helping colorblind users, or for adjusting stress and depression levels.
    \item \textit{Textures and sensations}: Capture tactile features to virtualize the sense of touch and playback sensations following the user interaction with the object. An example is to be able to sense the texture of clothes in fashion stores inside the Metaverse.
    \item \textit{Non-static environments}: Environments that change over time as a result of object manipulation and user interaction. Specifically, the virtual world should resemble the physical life, therefore objects are affected by metabolism and metamorphosis depending on the environment and surrounding conditions.
\end{itemize}

\textbf{AI.}
In the Metaverse, the environment is composed of virtual objects, that are either built manually using advanced visualization software or constructed using AI. Following the advancements in AI, it is now possible to build meshes of different objects in the Metaverse \cite{huynh2023artificial}. These objects (even human body parts) are rendered and manipulated by the users, thus the resulting shape, rotation, direction, and movement can also be handled using AI solutions through an automated and intelligent workflow. Nvidia proposed GANverse3D, which takes images of objects in the physical world and transforms them into 3D virtual objects with impressive accuracy \cite{zhangimage}. There also exist other attempts to handle object manipulation through shape update and texture learning using generative models \cite{ehsani2021manipulathor}.
In the same context of object rendering and manipulation in the Metaverse, DT requires synchronization between the physical and virtual worlds depending on the action applied to the object in either direction. Using AI and DT, it is possible to achieve this synchronization through state analysis, movement prediction, task learning and autonomous completion, risk reduction, and predictive maintenance \cite{groshev2021toward}.\\
Furthermore, businesses focus on the development of DT tools and frameworks, which are empowered by AI for object modeling and simulation. For instance, DT is utilized in the manufacturing industry through the Siemens Digital Twin tool \footnote{\url{https://www.plm.automation.siemens.com/global/en/our-story/glossary/digital-twin/24465}}. This tool offers means for simulation, modeling, data analytics, and visualization. Similarly, Microsoft Azure Digital Twin\footnote{\url{https://azure.microsoft.com/en-us/products/digital-twins/}} offers a cloud-based DT creation and  integration for object modeling and simulation. Besides, the Digital Twin sotware by GE \footnote{\url{https://www.ge.com/digital/applications/digital-twin}} is another platform offering DT solutions for managing models of industrial assets.

\textbf{Blockchain}: Providing immutability is also one of the main pillars to have a reliable environment. Blockchain is envisioned to be integrated within the core of the Metaverse even at the level where users are interacting with objects. NFTs on the one hand are derived from smart contracts to appear as tradeable objects that are preserved by Ethereum. It offers immutability, decentralization, and interoperability to the objects to be utilized across platforms \cite{wang2021non}. Furthermore, many concerns arose from the centralization of different objects, such as games and their rules, where privacy, latency, and rules manipulation were not accepted to be handled by central authorities. These concerns motivated the research industry to investigate a decentralized computation and token management infrastructure empowered by Blockchain \cite{muthe2020blockchain}. They also claimed that the proposed framework can give further consistency to the NFT trading system across players and games.

\begin{figure}
    \centering
    \includegraphics[width=.8\linewidth]{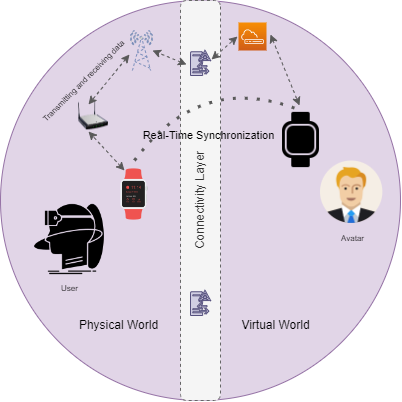}
    \caption{Digitally entangled IoT device}
    \label{fig:interaction_IoT}
\end{figure}

\textbf{Networking.} Creating a seamless and integrated experience inside the Metaverse is a step toward achieving immersion. A DT of a certain object may truly reflect its real-world counterpart if it was digitally entangled. A user-to-object interaction provides a sense of sight and control to the digital replica of a certain IoT device. DT Network relies on advanced telecommunication technologies to achieve seamless connectivity among multiple objects and their digital replicas \cite{wu2021digital}. The entanglement of such devices must rely on several factors to operate properly. For instance, the physical device needs to be connected through either (1) Bluetooth 5, (2) Wifi 6, or (3) LoRa in order to transmit data. In parallel, using high-speed network links (e.g. optical fiber) is essential while relying on optimal data compression techniques in order to enable real-time data communication and reduce latency \cite{mashaly2021connecting}. In such a manner, alternating the state of the object, digitally or physically, may result in real-time changes to its counterpart. Figure \ref{fig:interaction_IoT} reflects how IoT devices can be connected through the network. \\

\textbf{Computing.}
More objects to render results in the need for more computing power to achieve real-time immersive interaction between users and the surrounding environment. The creation of the environment at first requires powerful machines to render the whole scene, which is usually done by cloud computing. Following user-to-object interactions, fog and edge computing can be used to offer sub-rendering as a consequence of object manipulation \cite{lim2022realizing}. Furthermore, fog and edge computing can be used to empower personalized experiences which require separate computing power for the user that usually resides nearby. There exist some efforts in the literature to pre-render the scenes before the users ask them, such as loading the scene by predicting the next angle from the field of view or viewport the user would look at from the headset \cite{chao2022privacy}. This mechanism can reduce the delay that the user should wait to have the environment rendered. In addition, considering different preferences or customized fields of view for users raises the burden on computing requirements to handle the increasing volume of requests \cite{mourtzis2022human}. Moreover, Data generated following object manipulation that is either shared or owned by users requires real-time digestion and inference using AI. Hence, more computing and storage are needed to handle the data. \\

\textbf{Business.} Trading virtual objects is something very common nowadays. There exist businesses that function on the idea of advertising and trading for NFTs which can vary from virtual real estate to in-game items. One of the well-known companies is \textit{NBA Top Shot} \footnote{https://nbatopshot.com/} which offers a platform to trade officially licensed highlights from the NBA. A lot of users are already supporting the NFT concept to the extent that some items are being sold for millions of dollars. As of 2021, an artist named \textit{Pak} sold his artwork for \$91.8M, the most expensive NFT to ever be purchased \cite{xboenfrj_2022}. \\

\textbf{Ethics \& Social.} Users and businesses of the Metaverse should follow certain rules when it comes to object creation, rendering, and personalization. For instance, personalizing the experience for the user based on some preferences could affect this person's personality and demands for the same objects or experiences in real-life \cite{fernandez2022life}. If every user in the Metaverse has the power to control the objects, then there is no control over the whole environment due to the lack of proper policies. Proper control over the types of objects and manipulation permitted should take place, which requires a careful study of the implications on the psychology of users, in the long run \cite{kshetri2022policy}. \\

\textbf{Security \& Privacy}
Just as in real life, the Metaverse is assumed to offer freedom for interacting with objects while conserving the physical laws as much as possible. However, owned objects in the Metaverse should not be stolen or sabotaged by others. NFTs can empower such a feature while preserving the state of the object in case it was modified by a nonauthorized user. Furthermore, Federated Learning can also be applied to this field to study how users are reacting to object manipulation, and to extract knowledge from their sensors in a secure manner.


\section{Existing Challenges \& Research Directions}
\begin{figure}
    \centering
    \includegraphics[width=0.9\linewidth]{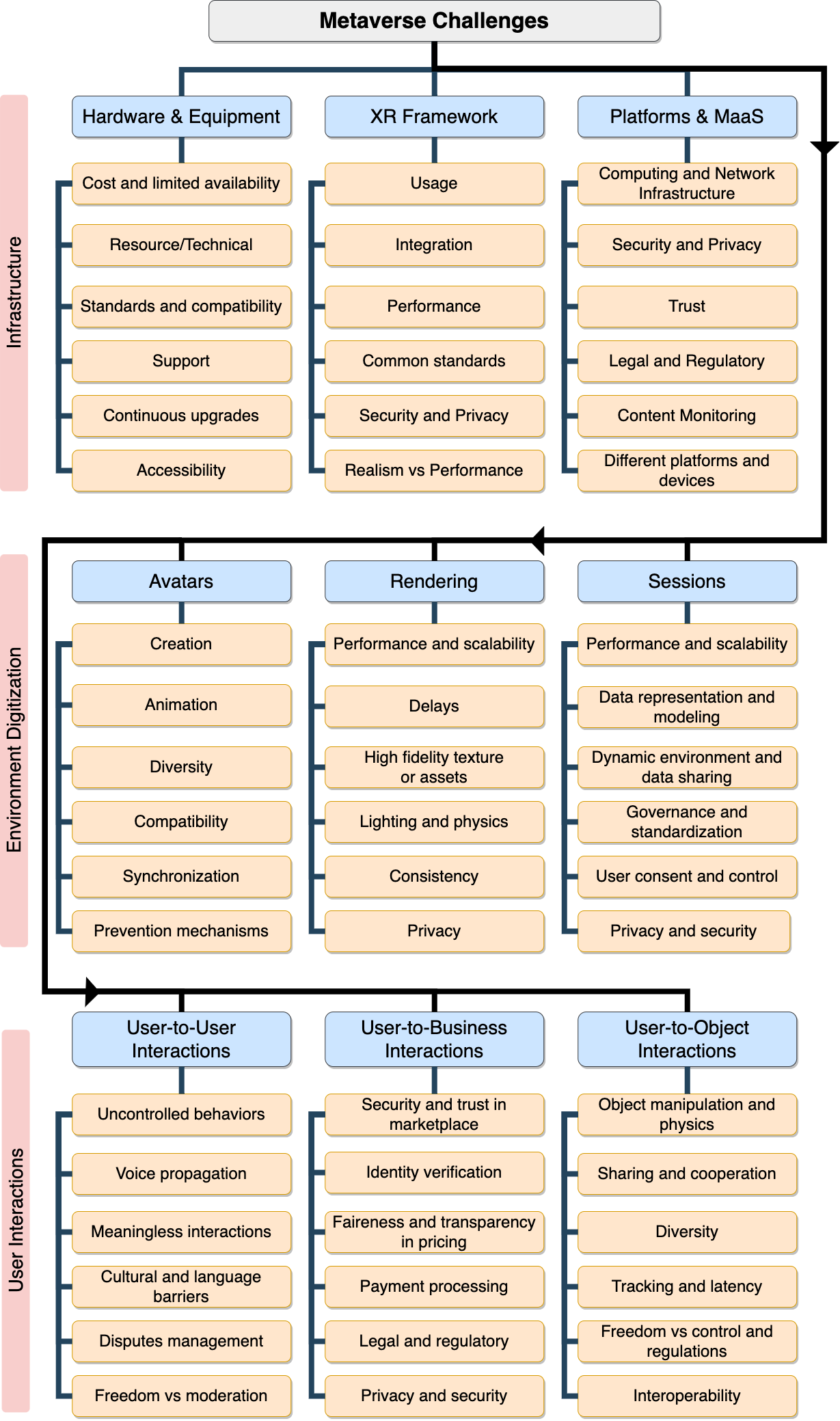}
    \caption{List of challenges per component in the Metaverse ecosystem}
    \label{fig:metaverse_challenges}
\end{figure}

The world is not yet ready for the full adoption of the Metaverse. The literature still requires a careful study of limitations and requirements to elevate the user experience and achieve the Metaverse objectives. Following our detailed literature review for each of its components, we study in this section the existing challenges of academia and industry facing the Metaverse realization. In Figure \ref{fig:metaverse_challenges}, we summarize the list of challenges per component in the proposed pipeline and multi-layered Metaverse architecture.

\begin{figure}
    \centering
    \includegraphics[width=0.9\linewidth]{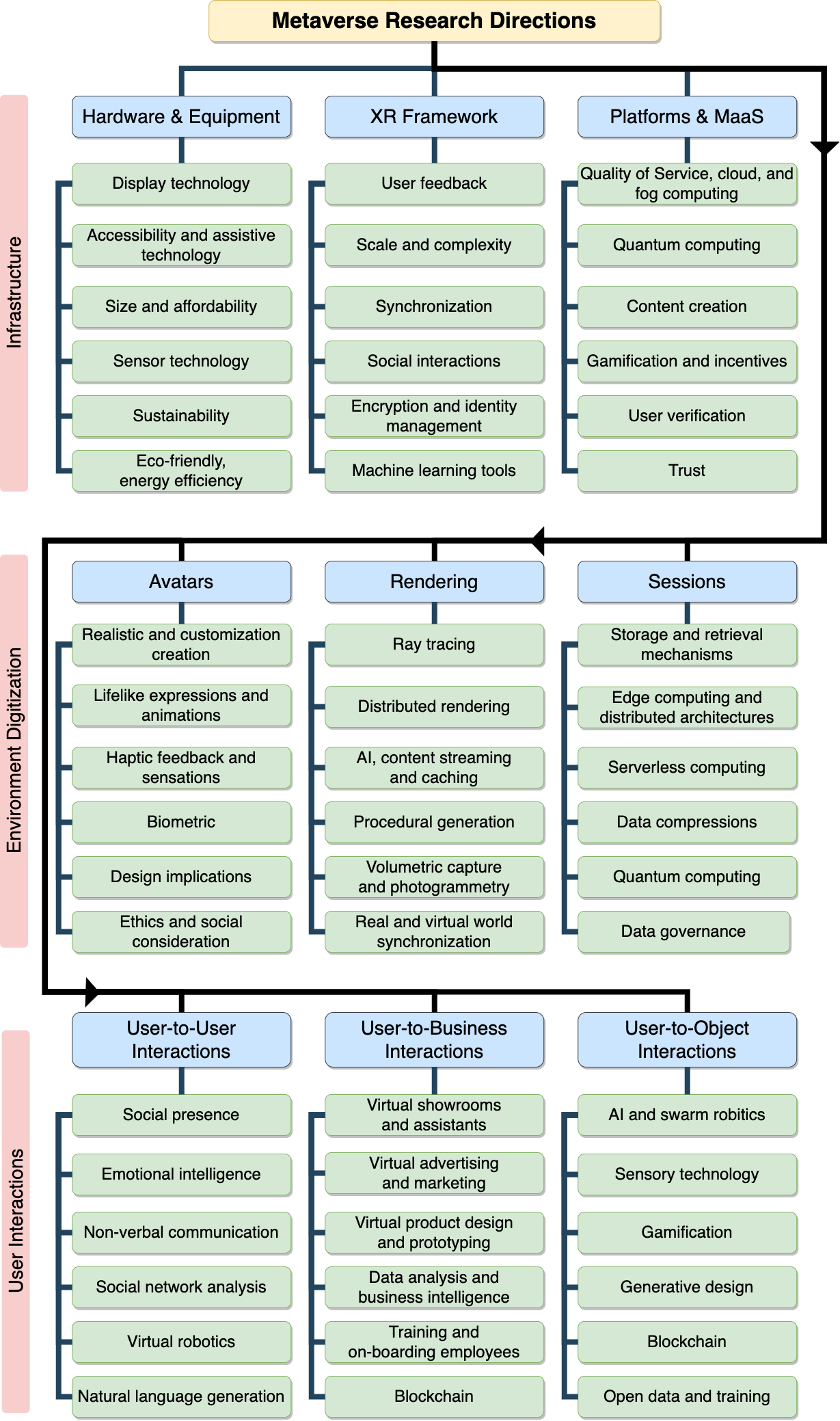}
    \caption{Future directions per component in the Metaverse ecosystem}
    \label{fig:metaverse_directions}
\end{figure}

For each set of challenges per component, we present a detailed study of potential impactful and effective research directions. Our presented directions are a result of a thorough study of the current demands to empower immersiveness and realism in the Metaverse ecosystem. In Figure \ref{fig:metaverse_directions}, we summarize these directions.

In the sequel, each section contains a list of challenges followed by potential research directions.
\subsection{Hardware \& Equipment} 
Despite the major benefits that the hardware and equipment enable the end users, entities, and organizations, there are some challenges that need to be tackled in the future. One of those challenges is lowering the production costs of these devices so that they can be more affordable to the public, especially since Metaverse-related applications rely heavily on these devices to provide richer experiences to end users. Another challenge is to encompass or fit more computing and networking power as well as more graphical abilities alongside more features while keeping the device size as compact as possible. Furthermore, existing hardware still faces many limitations in supporting technological advancements. These limitations include the field of view, rendering resolution, and battery life \cite{masnadi2022effects}. Some hardware also still lack the support for advanced features that empower immersiveness, such as eye tracking and haptic feedback capabilities. In terms of compatibility and cross-platform support, most of the hardware and equipment are not compatible when deployed for supporting cross-platforms or applications across different Metaverse environments. In addition, these devices require constant updates to cope with the technological and software-based advancements to adjust and upgrade their capacities. Such upgrades require constant maintenance and are expensive and time-consuming. \\

Existing Hardware and Equipment for supporting the creation, development, and deployment of the Metaverse are still in their infancy. Research is required to develop new display technologies, such as holographic and light-field displays, to improve realism and immersiveness using VR and AR technologies \cite{chen2019overview}. In addition, research on new technologies for building input devices is required, such as gloves or full-body suits. In addition, accessibility should be addressed by research on new techniques to support people with disabilities to use the Metaverse. This is possible through improving haptic feedback and utilizing eye-tracking technologies. It is also important to support voice recognition and control by improving its accuracy and responsiveness. Moreover, sensors on hardware and equipment require additional improvement to maintain stable accuracy by improving calibration algorithms and utilizing machine learning. Furthermore, the sensors' range and flexibility should be improved, which is important for motion-capturing devices. Range and flexibility can be increased by improving the speed and quality of the signal processing algorithms while incorporating and optimizing the use of additional sensors \cite{chen2019study}. Concerning manufacturing, hardware and equipment production must adhere to sustainability and eco-friendly requirements. Moreover, the energy utilization of devices should be further studied by investigating the use of low-power components, such as processors and displays. It is also possible to reduce the energy usage on the devices by resorting to the cloud or fog for faster computing and developing power management software on these devices. 
\subsection{XR Frameworks}
Several aspects are negatively reflected regarding the challenges of creating and adopting an XR framework. There exist several challenges in relation to the use of these frameworks, due to the limited features and capabilities offered. These limitations are represented as the inability to access and use the XR experience since XR frameworks may not operate with all devices and operating systems, causing communication or interaction delays. In addition, the overall complexity of XR frameworks for the users would make them less likely to use or interact with the XR experience. Furthermore, in terms of communication, the XR experience's reliance on networked communication to enable real-time interactions could impair its quality, restricting its ability to gather and analyze data in real time \cite{ratcliffe2021extended}. These challenges limit the possibility of integration between the XR frameworks and any of the pipeline components related to the Metaverse development ecosystem. This includes limited support for rendering and user interactions. In addition, applications offered or supported by these frameworks are exposed to various limitations affecting user immersiveness. For instance, existing XR frameworks have poor tracking and spatial understanding, leading to imperfect rendering and processing of the surroundings. On the other end, XR frameworks are subject to performance issues leading to latency or drops in frame rates. This can be a result of poor development quality or the lack of compatibility on multiple hardware and equipment, platforms, and Metaverse environments. With regard to existing guidelines, there are no standards that can regulate the deployment, development, and compatibility of the XR frameworks, which may lead to fragmentation problems. Additionally, the XR frameworks may communicate private information, proprietary information, and location data across the network, exposing them to security risks, including hacking and data leaks, which may affect the user's confidence in using the XR experience. 
Finally, there is a challenging trade-off that existing XR frameworks should study, which is the balance between improving performance and realism. In other words, improving and optimizing the performance of these frameworks could lead to reducing the immersiveness, quality of rendering, spatial understanding, and realism, and vice-versa.\\

In order to reflect on the challenges discussed over the XR frameworks, several aspects of potential directions and measures can be taken to lessen them and suggest potential solutions to ensure a positive experience for both developers and end users. An important direction is to adopt standards and protocols, ensuring that various XR frameworks and systems may communicate and interact with one another through developing and creating open-source libraries or APIs. Another direction is to create XR experiences that are simple and intuitive for users to understand by incorporating user feedback into the design and development process. Furthermore, creating new networking protocols and technologies that are more effective and resistant to latency and bandwidth limitations could improve network performance \cite{cheng2022will}. Moreover, the XR framework should offer solutions for the increased complexity of supported applications through dynamic scaling approaches and improving compatibility measures. Besides, real-time movements and tracking synchronization are vital for supporting interoperability, which is possible by developing standards to be adapted across these framework creators. In addition, maintaining the security and privacy of users' data by implementing a security mechanism that limits access to adversaries, such as encryption and authentication procedures, and ensures compliance with security legislation. Finally, creating AI models and algorithms that are transparent and accountable, and take into account diversity and inclusion during the data collection and training processes \cite{cao2022decentralized}. AI-based solutions should be developed to support social interactions by developing advanced natural language models and tools. Therefore, collaboration among specialists from different fields is vital in improving the XR framework development and usage.

\subsection{Platforms and MaaS}
As a result of the size and complexity of Metaverse Platforms, their deployment and development are being negatively impacted by a variety of factors. The interactions amongst avatars necessitate tight latency and network dependability requirements. 6G is indeed a crucial deployment solution for the Metaverse \cite{chang20226g}. However, the communication protocols and optimization techniques must be thoroughly explored to match the anticipated needs. When speaking of Metaverse platforms, enormous computing power and resources are necessary. The availability and efficiency of existing algorithms to compile such voluminous amounts of generated data are challenged by these factors. In addition, improvements in the performance of distributed computing are required to improve the operation and efficiency \cite{hammoud2018detection}. Besides, users will be subject to security and privacy risks concerning identity theft and cyberbullying within the platforms \cite{tugtekin2023dark}. Moreover, a high degree of trust must be maintained between users and the platform to provide a positive and secure user experience. Numerous aspects emphasize the necessity for strong procedures to regulate the conduct of avatars in the Metaverse to prevent criminal and immoral activities, which is still missing and requires careful investigation. In terms of regulation, there is no content monitoring in existing platforms to avoid harmful or offensive behaviors in the Metaverse. Besides, there is a need for a monetization model that controls the platforms and MaaS providers. In this context, it is challenging to balance profitability with user experience and fairness. Finally, there is no support for interoperability between different Metaverse platforms \cite{rawal2022rise}. In addition, the integration between various platforms within the same complex Metaverse environment is not yet supported.\\

The back end forming the backbone of Metaverse platforms and services requires flexibility and scalability in terms of computing and networking performance. Therefore, the Quality of Service (QoS) can be improved by integrating the use of cloud and fog computing that utilize containers and micro-services architectures for seamless on-demand deployment \cite{sami2020dynamic, sami2021demand}. Through a distributed on-demand fog and edge computing architecture, the computing and networking load are reduced between Metaverse users and the cloud, relaxing resource utilization. Furthermore, it is important to develop resource management solutions, ensuring effective and efficient dynamic and scalable service placement, host selection, and horizontal and vertical resource scalability \cite{sami2021ai, shamseddine2020novel, hammoud2021stable}. Resource management for supporting the Metaverse requires proactive and demand-driven decisions. As a potential direction, Reinforcement Learning (RL) and reward shaping solutions can be developed on top of an environment modeling design to meet the resource management requirement for Metaverse platforms and MaaS. Moreover, networking protocols should be offered and adapted to the platforms and services to ensure packet delivery by developing advanced traffic engineering, network slicing, and quality-based routing mechanisms. Furthermore, reducing latency and improving content delivery through caching mechanisms are of immense importance. Therefore, additional resources must be dedicated to developing Content Delivery Network (CDN) solutions for faster and more reliable delivery, by caching the content closer to users \cite{zolfaghari2020content}. Using CDN, requests are routed to the nearest server in the network that has a copy of the content. On another end, applications on platforms and algorithms running behind the services can be improved by utilizing Quantum Computing. Applications of Quantum Computing as tools to support the Metaverse platforms and MaaS developments include: (1) optimization towards rendering; (2) physics and environment simulation to study and improve the user experience; (3) advancement and security in computing and networking architectures; and (4) Quantum AI can be used to optimize the machine learning solutions and autonomous agents for a more immersive experience. Moreover, providers of the Metaverse platforms and services should manage content creation and offer curation tools to find and share relevant content that is more user-friendly. This is possible by offering solutions for digital assets management, content discovery, social sharing, and collaborative curation. With regards to controlling user behaviors, research can focus on developing gamification solutions that can incentivize virtual users or avatars to promote positive behavior and engage in the environment. Furthermore, new research directions should focus on user identification and building trust in cross-platform environments, where users can interpolate platforms to immersive different experiences \cite{wang2022survey}. This is possible through (1) developing distributed cross-chain solutions, (2) utilizing a reputation mechanism, (3) utilizing multi-factor authentication (MFA), and (4) partnering with trust identity providers, such as government agencies or banks.
\subsection{Avatars}
Various challenges are yet to be solved when considering avatar modeling for the Metaverse. For instance, realism is one of the primary directions during the avatar creation in which a regular, non-skilled, Metaverse user expects a high-quality representation of their body, skin, face, and voice tone. However, till now, none of the mentioned approaches in avatar modeling could generate a highly flexible and representative model of their users. Even though manual avatar creation can achieve some level of realism, it requires a fair amount of knowledge, time, and experience, which is absent for the average user. Creating avatars using AI techniques has the potential to produce realistic results, but it also presents several challenges that must be solved. Such techniques still need to be improved in critical areas when considering an interactive virtual environment expected from a Metaverse application \cite{genay2021being}. Thus, we highlight two key challenges: realism and interaction. Realism is affected by the narrow shape variation that limits the creation of highly realistic avatars for its users. It affects the accuracy of generating correctly shaped avatars, especially considering users’ weight. In this regard, avatar creation models should be able to correctly represent their users while considering the limitations in terms of collected information. Moreover, facial reconstruction and expression are affected, and none of the available approaches can achieve realism when considering facial representations. Furthermore, these models are limited in their interaction capabilities, posing constraints on users’ movements and engagement with the environment. Various approaches aim to solve a distinct aspect of each part of the model. However, such a domain still lacks a combinatory solution that integrates multiple models into one. Furthermore, the creation of animations and lifelike expressions, such as realistic movements and emotions, are not yet supported in existing avatar creation and animation solutions, which limits the user experience and engagement with others.\\
With regards to cross-platforms and as a result of interpolation across environments, the support for diversity in the avatar representation is important and not yet supported. For instance, a diverse representation of identity depending on the culture is necessary, which requires a large set of libraries and assets to create the personalized model representation. Furthermore, moving from one platform or environment to another raises compatibility issues for the avatar creation or rendering tools, which has not been addressed in the literature. In the same context, synchronizing avatars' movements and interactions across platforms is needed. In addition to the technical challenges of creating tailored avatar models, security and ethics pose major concerns. For instance, creating and animating avatars often involves the collection and use of personal data, such as images, sensory data, and behaviors. Access to this sensitive information can make users hesitant to enter the Metaverse, as they may be concerned about the publicity of their raw data. 
Finally, from the social and ethical aspects, there is no study that presents effective prevention measures to avoid harmful avatar behaviors, hate speech, or virtual harassment \cite{cheong2022avatars}.\\

In the future, it is expected that users might care more about their appearances in the Metaverse vice the real or physical world. The first impressions of users entering the Metaverse are driven by their look and appearance as avatars. Therefore, it is necessary to address the existing challenges related to avatars creation, movement, and management to enable the basic requirements for users as part of their Metaverse experience and onboarding. A research direction concerning avatar creation and modeling is to use machine learning generative models for more customizable and realistic avatars \cite{harshvardhan2020comprehensive}. This includes learning models for creating faces, textures, clothing, and voice tone adjustments. Existing advancements in convolutional and recurrent neural network design are a promising start for building such solutions. Furthermore, 3D scanning of the physical appearance of the user can be used for realistic avatar creation \cite{dixit2019tool}. The first step in 3D scanning for avatar creation is to perform scanning of physical appearance using photogrammetry software or just smartphones. The second step involves modeling, where scanned users are transformed into 3D digital representations subject to customization and adjustments. The last two steps include texturing (skin, hair, and clothes) and animation of the avatars to simulate realistic movements and interactions. Animation can be performed with the support of AI that utilizes motion capture to develop procedural animation systems, lip-syncing with mouth movements, and facial animation and expressions. In terms of animation, using advanced device sensors involves transmitting the sense of touch in addition to visual or audio feedback. The sensation for improving animation requires further integration and support of haptic feedback, tactile, and temperature sensors in the platforms, In addition, using biometric data, avatars become more realistic with automated personalization features \cite{gavrilova2011applying}. For instance, facial features can be accurately represented on the avatar face by using face recognition mechanisms, which result in real-time face animation. Moreover, the user voice can be analyzed for identity verification and the development of self-learning and adaptive AI models to create avatars that speak like the real user while controlling expressions and mouth movements. Furthermore, integrated biometric sensors in input devices can be used to capture physical states or movements. For instance, reading heart rate can be reflected in the avatar state in the Metaverse. 
In terms of security and privacy aspects, it is important for regulation on personal data acquisition to be updated accordingly. Additionally, applications should implement robust security measures to protect personal data, as well as be transparent about how the data is being used \cite{buck2022security}. 
Finally, ethical and social implications must be carefully investigated through a set of rules and regulations to avatar design implications and behaviors in the Metaverse.

\subsection{Rendering}
Despite all the benefits that rendering engines provide to users and Metaverse platform providers, there still exist challenges and problems related to performance, quality, and consistency. 
One of the main challenges related to rendering engine performance is the limited capability to render large and complex environments, such as large cities or natural landscapes, in real time \cite{el2019survey}. Real-time rendering is essential for producing sensory images while forming a continuous flow rather than discrete events. Real-time rendering also entails the construction of 3D worlds that communicate with avatars, thus requiring quick environment reflections of the consequences of such interactions. Due to the increased latency and complexity of existing rendering engines, latency is increased, thus affecting the immersiveness and quality of experience for Metaverse users. In addition, there is a scalability problem. Numerous users will coexist in the virtual worlds of the Metaverse and interact with each other there. Real-time user interactions and the rendering of 3D environments will require computation-intensive calculations in addition to high-performance communication networks \cite{xu2022full}. In the same context, rending high-fidelity textures and assets without exceeding the computing and networking limitations of the infrastructure is challenging. Besides, simulating realistic lightning is challenging and computationally expensive. Lightning includes simulating the behavior of light when affected by different atmospheric conditions, requiring accurate simulations of materials and surfaces \cite{xiong2021augmented}. This increase in complexity also applies to simulating reflections, shadows, and physics in general, which are not fully supported in existing rendering engines. Therefore, there is a major challenge in balancing the trade-off of realism vice performance and scalability. Moreover, there is a challenge in achieving consistency in rendering across various platforms and applications, requiring careful integration and compatibility with other components of the development pipeline. Finally, there is a privacy challenge in the emergence of AR/VR rendering because data can be collected in new modalities. For instance, eye-tracking may be captured when using HMDs. While this information may be crucial for enhancing the efficiency of rendering, companies can also utilize it to determine consumer attention spans to better promote their products.\\

To address the challenges posed by existing rendering engines in the context of the Metaverse, we present a list of directions that can improve the quality, speed, and performance of rendering solutions. For addressing the need for realistic rendering of light in the Metaverse, the ray tracing rendering technique can be used and augmented for supporting the high complexity in the Metaverse environments. Ray tracing is computationally expensive but results in realistic light simulation by respecting the underlying physics of different materials and surfaces \cite{he2018design}. For addressing the issue of performance and increased delays by rendering engines in the Metaverse, research should focus on advancing existing hardware technology and consider performing distributed rendering. More specifically, multiple computing machines can be used to render a single scene, which helps reduce the amount of time and resources required for rending. Another direction can be developed by relying on AI techniques to support scene rendering in the Metaverse \cite{tewari2020state, jing2019neural}. In this regard, AI can be used to optimize the rendering settings and resources required without sacrificing the quality. Besides, AI and heuristics can be used to cache content for streaming by studying the virtual user behavior and predicting the need for rendering to act proactively and avoid latency. Furthermore, rendering engine performance can be improved by investigating the use of procedural generation in the context of Metaverse, where content is generated on the fly \cite{liu2021deep}. For instance, procedural generation can be used to generate textures and models, which reduces the need for large storage or high computation power. Moreover, photogrammetry can be employed to create 3d models of objects with enhanced quality textures and details.  As a result, effective resource allocation is necessary to maximize service delivery performance to a large number of users at the edge. 

\subsection{Sessions}
Despite the importance that sessions support the Metaverse providers and users, some problems related to session resource optimization, scalability, data collection, and usability still need to be addressed. In this regard, several Metaverse questions arise related to the session timing, resource allocation and release, and the time the data should be stored, used, and maintained on the servers. One of the main challenges affecting sessions and data management in the context of Metaverse is related to performance and scalability issues. Particularly, as millions of concurrent users join the same Metaverse environment, which already requires extensive resources concerning the prior pipeline components (e.g., rendering, XR frameworks), session management increases the burden on resource utilization, including computation and networking resources. Consequently, the infrastructure faces a constant increase in the volume of data and user traffic from environments as a result of a set of user interactions and activities. With the increase in the volume of generated data, another challenge arises related to data representation and modeling. Therefore, scalable, efficient, and effective storage, processing, and analysis mechanisms still need to be investigated, especially in the context of the Metaverse. In addition, Metaverse environments are dynamic and require fast adaptation to changes, thus a dynamic change in the session management mechanism to read, store, process, and analyze the information. Furthermore, the literature is still missing an effective data-sharing mechanism inside the Metaverse environment, where various entities cooperate to achieve a task.\\
The availability of several platforms that provide Metaverse services, and the possibility of moving between these platforms (interoperability), raises concerns related to the mechanism used to move data between sessions across different platforms \cite{chen2022cross}. In this context, cross-platform session management is challenging to manage and synchronize the session state in the different environments effectively and in real time. To this end, session management mechanisms must be compatible with different environments, platforms, and Metaverse services. Additionally, there is no common standardization or governance over the session management mechanisms in the Metaverse. For instance, there is no agreement on how data can be managed and stored in a single or cross-platform environment. However, users should be given control over their session data, in addition to requiring consent to access or analyze it.\\
Finally, session data storage and exchange are subject to increasing security and privacy risks in the Metaverse, which requires further investigation and elaborated protections and countermeasures.\\

In the context of sessions and data management for the complex Metaverse environment handling millions of concurrent users, some technological contributions need to be explored to address existing challenges. The first direction would include investigating a set of distributed data storage mechanisms for managing sessions in highly loaded environments. In this context, data will be stored across distributed servers to improve accessibility and redundancy. Consequently, data retrieval can be improved as a result of parallel queries from distributed storage. In this regard, effective distributed storage and retrieval mechanism should be investigated, where AI mechanisms can be effective. Solutions to achieve distributed storage are to use distributed (1) cloud computing, (2) edge and fog computing, (3) Content Delivery Networks (CDN), and (4) Blockchain-based storage. Cloud computing offers automatic scaling and load balancing of storage and data analysis compared to on-premise servers (i.e., edge and fog servers). Furthermore, edge and fog servers offer reduced network utilization and latency, as well as fast analysis and reduced loads on external servers. 
To this end, a hybrid architecture is most likely to be the effective approach; however, several proposals should be designed per solution to obtain its full potential (i.e., divide and conquer strategy). Additionally, CDN can be used to deliver session-related information to users to reduce latency and computation on the Metaverse infrastructure. Besides, Blockchain is a distributed ledger technology that uses distributed storage by nature and requires a consensus mechanism for approving the addition of data in the form of transaction blocks. Thus, the use of Blockchain for managing sessions would improve distributed storage and retrieval, in addition to adding a robust security layer to protect user data \cite{cao2022decentralized}. The role of AI is major in managing the computing infrastructure of each of these solutions. In this regard, AI can be used for automatically managing resources on-demand by studying the change in loads and acting proactively for assigning storage and distributing analysis tasks while performing intelligent vertical and horizontal scaling and load balancing \cite{sami2021ai}. Furthermore, AI is the primary tool adapted to create personalized experiences by offering user sessions that adapt to user preferences. For instance, reinforcement learning and recurrent neural network can be used to create adaptable models that study the change in preferences over time. Besides, serverless Metaverse services bring additional flexibility and performance improvement due to the ability to create and delete instances on the fly, assign additional storage capacity, and reduce the cost of management \cite{li2022serverless}. Hence, serverless computing and storage architectures are promising and should be properly investigated.\\
To further improve the efficiency and cost of data storage, data compression mechanisms can be employed \cite{jayasankar2021survey}. Data compression, when effective, is guaranteed to reduce latency and improve the performance of data flow management and transmission. On another note, quantum computing combined with AI (i.e., Quantum AI) is promising to reduce and optimize the computation through faster processing of session data.\\
Finally, advanced networking architecture is required to support data management and transmission. Therefore, the 6G network perspective and contributions can be used to examine the resource heterogeneity problem between different devices, while supporting session synchronization and reliability.

\subsection{User-to-User Interactions}
The user-to-user interactions still lack the maturity to express straightforward guidelines on how to interact with others \cite{oh2023social}. There are still many questions to be asked regarding this matter. First of all, uncontrolled behaviors of avatars inside the Metaverse are one of the main challenges, such as toxic behaviors or harassment. For example, stalking in real life can be facilitated as the stalker would be able to collect information about the victim’s daily life. On the other hand, digital stalking is a topic of no less importance. Digital stalkers might follow the victim’s avatar and keep track of his virtual activities and behaviors. Furthermore, the concept of blacklisting people in the virtual world is important. Currently, there is no specific guideline for application providers to follow on how to restrict people from encountering others.
Besides, some of the concerns are also relevant to Voice Propagation \cite{jot2021rendering}. Specifically, voice propagated between avatar representations of users is subject to multiple existing issues, including (1) quality of audio, (2) latency, (3) spatial audio, and (4) accessibility. Spatial audio should allow the avatar to hear from the direction of the speaker. Furthermore, accessibility is not yet addressed in the literature, allowing users with impairment or disability to have access to this feature. Furthermore, voice propagation is subject to privacy and security concerns, as conversations might be interrupted by others.
In terms of guidelines and regulations, there is no common scheme for defining the type of applications or interactions in the Metaverse, which should ensure meaningful interactions for benefiting from their presence in the Metaverse. The quality of these interactions between users is not currently monitored by any tool or solution. Furthermore, cultural and language barriers must be avoided to expand the Metaverse's capabilities and overcome communication limitations. Furthermore, there is no existing mechanism for managing user disputes, allowing reporting and addressability.
Finally, there is a major trade-off of freedom vice moderation and control, which should be heavily studied to achieve a well-balanced interaction between users.\\

For addressing the existing challenges entailing user-to-user interactions in the Metaverse, there is a need for collaboration among multiple companies and regulators. Furthermore, efforts are required to state the guidelines for such a type of interaction, especially when hosting an environment for multicultural users. Empowering social presence among avatars in the Metaverse is of immense importance for achieving immersiveness and engagement in the virtual space. Social presence can be achieved in different ways. One of them is implementing spatial audio to improve interactions and create a sense of presence \cite{kim2019immersive}. There are several steps to implement spatial audio, which include (1) capturing data, (2) creating a virtual audio environment, (3) assigning audio sources, (4) simulating sound propagation, and (5) rendering the audio stream in the virtual space. In addition, social network analysis and optimization can be achieved using AI technologies for learning and improving past and undergoing users' interactions in the environments, thus leading to an increased quality of experience \cite{laurell2019exploring}. Furthermore, utilizing avatars that mimic emotions and expressions enriches the social presence experience for Metaverse users. Another important research direction is to investigate non-verbal communication techniques, such as text or sign language, for voice propagation accessibility.\\
Tracking and sensing technologies can also improve user-user interactions and immersiveness by achieving accurate and responsive rendering feedback during communications. Investigating advanced tracking and sensing technologies leads to improving body tracking, gesture recognition, eye tracking, and haptic feedback.\\
Another method to improve user-to-user interactions is by creating virtual agents or assistants and utilizing natural language generation (NLG) \cite{gatt2018survey}. In specific, virtual assistants support avatars and guide them to navigate the Metaverse and perform actions. NLG and agents can also be used for real-time translation to overcome the barrier of multi-cultures and communications. NLG can be used to generate natural-sounding speech based on the interaction context, thus empowering more engagement and dynamicity.

\subsection{User-to-Business Interactions}
Businesses face challenges concerning interactions occurring in the Metaverse following the services and applications offered to users in the virtual space \cite{pozniak2022could}. First, the computing and networking infrastructure is limited per the Metaverse environment. With many businesses sharing the same environment, it is challenging to assign and scale computing resources for each application per business while considering the fairness factor. Besides, the marketplace composed of businesses should be trustworthy for customers to ensure smooth buy and sell operations. In particular, applications should be tested for various issues that may arise from using AI, Blockchain, overloading networking, and computing resources, and most importantly, ensuring the maximum possible protection against attacks and malicious users. Furthermore, no solution can let businesses verify users' identities for building trust during business interactions. Moreover, in virtual economies, there is no mechanism that ensures a fair and transparent pricing system and currency exchange rates. Besides, one of the challenges facing user-to-business interaction is the lack of a reliable and efficient payment processing mechanism, including a transaction tracking and management lifecycle to support users in receiving a higher quality of experience from businesses in the Metaverse.\\
Every business must account for being responsible for damaging or exposing user information and details by following some guidelines to make sure that the minimum requirements are met before deployment in the Metaverse. Henceforth, one of the main challenges arising from such type of interaction is to achieve and control a balance between protecting the user privacy and security, vice the need for personalized business services, and for forming targeted advertisements for promoting products \cite{kim2021advertising}.
No single entity owns the Metaverse; however, multiple mini-environments are more likely to be developed for different experiences, where multiple businesses can share applications. To this end, there are currently no guidelines for businesses to follow when deciding to join the Metaverse to offer their customers an immersive virtual experience by promoting their products.\\

Following the list of challenges facing User-to-Business interactions, we present a list of research directions that can potentially address some of these challenges. Starting with the infrastructure limitations, there is no clear mechanism for identifying the available computing and networking infrastructure and how to use it by each. Henceforth, a clear study about the management and scheduling of such resources is essential, while considering fairness and trust among applications and customers. Furthermore, the Metaverse is hungry for resources, thus providing fair allocation and innovative coordination between the physical and virtual worlds for sharing network and computing resources is necessary. More specifically, computing resources should be allocated fairly to cope with the increase in demands as users start relying more on Metaverse to engage socially and perform daily tasks \cite{sami2021demand}. Moreover, the use of cloud, fog, edge, and CDN resources and technologies would essentially improve the support for businesses in the Metaverse. Through business-to-business collaboration, various challenges can be conquered.\\
From a user perspective, a research direction should consider building and managing the virtual economy of businesses, as well as transactions in the Metaverse \cite{popescu2022augmented}. One step is to motivate the creation of virtual showrooms for shopping, browsing virtual products, and hosting events for brand management. Furthermore, virtual assistants can be created through the help of AI with customized user experience to help create a trusted bond and relationship between customers and businesses \cite{lee2022toward}. Those assistants can also be used as customer support agents for managing users' complaints. Virtual marketing and immersive advertisement strategies are also important for supporting virtual businesses and building a powerful economy. This also entails using the Metaverse for product design and prototyping \cite{escalada2019design}. Powerful rendering engines are also promising for managing the complete supply chain in the Metaverse, which has a positive impact on businesses. For instance, supply chain management in the Metaverse can include tracking inventory and performing required logistics in real-time. Furthermore, the Metaverse can help businesses in expediting training and onboarding sessions for new employees while facilitating team building and enabling remote work opportunities \cite{kral2022virtual}.\\
On another note, the use of Blockchain technology by businesses empowers secure and transparent digital commerce to motivate the actions of buying, selling, and trading virtual goods. Moreover, interoperability raises additional challenges, including ways to teleport required data across various environments, establish reliable connectivity with distant servers, monitor and maintain access, and standards for financial exchange. In this regard, the Blockchain can be used to address some of the interoperability challenges by employing a cross-chain technology with a consortium Blockchain between partners.
\subsection{User-to-Object Interactions}
This type of interaction entails challenges that could hinder the user experience, including reliability, response time, accessibility, interoperability, and security. First, the user-object interactions are subject to an increase in the computing and networking resources of the Metaverse. As the number of users inside the same environment rise, interactions with objects increase while still expecting a reliable and real-time response. Due to the increase in resource demands, the underlying infrastructure can become overloaded and might not account for all the requests, leading to service unavailability and degraded quality of experience. Furthermore, users might request personalized experiences that will increase the load in rendering separate objects and sub-environments per a single or group of users. Besides, the issue of scalability and increased resource usage arises when creating a DT of objects inside the Metaverse to replicate real-world objects in real-time and vice-versa \cite{van2022edge}. Examples of existing DT frameworks and tools from the industry are 3D Experience platform by Dassault Systems \footnote{\url{https://www.3ds.com/3dexperience}} and Anasys Twin Builder \footnote{\url{https://www.ansys.com/products/digital-twin/ansys-twin-builder}}, which help in creating and managing DTs through data modeling, simulations, and data analysis. In this scenario and following the features offered by these frameworks, DT requires an increase in network usage to read and transfer data between the virtual and physical worlds in real time, in addition to the computational resources needed to maintain a real-time rendering scheme for the object.\\
Due to the immersiveness requirement, manipulating objects should be a natural and intuitive action, such as picking up or moving objects, which requires additional research efforts in the computer graphics field \cite{choi2022study}. Similarly, object physics and interactions should be natural and close to real life, which is also challenging to achieve, especially across multiple platforms. In many environments, achieving consistent object manipulation is challenging. Furthermore, object interaction should be supported for different types, sizes, or weights of objects forming an interactive environment. In terms of hardware support and sensors, latency in input devices and a decreased accuracy of object tracking degrade the system performance, especially as the number of concurrent interactions increases. In addition, there is a need for a mechanism for managing ownership and sharing objects in collaborative settings.  To this end, there is a challenging balance between guaranteeing freedom for users with regard to interactions with objects, and the increase in control with additional limiting guidelines. Achieving a balance for this trade-off is problematic.\\
On another end, there will be various Metaverse owners, which leads to the problem of interoperability \cite{chen2022cross}. With regards to user-object interactions, owning digital assets raises security and theft concerns due to their exposure to the public internet and management by various operators. Furthermore,  sharing the owned objects across environments is not straightforward and requires some guidelines and regulations. In the same context, guaranteeing accessibility to objects and features by anyone and everyone is mandatory, mandating a set of rules and regulations which have not yet been studied and investigated.\\

Promising potential solutions to the existing limitations as a result of user-to-object interactions are many folds. First, developing on-demand fog and edge computing layers next to users leads to rendering real-time personalized experiences, this accounting for DT requirements. Consequently, fewer network and computing resources are consumed by transferring the load to neighboring servers. Furthermore, it is essential to integrate advanced, fast, and scalable AI techniques to decide on the right time and place to deploy the on-demand fog servers and perform the required resource scheduling \cite{sami2021ai, hammoud2022demand}. In addition, AI solutions can advance the rendering capabilities by pre-rendering interactions with objects by studying users' behaviors and predicting the next needed object manipulation inside the Metaverse. Besides resource management and proactive decision-making, AI has great potential in providing assistance for users to facilitate user-to-object interactions. This includes the role of AI to process voice commands and provide personalized interaction experiences. Furthermore, an improvement in sensory technology combined with the power of AI can improve the accuracy of object tracking to reduce latency in manipulation. Furthermore, combined with Blockchain, AI can be used to create a smart distributed ownership management system, adding an additional layer of security and transaction management mechanism \cite{cao2022decentralized}. Additionally, using AI, utilizing swarm robots is another promising solution for elevating the interactions with objects, where robots can collaborate together to process and handle users' commands and feedback. Furthermore, gamification can be utilized with an AI synergy in order to motivate users through rewards to interact more with the surrounding, thus increasing engagement inside the Metaverse.\\
Achieving real-time synchronization of objects and environment manipulation between the real and physical world, in addition to studying the surroundings in the case of augmented reality requires additional research. Potential research directions consist of developing mechanisms for (1) accurate tracking, (2) real-time mapping, and (3) dynamic object placement \cite{park2020augmented}. For advancing tracking accuracy, techniques such as computer vision and sensor fusion can be used. Moreover, real-time mapping can be achieved using advanced sensors that can capture detailed information about the environment, including Li-DAR and depth cameras.\\
Another promising track of research is the use of generative design to enhance user-to-object interaction \cite{keshavarzi2020v, jennings2022generativr}. Using the generative design, the following features are guaranteed: (1) customization, (2) efficiency, (3) realism, and (4) innovation. In particular, generative models empower personalized user experiences by providing a tool for customizing objects in relation to preferences. Besides, efficiency is guaranteed by optimizing the design of the objects in the Metaverse, leading to less resource utilization for rendering and post-interaction manipulation. In addition, generative models enable realistic object design with an increased margin of innovation by providing a wider range of implementing designs. Therefore, generative design is promising and can push the boundaries of efficiency, realism, and innovation.\\
Furthermore, it is essential to integrate the use of Blockchain, which is integrated and shared across Metaverse environments, where users can easily share and transfer objects ownerships (e.g. using the cross-chain concept). Furthermore, a set of regulations and guidelines must be outlined, shared, and adapted by all Metaverse owners or managers to make sure that all objects and features for available interactions and services are available for everyone and not limited to a group of users.\\
Finally, with the increase in diversity and need for additional object designs, it is essential to share a public repository including large libraries of design and manipulation options. Moreover, the Metaverse can be used to provide training for new users joining and teach them about the existing types of interactions and experiences.


\section{Conclusion}

A Metaverse is a virtual place that allows people to become the masters of their own world and businesses giving the chance to embark on new potentials and widen their visions.
In this survey, we present a detailed study of the Metaverse development ecosystem by devising a novel multi-layered pipeline ecosystem composed of: (1) Infrastructure, (2) Environment Digitization, and (3) User Interaction. 
Furthermore, we present a multi-level classification of each component within each layer in the pipeline with respect to a set of the most recent and state-of-the-art literature work contributing to its development and success. The Infrastructure layer of the pipeline is composed of a study related to the Hardware, XR Frameworks, and Platforms and MaaS. With regards to the Environment Digitization, the components are planned as follows: Avatar Modeling, Rendering, and Session Management. Finally, the User Interaction layer is the application layer of the Metaverse where users can interact with their surroundings, which is composed of User-User Interaction, User-Business Interaction, and User-Object Interactions. The advancements within each component are investigated against a set of technologies and social enablers, including AI, Blockchain, computing, networking and communications, privacy \& security, ethics, sociopsychology, and business aspects. Furthermore, our survey presents a list of existing challenges behind each component, followed by desirable criteria that help the research community devise academic and business-oriented solutions. 

We believe that this survey is the first to comprehensively present the full development ecosystem of the Metaverse against a set of technologies, empowering domains, and social enablers for the academic and business aspects. In addition, our survey is the first that devise future directions for each challenge entailed by the components of the Metaverse ecosystem.

\addcontentsline{toc}{section}{Bibliography}
\bibliographystyle{unsrt}
{\small
\bibliography{references-all}}
\end{document}